\documentclass[opre]{informs3}
\RequirePackage{endnotes}

\OneAndAHalfSpacedXI

\usepackage{booktabs} 
\usepackage[ruled]{algorithm2e} 

\SetAlFnt{\small}
\SetAlCapFnt{\small}
\SetAlCapNameFnt{\small}
\SetAlCapHSkip{0pt}
\IncMargin{-\parindent}

\usepackage{subcaption}

\usepackage[capitalise]{cleveref}

\usepackage{amsmath}
\usepackage{amsfonts}
\usepackage{aligned-overset}
\usepackage{mdframed}
\usepackage{mathtools}
\usepackage{bbm}
\usepackage{comment}
\usepackage{dsfont}
\usepackage[T1]{fontenc}
\allowdisplaybreaks
\usepackage{enumitem}
\setlist[enumerate]{label=(\roman*)}
\usepackage{wrapfig}
\usepackage{xcolor,colortbl}
\usepackage{caption}
\crefname{assumption}{Assumption}{Assumptions}
\Crefname{assumption}{Assumption}{Assumptions}

\usepackage{microtype}
\renewcommand{\endproof}{\hfill$\square$\smallskip}

\usepackage{multirow,array}
\usepackage{tabularx}
\newcolumntype{C}[1]{>{\centering\arraybackslash}p{#1}}
\newcolumntype{L}[1]{>{\raggedright\arraybackslash}p{#1}}
\newcolumntype{R}[1]{>{\raggedleft\arraybackslash}p{#1}}

\usepackage{booktabs}

\renewenvironment{quote}
  {\par\vspace{0.4em}            
   \leftskip=3em           
   \rightskip=3em
   \noindent\ignorespaces}
  {\par\vspace{0.4em}}           

\usepackage{tikz,pgfplots,pgfplotstable,statistics}
\usetikzlibrary{intersections, external, decorations.markings, calc}
\usepgfplotslibrary{fillbetween,groupplots}
\pgfplotsset{
    xlabel style={font=\small},
    ylabel style={font=\small},
    tick label style={font=\footnotesize},
    legend style={font=\scriptsize, inner sep=1pt, row sep=-2pt, column sep=2pt},
    compat=newest
}
\tikzset{mid arrow/.style={
        decoration={markings, mark=at position 0.6 with {\arrow[scale=1.5]{stealth}}},
        postaction={decorate},
    }}
\tikzset{mid arrow short/.style={
        decoration={markings, mark=at position 0.9 with {\arrow[scale=1]{stealth}}},
        postaction={decorate},
    }}
\pgfplotsset{
      myAxisStyle/.style={
        width=4.5cm, height=4.6cm,
        xlabel=$\varepsilon$,
        ylabel near ticks,
        xlabel near ticks,
        grid=both,
        xmin=0,xmax=1,
        legend style={
          font=\scriptsize,
        },
      }
    }
\usepackage{todonotes}
\setuptodonotes{inline}

\definecolor{ETHblue}{RGB}{20,80,240}
\definecolor{ETHpetrol}{RGB}{0,120,148}	
\definecolor{ETHgreen}{RGB}{98,115,19}
\definecolor{ETHbronze}{RGB}{142,103,19}
\definecolor{ETHred}{RGB}{183,53,45}
\definecolor{ETHpurple}{RGB}{163,7,116}	
\definecolor{ETHgray}{RGB}{111,111,111}

\let\argmax\relax
\DeclareMathOperator*{\argmax}{argmax}

\DeclareMathOperator{\diag}{diag}
\newcommand{\ppad}{{\sf PPAD}}

\newcommand{\cardinality}[1]{|#1|}

\newcommand{\naturals}{\mathbb{N}}
\newcommand{\reals}{\ensuremath{\mathbb{R}}}
\newcommand{\posReals}{\reals_{>0}} 
\newcommand{\nonNegReals}{\reals_{\geq0}}

\newcommand{\realVec}[1]{\reals^{#1}}

\newcommand{\nonNegRealVec}[1]{\nonNegReals^{#1}}
\newcommand{\onesVec}[1]{\mathbbm{1}_{#1}}
\newcommand{\zerosVec}[1]{\mathds{O}_{#1}}

\newcommand{\realMat}[2]{\reals^{\cartProd{#1}{#2}}}

\newcommand{\Cb}[1]{C_b(#1)}

\newcommand{\spaceA}{X}
\newcommand{\elA}{x}
\newcommand{\spaceB}{Y}

\newcommand{\spaceProbMeasures}[1]{\mathcal{P}(#1)}

\newcommand{\probMeasureA}{\mu}
\newcommand{\probMeasureASeqEl}[1][n]{\probMeasureA_{#1}}
\newcommand{\probMeasureASeq}{\left(\probMeasureASeqEl\right)_{n\in\naturals}}
\newcommand{\probMeasureASubSeqEl}[1][n]{\probMeasureA_{#1_k}}
\newcommand{\probMeasureASubSeq}{\left(\probMeasureASubSeqEl\right)_{k\in\naturals}}
\newcommand{\dprobMeasureA}{\mathrm{d}\probMeasureA}
\newcommand{\dprobMeasureAArgs}[1]{\dprobMeasureA\left(#1\right)}
\newcommand{\dprobMeasureASeqEl}{\mathrm{d}\probMeasureASeqEl}
\newcommand{\dprobMeasureASeqElArgs}[1]{\dprobMeasureASeqEl\left(#1\right)}

\newcommand{\probMeasureB}{\nu}
\newcommand{\probMeasureBSeqEl}[1][n]{\probMeasureB^{#1}}
\newcommand{\probMeasureBSeq}{\left(\probMeasureBSeqEl\right)_{n\in\naturals}}

\newcommand{\narrowconvergence}{\rightharpoonup}

\newcommand{\wassnorm}{s}
\newcommand{\wasserstein}[1][\wassnorm]{W_{#1}}
\newcommand{\wassersteinCorrespondence}[1][\wassnorm]{w}
\newcommand{\wassersteinOpenCorrespondence}[1][\wassnorm]{\mathring{w}}
\newcommand{\distance}{\mathbf{d}}

\newcommand{\distancePlayer}[1]{\distance^{#1}}

\newcommand{\distancePlayerB}{\distancePlayer{2}}

\newcommand{\distanceArgsPlayer}[3]{\distance^{#1}(#2,#3)}

\newcommand{\distanceArgsPlayerB}[2]{\distanceArgsPlayer{2}{#1}{#2}}
\newcommand{\distanceArgsPlayerI}[2]{\distanceArgsPlayer{i}{#1}{#2}}
\newcommand{\Distance}{D}
\newcommand{\DistancePlayer}[1]{\Distance^{#1}}

\newcommand{\DistancePlayerB}{\DistancePlayer{2}}

\newcommand{\DistancePlayerIMinus}{\DistancePlayer{-i}}

\newcommand{\wassersteinBallOpen}[2][\varepsilon]{\ensuremath{\mathring{\mathcal{B}}_{#1}(#2)}}

\newcommand{\wassersteinBallOne}[2][\varepsilon]{\ensuremath{\mathcal{B}^{\wasserstein[1]}_{#1}(#2)}}
\newcommand{\ambiguitySet}[3]{\ensuremath{\mathcal{B}_{#2}^{#1}(#3)}}
\newcommand{\ambiguitySetOpen}[3]{\ensuremath{\mathring{\mathcal{B}}_{#2}^{#1}(#3)}}   
\newcommand{\ambiguitySetMap}[2]{\ensuremath{\mathcal{B}_{#2}^{#1}}}

\newcommand{\setPlans}[2]{\Gamma\left(#1,#2\right)}

\newcommand{\expectedValue}[2]{\mathbb{E}^{#1}\left[#2\right]}

\newcommand{\game}{\mathcal{G}(\actionSpace, \utilityMap)}

\newcommand{\actionSpace}{\mathcal{A}}
\newcommand{\actionSpacePlayer}[1]{\actionSpace^{#1}}
\newcommand{\surActionSpacePlayer}[1]{\tilde\actionSpace^{#1}_\varepsilon}
\newcommand{\actionSpacePlayerI}{\actionSpacePlayer{i}}
\newcommand{\actionSpacePlayerA}{\actionSpacePlayer{1}}
\newcommand{\actionSpacePlayerB}{\actionSpacePlayer{2}}
\newcommand{\actionSpacePlayerIMinus}{\actionSpacePlayer{-i}}
\newcommand{\actions}{a}
\newcommand{\actionPlayer}[1]{\actions^{#1}}
\newcommand{\actionPlayerA}{\actionPlayer{1}}
\newcommand{\actionPlayerB}{\actionPlayer{2}}
\newcommand{\actionPlayerI}{\actionPlayer{i}}
\newcommand{\actionPlayerIMinus}{\actionPlayer{-i}}
\newcommand{\surActionPlayerIMinus}{\hat{\actions}^{-i}}
\newcommand{\optAction}{\Bar{\actions}}
\newcommand{\optActionPlayer}[1]{\optAction^{#1}}
\newcommand{\strategySpace}{\Delta}
\newcommand{\strategySpacePlayer}[1]{\strategySpace^{#1}}
\newcommand{\strategySpacePlayerA}{\strategySpacePlayer{1}}
\newcommand{\strategySpacePlayerI}{\strategySpacePlayer{i}}
\newcommand{\strategySpacePlayerIMinus}{\strategySpacePlayer{-i}}
\newcommand{\strat}{p}
\newcommand{\stratPlayer}[1]{\strat^{#1}}
\newcommand{\stratPlayerA}{\stratPlayer{1}}
\newcommand{\stratPlayerB}{\stratPlayer{2}}
\newcommand{\stratPlayerI}{\stratPlayer{i}}
\newcommand{\stratPlayerIMinus}{\stratPlayer{-i}}

\newcommand{\optStrat}{\bar{\strat}}
\newcommand{\optStratPlayer}[1]{\optStrat^{#1}}
\newcommand{\optStratPlayerA}{\optStratPlayer{1}}

\newcommand{\optStratPlayerI}{\optStratPlayer{i}}
\newcommand{\optStratPlayerIMinus}{\optStratPlayer{-i}}
\newcommand{\stratWasserstein}{\sigma}
\newcommand{\stratWassersteinPlayer}[1]{\stratWasserstein^{#1}}

\newcommand{\stratWassersteinPlayerB}{\stratWassersteinPlayer{2}}

\newcommand{\stratWassersteinPlayerIMinus}{\stratWassersteinPlayer{-i}}

\newcommand{\securityLevel}{S}
\newcommand{\securityLevelPlayer}[1]{\securityLevel^{i}}

\newcommand{\utilityMap}{\textup{u}}
\newcommand{\utilityPlayer}[1]{\utilityMap^{#1}}
\newcommand{\utilityPlayerI}{\utilityPlayer{i}}
\newcommand{\surUtilityPlayerI}{\tilde\utilityMap^{i}}

\newcommand{\utilityArgsPlayer}[3]{\utilityPlayer{#1}(#2,#3)}
\newcommand{\utilityArgsPlayerI}{\utilityArgsPlayer{i}{\actionPlayerI}{\actionPlayerIMinus}}
\newcommand{\utilityExpMap}{\textup{U}}
\newcommand{\utilityExpPlayer}[1]{\utilityExpMap^{#1}}
\newcommand{\utilityExpPlayerI}{\utilityExpPlayer{i}}

\newcommand{\utilityExpArgsPlayer}[3]{\utilityExpPlayer{#1}(#2,#3)}

\newcommand{\utilityExpArgsPlayerIW}{\utilityExpArgsPlayer{i}{\stratPlayerI}{\stratWassersteinPlayerIMinus}}
\newcommand{\utilityWasserstein}{\Bar{\utilityExpMap}}
\newcommand{\utilityWassersteinPlayer}[1]{\utilityWasserstein^{#1}}
\newcommand{\utilityWassersteinArgsPlayer}[3]{\utilityWasserstein^{#1}(#2,#3)}
\newcommand{\utilityWassersteinArgsPlayerI}{\utilityWassersteinArgsPlayer{i}{\stratPlayerI}{\stratPlayerIMinus}}
\newcommand{\utilityMatA}{\utilityExpPlayer{1}}

\newcommand{\utilityMatI}{\utilityExpPlayer{i}}

\newcommand{\bestResponseMap}{r}

\newcommand{\stratRobustBestResponseMap}{\bestResponseMap_\mathrm{SR}}
\newcommand{\stratRobustBestResponseMapPlayer}[1]{\stratRobustBestResponseMap^{#1}}
\newcommand{\stratRobustBestResponseMapPlayerI}{\stratRobustBestResponseMapPlayer{i}}
\newcommand{\stratRobustBestResponseArgs}[1]{\stratRobustBestResponseMap(#1)}

\newcommand{\abs}[1]{\left|{#1}\right|}

\newcommand{\norm}[1]{\left\lVert #1 \right\rVert}

\newcommand{\cartProd}[2]{{#1}\times {#2}}

\newcommand{\powerSet}[1]{2^{#1}}
\newcommand{\probSpace}[1]{\mathcal{P}(#1)}

\newcommand{\goesto}{\rightarrow}

\def\aMaintain{\textsc{Maintain}}
\def\aDecelerate{\textsc{Decelerate}}
\def\aStop{\textsc{Stop}}
\def\aWait{\textsc{Wait}}
\def\aCross{\textsc{Cross}}

\usepackage{natbib}
 \bibpunct[, ]{(}{)}{,}{a}{}{,}%
 \def\bibfont{\small}%
\EquationsNumberedThrough    

\TheoremsNumberedThrough     
\ECRepeatTheorems  %

\MANUSCRIPTNO{MOOR-0001-2024.00}

\begin{document}


\RUNAUTHOR{Lanzetti, Fricker, Bolognani, Dörfler, Paccagnan}

\RUNTITLE{Strategically Robust Game Theory}

\TITLE{Strategically Robust Game Theory \\ via Optimal Transport}

\ARTICLEAUTHORS{%
\AUTHOR{Nicolas Lanzetti}
\AFF{Automatic Control Laboratory,
ETH Zürich, \EMAIL{lnicolas@ethz.ch}}

\AUTHOR{Sylvain Fricker}
\AFF{Automatic Control Laboratory,
ETH Zürich, \EMAIL{sylvain.fricker@protonmail.com}}

\AUTHOR{Saverio Bolognani}
\AFF{Automatic Control Laboratory,
ETH Zürich, \EMAIL{bsaverio@ethz.ch}}

\AUTHOR{Florian Dörfler}
\AFF{Automatic Control Laboratory,
ETH Zürich, \EMAIL{dorfler@ethz.ch}}

\AUTHOR{Dario Paccagnan}
\AFF{Department of Computing,
Imperial College London, \EMAIL{d.paccagnan@imperial.ac.uk}}

} 

\ABSTRACT{%

In many game-theoretic settings, agents are challenged with taking decisions against the uncertain behavior exhibited by others. Often, this uncertainty arises from multiple sources, e.g., incomplete information, limited computation, bounded rationality. While it may be possible to guide the agents’ decisions by modeling each source, their \emph{joint} presence makes this task particularly daunting. Toward this goal, it is natural for agents to seek protection against deviations around the emergent behavior itself, which is ultimately impacted by all the above sources of uncertainty. To do so, we propose that each agent takes decisions in face of the worst-case behavior contained in an ambiguity set of tunable size, centered at the emergent behavior so implicitly defined. This gives rise to a novel equilibrium notion, which we call \emph{strategically robust equilibrium}. Building on its definition, we show that, when judiciously operationalized via optimal transport, strategically robust equilibria (i) are guaranteed to exist under the same assumptions required for Nash equilibria; (ii) interpolate between Nash and security strategies; (iii) come at no additional computational cost compared to Nash equilibria. Through a variety of experiments, including bi-matrix games, congestion games, and Cournot competition, we show that strategic robustness protects against uncertainty in the opponents’ behavior and, surprisingly, often results in higher equilibrium payoffs - an effect we refer to as \emph{coordination via robustification}.

}%





\maketitle


\section{Introduction}\label{sec:introduction}

{\let\thefootnote\relax\footnote{

The first two authors contributed equally to this work. 

This work was supported as a part of NCCR Automation, a National Centre of Competence in Research, funded by the Swiss National Science Foundation (grant number 51NF40\_225155). D. Paccagnan was supported by the EPSRC grant EP/Y001001/1, funded by the International Science Partnerships Fund and UKRI, and by the MIT-Imperial Seed Fund.}}%
Building upon von Neumann's and Nash's seminal work, game theory has emerged as a set of tools for studying decision-making in the presence of strategic agents, especially in non-cooperative scenarios.
%
Its foundations have had a lasting impact on the field of Operations Research, driving advances across core areas such as transportation and network optimization \citep{cominetti2009impact,gairing2023congestion}, 
economics \citep{hobbs2007nash,sherali1983stackelberg},
hierarchical decision making \citep{luo1996mathematical,facchinei1999smoothing},
mechanism design and auctions \citep{balseiro2024mechanism,han2025optimal},
revenue management \citep{netessine2005revenue},
security \citep{bagchi2014optimal}, and supply chain \citep{ha2011sharing}, to name a few.  
%
In many of these settings, agents find themselves needing to make decisions while being \emph{uncertain} about what behavior of the others to expect. 
Such uncertainty is often the results of individual factors including limited information on the game being played~\citep{Harsanyi1967BayesianGamesPart1, Harsanyi1968BayesianGamesPart2}, on the agents' rationality~\citep{Rosenthal1989BoundedRationalityEquilibrium, Stahl1995ModelsOtherPlayers}, on computational capabilities~\citep{horvitz1987computationConstraints}, and many others. Naturally, each and any of these factors has received significant attention in the literature, for example through the lens of behavioral game theory~\citep{camerer2011behavioral} or with the development of algorithmic game theory \citep{roughgarden2010algorithmic}.
However, real-world settings often entail the \emph{joint} presence of many such sources of uncertainty, making it difficult -- if not impossible -- to adopt a ``white-box'' approach for their modeling. 
%
In these settings, it is natural for the agents to seek decisions that protect themselves against deviations around the emergent behavior \emph{itself}, rather than trying to explicitly model each and every type of uncertainty at the source.%
\medskip

\noindent
In this work, we take this point of view. 
We do so by proposing that each agent takes decisions in the face of the worst-case behavior others could exhibit within a suitably-defined ambiguity set, which is itself centered around the emergent behavior that is so implicitly defined. 
In doing so, we construct the ambiguity sets to contain all behaviors that are not too dissimilar, in a sense to be defined later, from the emergent one, up to a \emph{tunable} threshold. We refer to this novel concept as \emph{strategically robust equilibrium}.
As we will show in this work, when judiciously operationalized, strategically robust equilibria have three appealing features: (i) They interpolate in a principled way between Nash equilibria and security strategies; (ii) They come at no additional computational cost; (iii) They result in more robust decisions %
which often produce equilibrium payoffs that are higher than those received in the nominal model for all agents.
For a detailed discussion on how our work connects to the literature, we refer the reader to~\cref{subsec:related work}.
%
%
%
\medskip

\noindent
In shaping the ambiguity sets, we observe that the emergent behavior is described through a collection of mixed strategies, that is, a probability distribution over the joint action space. 
It is therefore convenient to define ambiguity sets as sets of probability distributions that are not too dissimilar from the emergent behavior, up to a tunable threshold, as measured through a notion of ``similarity'' between probability distributions.
While many such notions exist, our choice is guided by the following three principles. 
First, we wish ambiguity sets to be \emph{highly expressive}. In particular, we require ambiguity sets to contain non-parametric families of distributions mixing actions not used at the emergent behavior. This allows to protect agents against out-of-equilibrium play. 
Second, we require ambiguity sets to \emph{grow monotonically} as a function of the chosen threshold, to collapse to their center when such threshold is set to zero, and to inflate to the full probability space when the threshold approaches infinity. This allows strategically robust equilibria to interpolate, in a principled way, between Nash equilibria and security strategy depending on the chosen threshold, which controls the level of robustness.
Third, we wish to employ ambiguity sets that, while protecting against out-of-equilibrium play, do not do so at the price of significantly increasing their \emph{computational complexity}. This allows strategically robust equilibria to be of practical relevance. 
\medskip

\noindent
Guided by these three principles, we propose to model ambiguity sets via \emph{optimal transport}~\citep{villani2009optimal}. Specifically, we model ambiguity sets as balls of mixed strategies constructed using an optimal transport distance between probability distributions, e.g., the celebrated Wasserstein distance, and centered at the emergent behavior. Interestingly, such balls embody all the above principles. Indeed, they (i) contain non-parametric families of distributions whose support is not limited to their center \citep{villani2009optimal,kuhn2019wasserstein}, (ii) grow monotonically from a singleton to the full set of probability distributions as a function of their radius, and (iii) result in strategically robust equilibria whose computation, as we will unveil in this work, is no harder than that of Nash equilibria.
We are not aware of another notion of similarity for probability distributions with these properties. For instance, the Kullback-Leibler divergence does not allow to compare mixed strategies that do not share the same support, leading to equilibria that do not protect against actions that are not mixed at the equilibrium itself. Ambiguity sets based on (finitely many) moments 
do not collapse to their center when their radius is set to zero, and thus can not be used to recover Nash equilibria. Moreover, we wish not to use ambiguity sets based on parametric families of distributions, e.g., by contemplating $\varepsilon$-deviations to a uniform distribution, due to their poor expressivity and dubious motivation.
\medskip

\noindent
With this formulation, strategically robust equilibria correspond to fixed points of optimal-transport-based distributionally robust optimization problems with \emph{decision-dependent} ambiguity sets, which raises significant analytical and computational challenges that we untangle in the ensuing section.%
\subsection{Our Results}
As anticipated, at the core of our work is a novel equilibrium notion, termed \emph{strategically robust equilibrium}, to tackle the commonly-encountered scenario where decision-makers are unsure about each other's behavior. Our most central contributions are discussed below.

\begin{mdframed}[hidealllines=true,backgroundcolor=blue!5,skipabove=2.5ex]
\OneAndAHalfSpacedXI
\textbf{C1)}
    First, we define strategically robust equilibria as fixed points of a best response map where each agent best responds to the worst-case behaviors others could exhibit within an 
    ambiguity set centered around the equilibrium itself (\cref{def:SRE}).
    Second, we show the existence of strategically robust equilibria under minimal regularity assumptions on the cost functions, action sets, and ambiguity sets (\cref{thm:existence}).
    We do so in general settings, i.e., we do not specify the ``shape'' of the ambiguity sets, nor require action sets to be finite. 
    Third, we leverage the previous result to show that strategically robust equilibria based on optimal transport exist under the same assumptions ensuring existence of mixed Nash equilibria~(\cref{corollary:existence:optimal_transport}). 
\end{mdframed}

\vspace{1.5ex}

\noindent
In order to achieve these objectives, we need to overcome a number of technical challenges. At its heart, these challenges stem from the fact that the agents' ambiguity sets are \emph{decision-dependent}, i.e., they depend on the decisions taken by all other agents -- a challenge compounded by the fact that we do not require action sets to be finite. Therefore, the existence of fixed points of the best response map (and thus equilibria) hinges on carefully analyzing the properties of this map. 
For general ambiguity sets whose ``shape'' is not specified, we show that hemicontinuity of the ambiguity sets with respect to their center is sufficient to conclude, therefore providing a result that may guide others in studying alternative ambiguity sets. 
Interestingly, however, we then show that such hemicontinuity property is \emph{always} satisfied by optimal transport ambiguity sets --  a result we believe may be of independent interest. The upshot of our first contributions can thus be summarized as follows:
\begin{quote}
\emph{Existence of strategically robust equilibria based on optimal transport is guaranteed under the very same assumptions that ensure existence of mixed Nash equilibria.}
\end{quote}
%

\noindent
Building on this, we then focus on finite action games, addressing two key algorithmic questions. 

\begin{mdframed}[hidealllines=true,backgroundcolor=blue!5,skipabove=2.5ex]
\OneAndAHalfSpacedXI
\textbf{C2)}
For finite action games,
    we (i) settle the computational complexity of strategically robust equilibria and (ii) provide computation tools. Regarding (i), we show that the problem of computing strategically robust equilibria \emph{belongs} to the \ppad{} class, and therefore is no-harder than that of computing a mixed
 Nash equilibrium~(\cref{th:computationalComplexity}). 
    Regarding (ii), we show that computing strategically robust equilibria in $N$-player games amounts to solving a multilinear complementarity problem, which reduces to a linear complementarity problem in the case of 2 players (\cref{prop:finite:computation:N players}). 
    In either case, we can therefore deploy existing computational methods, also used for mixed Nash equilibria, to compute them. 
\end{mdframed}

\vspace{1.5ex}

\noindent
Settling the complexity of strategically robust equilibria and proposing tools for their computation is challenging for at least two reasons:  (i) optimal transport ambiguity sets contain infinitely many behaviors, which effectively prohibits the use of enumeration techniques, and (ii) the mere evaluation of the optimal transport distance between two mixed strategies entails solving a transportation problem. These observations seem to suggest that, from a computational standpoint, strategically robust equilibria may be more difficult to compute than mixed Nash equilibria. To the contrary, our second main result shows that this is \emph{not} the case. We prove this claim and address the above-mentioned challenges by leveraging duality theory to show that computing strategically robust equilibria is equivalent to computing Nash equilibria for a suitably defined concave game, jointly with recent results on the complexity of this class \citep{papadimitriou2022computational}. Finally, such dual reformulation also paves the way to equilibrium-computing algorithms akin to those used for mixed Nash equilibria. This is achieved by casting strategically robust equilibria as solutions to multilinear complementarity problems. In summary, our second contribution shows that: 
\begin{quote}
\emph{Strategically robust equilibria based on optimal transport belong to the same complexity class of Nash equilibria, and similar algorithms can be deployed for their computation.}
\end{quote}

\noindent
We then consider concave games, a celebrated class of continuous action games introduced in \cite{rosen1965existence} whose applicability spans multiple domains, e.g., Cournot competition.
%

\begin{mdframed}[hidealllines=true,backgroundcolor=blue!5,skipabove=2.5ex]
\OneAndAHalfSpacedXI
\textbf{C3)} For concave games, we first show that a \emph{pure} strategically robust equilibrium based on optimal transport exists (\cref{thm:continuous:existence pure strategically robust eq}). We note that this is a significantly stronger statement, and that, while being pure, such equilibria remain strategically robust against \emph{mixed} deviations.
As a consequence, we show that their computation is equivalent to that of a Nash equilibrium for a surrogate concave game with augmented decision space and suitably modified payoffs (\cref{prop:equivalent game}). 
    Finally, we specialize this result to quadratic games and showcase how 
    strategic robustness has an effect similar to that of \emph{regularizing} the agents' nominal payoffs. 
\end{mdframed}

\vspace{1.5ex}

\noindent
The continuous setting presents a number of unique challenges that add to those described above. Central to them is the fact that the set of probability distributions supported on the joint action space is now \emph{infinite} dimensional. Notably, the problem remains infinite-dimensional even when, motivated by applications, we focus on \emph{pure} equilibria for this class of problems.
This is because pure strategically robust equilibria are required to be robust against \emph{any} distribution and not just against \emph{pure} deviations. We overcome this difficulty by building upon recent results in distributionally robust optimization, which show that the worst-case distribution centered in a pure strategy is obtained by mixing finitely many pure strategies. This allows us to reduce the problem to finite dimension, and thus reformulate strategically robust equilibria as Nash equilibria of a concave game. In these settings, we also show that strategic robustness can be interpreted as an additional term that regularizes the agents' nominal payoff. In summary, the resounding message is that:
\begin{quote}
\emph{In concave games, pure strategically robust equilibria based on optimal transport exist, their computation can be made finite-dimensional, and the effect of strategic robustness is akin to that of regularizing the agents' nominal payoff.}
\end{quote}

\noindent
Finally, across a variety of experiments in discrete and continuous settings -- including classical bi-matrix games, congestion games, and Cournot competition -- we demonstrate that strategically robust equilibria (i) protect against the uncertain behavior exhibited by others, and (ii) consistently lead to higher payoffs for \emph{all agents}, underpinning a surprisingly ``\emph{coordination via robustification}'' effect.
\medskip

\noindent
In summary, our work shows that strategically robust equilibria allow decision-makers to achieve any desired level of robustness against deviations from the equilibrium play, and hence ripe the corresponding benefits, at no additional computational cost compared to (mixed) Nash equilibria.

\subsection{An Illustrative Example}\label{sec:motivating example}
\begin{figure}[tb]
\begin{minipage}[t]{0.49\textwidth}
    \ 
    
    \centering
    \includegraphics[width=0.65\textwidth]{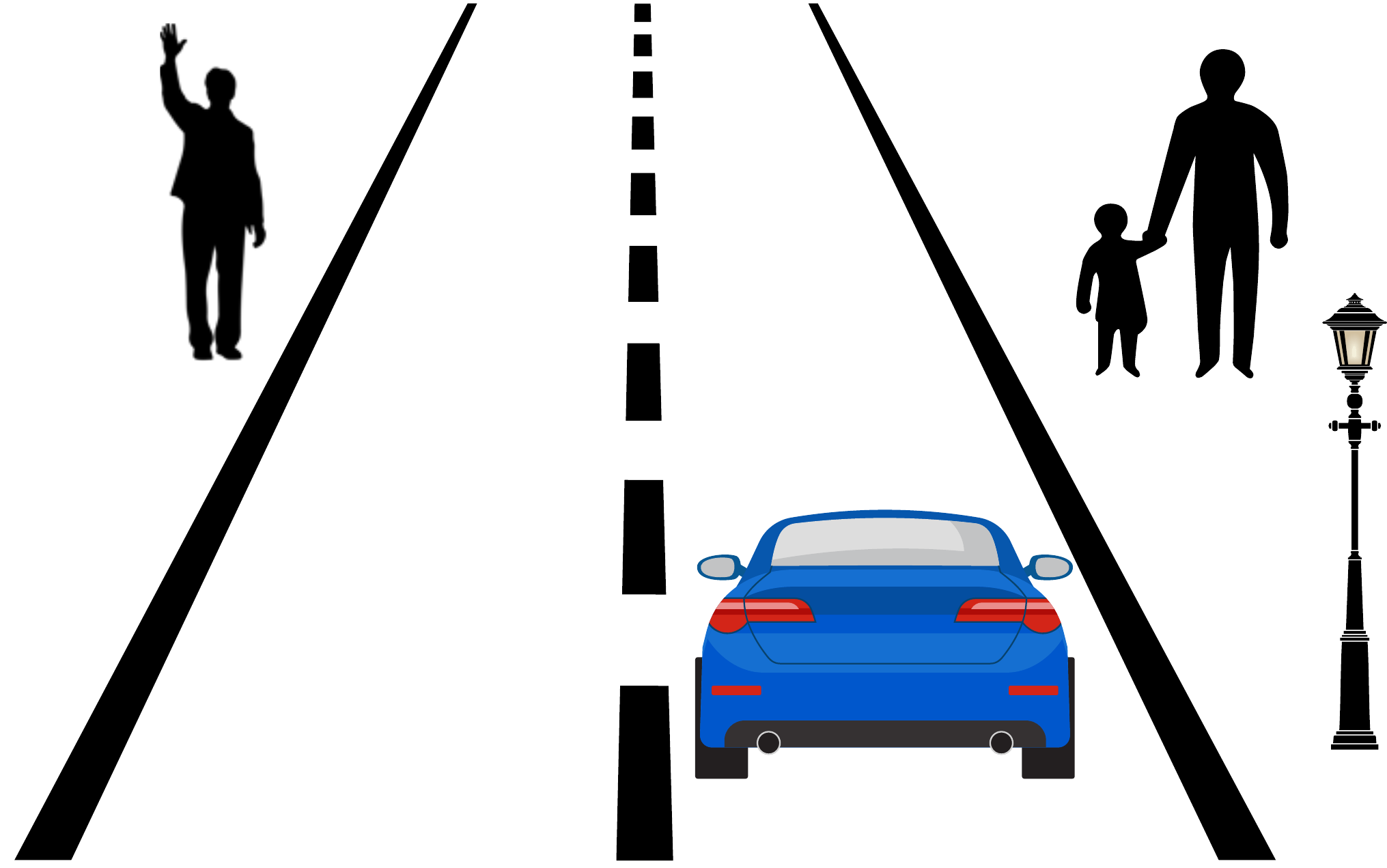}

    \medskip
    
    \setlength{\extrarowheight}{3pt}
    \begin{tabular}{C{1.8cm}|C{1.6cm}|C{1.6cm}|}
          \multicolumn{1}{c}{} & \multicolumn{1}{c}{\footnotesize \aWait}  & \multicolumn{1}{c}{\footnotesize \aCross} \\[3pt]
          \cline{2-3}
          \footnotesize \aMaintain & \footnotesize$(10,-1)$ & \footnotesize$(-50,-100)$ \\[3pt]
          \cline{2-3}
          \footnotesize \aDecelerate & \footnotesize$(9,-1)$ & \footnotesize$(-5,-10)$ \\[3pt]
          \cline{2-3}
          \footnotesize \aStop & \footnotesize$(0,-1)$ & \footnotesize$(0,10)$ \\[3pt]
          \cline{2-3}
    \end{tabular}
    \caption{A family comprised of parent and daughter are about to cross the street as an autonomous vehicle approaches. We represent this as a bi-matrix game.}
    \label{table:finite:pedestrian}
\end{minipage}
\hfill 
    \begin{minipage}[t]{0.49\textwidth}
        \centering
        \pgfplotstableread[col sep=comma]{results/pedestrian.csv}\datatable
        \begin{tikzpicture}[baseline=(current bounding box.north)]
            \begin{axis}[
            width=7cm, height=4.5cm,
            xlabel={\footnotesize Probability that the family crosses},
            ylabel={\footnotesize Vehicle's payoff},
            y label style={at={(-0.12,0.5)}},
            grid=both,
            scaled ticks=false,
            legend style={draw=none,fill=none},
            xmin=0, ymin=-10, xmax=0.2, ymax=25,
            xtick={0, 0.05, 0.1, 0.15, 0.2}, 
            xticklabels={0, 0.05, 0.1, 0.15, 0.2}, 
            ]
            \addplot[ETHbronze,ultra thick] table[x=f, y expr=\thisrow{meanNash}] {\datatable};
            \addplot[name path=us_top,ETHbronze,thin,draw=none,forget plot] table[x=f, y expr=\thisrow{nashPlus}] {\datatable};
            \addplot[name path=us_down,ETHbronze,thin,draw=none,forget plot] table[x=f, y expr=\thisrow{nashMinus}] {\datatable};
            \addplot[fill=ETHbronze!30,fill opacity=0.3,forget plot] fill between[of=us_top and us_down];
            \addplot[ETHpetrol, ultra thick] table[x=f, y expr=\thisrow{meanSecure}] {\datatable};
            \addlegendentry{NE (\textsc{Maintain}, \textsc{Wait})}
            \addlegendentry{Security strategy \textsc{Stop}}
        \end{axis}
        \end{tikzpicture}
        \caption{Vehicle's payoff attained by maintaining its speed (Nash equilibrium strategy \aMaintain), as a function of the probability that the family crosses the road (deviating from their Nash equilibrium strategy \aWait). The shaded area represents the expected payoff plus/minus one standard deviation to illustrate the risk that the agent is taking.
        The security strategy \aStop{} is shown as a robust baseline.}
        \label{fig:pedestrianGame:intro}
    \end{minipage}
     \vspace*{-4mm}
\end{figure}

Consider the setting of~\cref{table:finite:pedestrian} where a family comprised of father and daughter wishes to cross the street as an autonomous vehicle approaches.  
The vehicle, which seeks to move forward while avoiding a collision, has three options: \aMaintain{} its current speed (M), \aDecelerate{} (D), and \aStop{} completely (S). Similarly, father and daughter, 
have two available actions: \aWait{} (W) and \aCross{} (C). In total, there are therefore six possible outcomes, and we report the agents' payoffs in the table in~\cref{table:finite:pedestrian}.
The vehicle receives a positive reward when safely moving ahead, corresponding to the outcome (M, W). This reward diminishes if deceleration is needed, i.e., in outcome (D, W), and vanishes when stopping completely (\aStop{} row). 
The vehicle incurs a high cost when the collision occurs at a high speed, corresponding to (M, C). This cost considerably diminishes at lower speeds since emergency maneuvers remain feasible, corresponding to (D, C). Similarly, the father and the daughter receive a positive reward when safely crossing, experience a small cost when waiting, and incur an increasingly high cost when colliding, depending on the vehicle's speed.

The resulting game exhibits three Nash equilibria: (M, W), (S, C), and a mixed one. 
From the standpoint of the autonomous vehicle, the Nash equilibrium (M, W) is the most attractive. However, this Nash equilibrium lacks robustness and immediately deteriorates to the tragic outcome $(-50,-100)$ if the pedestrians cross the road, even if that happens with low probability (\textcolor{ETHbronze}{brown} curve in~\cref{fig:pedestrianGame:intro}). 
From the vehicle's perspective, this can model the chance of encountering a careless pedestrian. 
Similarly, the Nash equilibrium (S, C) ceases to be attractive if there is a chance that the vehicle does not stop, which may be caused by other sources of uncertainty, e.g., sensors' misdetection. The mixed Nash equilibrium displays a similar behavior, see Appendix~\ref{app:pedestrian game:additional results}.
Alternatively, the decision-makers might opt for the security strategies (S, W). However, while this configuration is robust against any deviation, it results in a gridlock where neither the family nor the vehicle make any progress (\textcolor{ETHpetrol}{petrol} curve in~\cref{fig:pedestrianGame:intro}).
Interestingly, a natural outcome such as (D, W), which drastically increases robustness for both agents at a small cost in terms of performance, is ruled out a priori by both Nash equilibria and security strategies. In \cref{subsec:finite:example}, we will show how this natural outcome is captured as a strategically robust equilibrium for an adequate level of robustness.

\subsection{Related Work}\label{subsec:related work}

Our work connects with several research streams in the literature, albeit with significant differences.%

\paragraph{Nash equilibria and security strategies}
While building upon the notions of \emph{Nash equilibrium} \citep{Nash1951EquilibriaNPlayerGames} and \emph{security strategies} \citep{von1947theory}, our approach recognizes that, oftentimes, such notions are either \emph{not robust} or \emph{too robust} to deviations from the equilibrium play. In the former case, agents may receive payoffs that are significantly lower than those at the purported Nash equilibrium since no protection to out-of-equilibrium play is incorporated. In the latter case, by protecting against \emph{any} play, security strategies may result in conservative decisions, which, ultimately, reduce the attainable payoff (see the illustrative example in~\cref{sec:motivating example}).
Contrary to that, strategically robust equilibria, which we propose here, allow us to interpolate between these two extremes.  This is achieved by controlling the size of the ambiguity sets through a single parameter, which can be used to tune the level of robustness depending on the specific application. 
%

\paragraph{Distributionally robust optimization}
In the setting of distributionally robust optimization, a single decision-maker seeks a decision that performs well under a class of probability distributions~\citep{kuhn2024distributionally,gilboa1989maxmin}.
While this is an area that has seen a surge of interest in recent years, thanks also to recent computational breakthroughs (e.g., see~\citet{blanchet2019quantifying,mohajerin2018data,gao2023distributionally}) and its link to regularization (e.g., see~\citet{gao2024wasserstein,shafieezadeh2019regularization}), our line of work is fundamentally different. Chiefly, in that strategically robust equilibria model \emph{multiagent} decision-making problems, where decision-makers require a desired degree of robustness against \emph{one another's behavior}.
Nonetheless, the proof of some of our technical results builds on existing duality results for distributionally robust optimization.

\paragraph{Behavioral game theory}
Motivated by the mismatch between predicted behavior 
and experimental data, behavioral game theory proposes alternative models to describe the agents’ decision-making, with celebrated examples including Camerer's cognitive hierarchy~\citep{camerer2004cognitive}, imperfect equilibria~\citep{beja1992imperfect}, and many others for which we refer the reader to~\citep{camerer2011behavioral}. Perhaps the most commonly encountered equilibrium concept introduced by this line of work is the so-called quantal response equilibrium, which is based on softening the optimality requirement so that each pure strategy is chosen with a probability positively related to its payoff, depending on a temperature parameter~\citep{mckelvey1995quantal}. Albeit related, such models significantly differ from our approach in that they still do \emph{not} protect against deviations from the corresponding equilibrium distribution, which, instead, is a distinguishing feature of our approach. Additionally, while quantal response equilibria interpolate between Nash equilibria and uniformly random responses, strategic robust equilibria crucially interpolate between Nash equilibria and \emph{security strategies}.


\paragraph{Other equilibrium notions}
There exist many notions of equilibrium geared towards incorporating robustness in game theory. Some of them refine the set of Nash equilibria, effectively discarding the equilibria that fail to satisfy some given conditions. Notable examples are trembling-hand (or perfect) equilibria~\citep{Selten1975PerfectionConcept}, proper equilibria \citep{myerson1978refinements}, and strategic stability properties~\citep{kohlberg1986strategic}. 
Contrary to these approaches, our work does \emph{not} advocate for the selection of an equilibrium over another, but rather constructs possibly different equilibria depending on the size of the ambiguity set. In particular, note that strategically robust equilibria need not be Nash equilibria (see example in \cref{subsec:finite:example}).
%
Another line of work, instead, introduces different equilibrium notions altogether, with examples ranging from Nash equilibria under uncertainty~\citep{dow1994nash,marinacci2000ambiguous} to minimax regret equilibria \citep{renou2010minimax}. 
Amongst these, interestingly, \citet{bich2019strategic} introduces prudent equilibria, defined as Nash equilibria of a modified game with regularized payoff.
{In this context, our work provides a principled approach to obtain regularized payoffs as a result of imposing strategic robustness via optimal transport. Beyond, it departs from \cite{bich2019strategic}, which focuses only on games with continuous actions.} 
%
Our work is also related to risk-aware game theory, which originates in prospect theory~\citep{kahneman2013prospect}. In these settings, agents do not optimize for the expected payoff but rather choose a different risk measure. For instance, \citet{yekkehkhany2020risk} defines an equilibrium as a point at which each agent maximizes the probability of achieving the largest reward and, in the setting of Markov games,~\citet{mazumdar2025tractable} combines bounded rationality and risk aversion to define risk-averse quantal response equilibria -- a class of equilibria that are computationally tractable when players have sufficient degrees of risk-aversion and bounded rationality. 
Strategically robust equilibria distinguish themselves from this line of work by describing agents that still consider the expected payoff but do so with respect to the mixed strategy representing the worst deviation within an ambiguity set centered at the equilibrium implicitly defined.
Finally, \citet{ganzfried2023safe} defines safe equilibria, whereby each player responds to opponents that behave rationally with a specified probability and adversarially with the remaining probability. This cannot lead to robustness against out-of-equilibrium play, which is instead the key distinguishing factor of strategically robust equilibria. Moreover, safe equilibria are defined only for two-players finite-action games and extensions to $N$-players games need to designate an artificial ``main'' player.

\paragraph{Games with exogenous uncertainty}
There is a vast literature on game-theoretic models where agents' payoffs are affected by \emph{exogenous} uncertainty, i.e., uncertainty which does not depend on the agents' decisions. If the agents have a belief for this uncertainty and maximize the resulting \emph{expected} payoff, we obtain Bayesian games~\citep{Harsanyi1967BayesianGamesPart1, Harsanyi1968BayesianGamesPart2}. If the agents, instead, adopt a \emph{worst-case} approach, we obtain robust games~\citep{crespi2017robust,crespi2025insights,aghassi2006robust,perchet2020finding} which seek for equilibria that are robust to the worst-case realization of a given \emph{exogenous} parameter defining the agents' payoffs. For instance, in \citet{aghassi2006robust}, a tuple of strategies is an equilibrium if each agent’s strategy is a best response to the other agents’ strategies, under the worst-case realization of the uncertain parameter. To reduce conservatism, robust equilibria were recently extended to distributionally robust equilibria~\citep{loizou2015distributionally,qu2017distributionally,liu2018distributionally}, whereby equilibrium strategies need to be robust to all distributions of the uncertain parameter contained in a prescribed ambiguity set. Strategically robust equilibria, which we introduce here, depart from these lines of work as they account for and protect against \emph{endogenously}-generated uncertainty, a distinguishing feature of our approach, not present in the above works.%

\paragraph{Games with subjective utility preferences.}
Our work departs from the literature on games with subjective utility preferences, whereby agents maximize an expected payoff based on possibly misspecified beliefs about the game, which results in equilibrium notions such as self-confirming equilibria~\citep{fudenberg1993self} and Nash-Berk equilibria~\citep{esponda2016berk}. In this context, a set of mixed strategies and of beliefs are at equilibrium if, in addition to the mixed strategies being a best response to the given belief, the agents' belief is consistent with data on the observed play. Strategically robust equilibria differ significantly from this setting in that their definition does not involve introducing a notion of belief, and, crucially, in that it models agents that seek protection from deviations in the equilibrium play.
\section{Strategically Robust Game Theory}\label{sec:strategically_robust_game_theory}

We consider a game with $N$ agents, where each agent $i$ is endowed with actions $\actionPlayerI{}\in\actionSpacePlayerI$, and aims at maximizing a payoff function $\utilityMap^i: \actionSpacePlayer{1}\times\dots\ \times\actionSpacePlayer{N}\rightarrow\mathbb{R}$ depending on all agents' choices. In doing so, we do not require action spaces to be finite, i.e., $\actionSpacePlayer{i}$ might well represent a subset of, e.g., $\mathbb{R}^{n}$. We denote this game by $\game$. Further, we let $\stratPlayerI$ be a mixed strategy for agent $i$, i.e., a probability distribution over $\actionSpacePlayerI$, and denote by $\strategySpacePlayerI=\probSpace{\actionSpacePlayerI}$ the space of mixed strategies, i.e., the probability space over $\actionSpacePlayerI$.
We denote with $\stratPlayerIMinus$ a collection of mixed strategies for all agents but~$i$. Similarly, $\actionSpacePlayerIMinus$ denotes the Cartesian product of the action sets of all agents but $i$.
    %
\medskip

\noindent
As previously discussed, given a mixed strategy profile $(\stratPlayerI,\stratPlayerIMinus)$, each agent $i$ wishes to protect itself against mixed strategies that other agents could select in the vicinity of $\stratPlayerIMinus$, as their behavior is uncertain. Toward this goal, agent $i$ constructs an ambiguity set $\ambiguitySet{i}{\varepsilon}{\stratPlayerIMinus}$ around $\stratPlayerIMinus$ in the space of probability distributions supported on $\actionSpacePlayerIMinus$ whose size is controlled by parameter $\varepsilon\in\mathbb{R}_{\ge0}$.\footnote{While we have used a single parameter $\varepsilon$ to control the size of the ambiguity sets across \emph{all} agents, our results extend readily to the case where $\varepsilon$ is agent-dependent. We do not pursue this direction, to ease the presentation.}
Concrete ways to construct this ambiguity set will be discussed later. However, they will ensure that if $\varepsilon=0$ the ambiguity set only contains $\stratPlayerIMinus$ (no ambiguity), while if $\varepsilon\rightarrow\infty$, the ambiguity set contains all probability distributions.
Each agent then selects a strategy that maximizes its expected payoff \mbox{in face of the worst deviation others could exhibit within this ambiguity set, as formalized next.}
%
\begin{definition}[Strategically Robust Equilibrium]\label{def:SRE}
Given a game $\game$, a strategy profile $(\optStratPlayer{1}, \dots, \optStratPlayer{N})$ is a \emph{strategically robust equilibrium} with robustness level $\varepsilon\in\nonNegReals$ if for all $i\in\{1,\ldots,N\}$ we have 
\begin{equation}
\label{eq:StrategicallyRobustEquibrium}
    \optStratPlayerI
    \in 
    \argmax_{\stratPlayerI \in \strategySpacePlayerI} \min_{\stratWassersteinPlayerIMinus \in \ambiguitySet{i}{\varepsilon}{\optStratPlayerIMinus}} \utilityExpArgsPlayerIW,
\end{equation}
where $\utilityExpArgsPlayerIW \coloneqq \expectedValue{(\actionPlayerI,\actionPlayerIMinus)\sim(\stratPlayerI,\stratWassersteinPlayerIMinus)}{\utilityArgsPlayerI}$ is the expected payoff of agent $i$.%
\footnote{Note that, at this point, $\stratWassersteinPlayerIMinus$ appearing in \eqref{eq:StrategicallyRobustEquibrium} represents a general distribution on the action space $\actionSpacePlayerIMinus$. As such, $\stratWassersteinPlayerIMinus$ can be correlated, i.e., it needs not be obtained by the product of independent distributions. We hence use the Greek letter $\stratWassersteinPlayerIMinus$ to distinguish this from $\stratPlayerIMinus$ introduced earlier. We will further discuss this aspect in~\cref{remark:product ambiguity sets}.}
\end{definition}

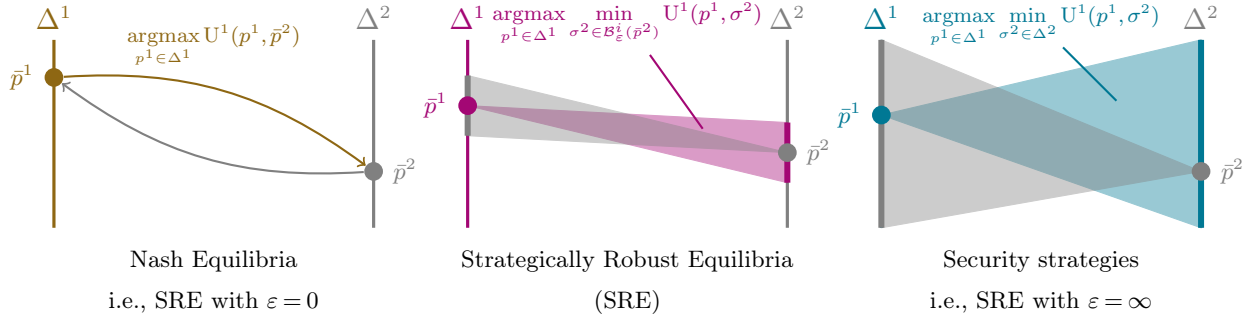
\begin{figure}[t!]
    \tikzexternaldisable
    \centering
    \newcommand{\axisSep}{4.25cm}
\newcommand{\panelSep}{5.5cm}
\newcommand{\myheight}{2.5}

\tikzstyle{mypoint}=[circle, fill, inner sep=2.5pt]
\tikzstyle{myaxis}=[axis, very thick]
\tikzstyle{myline}=[-,line width=0.08cm]

\begin{tikzpicture}[
    axis/.style = {thick},
    every label/.append style = {font=\footnotesize,label distance=0pt}
]

\begin{scope}[shift={(0,0)}]
  \draw[myaxis, ETHbronze] (0,0) -- (0,\myheight) node[above] {$\strategySpacePlayerA$};
  \draw[myaxis, gray] (\axisSep,0) -- (\axisSep,\myheight) node[above] {$\strategySpacePlayer{2}$};

  \node[mypoint, ETHbronze,label={west:\textcolor{ETHbronze}{$\optStratPlayer{1}$}}] (A1) at (0,0.8*\myheight) {};
  \node[mypoint, gray,label={east:\textcolor{gray}{$\optStratPlayer{2}$}}]  (B1) at (\axisSep,0.3*\myheight) {};

  \draw[->, ETHbronze, thick] (A1) to[bend left=20] (B1);
  \draw[->, gray, thick] (B1) to[bend left=20] (A1);

  \node[anchor=south, ETHbronze] (C)
    at (\axisSep/2,2.0) {\scriptsize
      $\displaystyle
      \argmax_{\stratPlayerA \in \strategySpacePlayerA} \utilityExpArgsPlayer{1}{\stratPlayerA}{\optStratPlayer{2}}$
    };

  \node[align=center, below=5pt, font=\footnotesize] at (\axisSep/2,0) {
    Nash Equilibria\\
    i.e., SRE with $\varepsilon = 0$
  };
\end{scope}

\begin{scope}[shift={(\panelSep,0)}]
    \draw[myaxis, ETHpurple] (0,0) -- (0,\myheight) node[above] {$\Delta^1$};
    \draw[myaxis, gray]    (\axisSep,0) -- (\axisSep,\myheight) node[above] {$\Delta^2$};
    
    \node[coordinate]    (A) at (0,0.65*\myheight) {};
    \node[coordinate]    (A1) at ($(A)+(0,0.4)$) {};
    \node[coordinate]    (A2) at ($(A)+(0,-0.4)$) {};
    
    \node[coordinate]    (B) at (\axisSep,0.4*\myheight) {};
    \node[coordinate]    (B1) at ($(B)+(0,0.4)$) {};
    \node[coordinate]    (B2) at ($(B)+(0,-0.4)$) {};
    
    \draw[myline, ETHpurple] (B1) -- (B2);
    \draw[myline, gray] (A1) -- (A2);

    \draw[fill=ETHpurple, draw=ETHpurple, opacity=0.4] (A) -- (B1) -- (B2) -- cycle;
    \draw[fill=gray, draw=gray, opacity=0.4] (B) -- (A1) -- (A2) -- cycle;

    \node[mypoint, ETHpurple, label={west:\textcolor{ETHpurple}{$\optStratPlayer{1}$}}] at (A) {};
    \node[mypoint, gray, label={east:\textcolor{gray}{$\optStratPlayer{2}$}}]    at (B) {};
    
    \node[align=center, below=5pt, font=\footnotesize] at (\axisSep/2,0) {
    Strategically Robust Equilibria\\
    (SRE)
    };

    \node[anchor=south, ETHpurple] (C)
    at (\axisSep/2,2.3) {\scriptsize
      $\displaystyle
      \argmax_{\stratPlayerA \in \strategySpacePlayerA} \min_{\stratWassersteinPlayer{2} \in \ambiguitySet{i}{\varepsilon}{\optStratPlayer{2}}} \utilityExpArgsPlayer{1}{\stratPlayerA}{\stratWassersteinPlayer{2}}$
    };
    \draw[ETHpurple,thick] (C) --++ (1,-1.4); 
\end{scope}

\begin{scope}[shift={(2*\panelSep,0)}]
    \draw[myaxis, ETHpetrol] (0,0) -- (0,\myheight) node[above] {$\Delta^1$};
    \draw[myaxis, gray] (\axisSep,0) -- (\axisSep,\myheight) node[above] {$\Delta^2$};

    \node[coordinate] (A) at (0,0.6*\myheight) {};
    \node[coordinate]    (A1) at (0,0) {};
    \node[coordinate]    (A2) at (0,\myheight) {};
    \node[coordinate] (B) at (\axisSep,0.3*\myheight) {};
    \node[coordinate]    (B1) at (\axisSep,0) {};
    \node[coordinate]    (B2) at (\axisSep,\myheight) {};

    \draw[myline, gray] (A1) -- (A2);
    \draw[myline, ETHpetrol] (B1) -- (B2);
    
    \draw[fill=ETHpetrol, draw=ETHpetrol, opacity=0.4] (A) -- (B1) -- (B2) -- cycle;
    \draw[fill=gray, draw=gray, opacity=0.4] (B) -- (A1) -- (A2) -- cycle;

    \node[mypoint, ETHpetrol, label={west:\textcolor{ETHpetrol}{$\optStratPlayer{1}$}}] at (A) {};
    \node[mypoint, gray, label={east:\textcolor{gray}{$\optStratPlayer{2}$}}] at (B) {};
    
    \node[align=center, below=5pt, font=\footnotesize] at (\axisSep/2,0) {
    Security strategies\\
    i.e., SRE with $\varepsilon = \infty$
    };
    
    \node[ETHpetrol, anchor=south, inner sep=0pt] (C)
    at (\axisSep/2,2.4) {\scriptsize
      $\displaystyle
      \argmax_{\stratPlayerA \in \strategySpacePlayerA} \min_{\stratWassersteinPlayer{2} \in \strategySpacePlayer{2}} \utilityExpArgsPlayer{1}{\stratPlayerA}{\stratWassersteinPlayer{2}}$
    };
    \draw[ETHpetrol,thick] (C) --++ (1,-1);
\end{scope}

\end{tikzpicture}
    \vspace*{-8mm}
    \caption{Representation of the concepts of Nash Equilibrium ($\varepsilon=0$), Strategically Robust Equilibrium \eqref{eq:StrategicallyRobustEquibrium}, and Security Strategy ($\varepsilon=+\infty$) in the case of two agents. The colored part of the diagrams represents the perspective of Player 1. The shaded areas (or the arrows) point
    at the set of strategies against which the agent's strategy is robustly optimal.}
    \label{fig:equilibria}
    \vspace*{-3mm}
\end{figure}

Three observations are in order. First, note that, at a strategically robust equilibrium, each agent $i$ is guaranteed a payoff no-lower than 
$\min_{\stratWassersteinPlayerIMinus \in \ambiguitySet{i}{\varepsilon}{\optStratPlayerIMinus}} \utilityExpMap^i(\optStratPlayer{i},\stratWassersteinPlayerIMinus)$, not only if the other agents play $\optStratPlayer{-i}$, but \emph{also} if they deviate and select any other strategy in the ambiguity set $\ambiguitySet{i}{\varepsilon}{\optStratPlayerIMinus}$. For this reason, ambiguity sets of larger size provide agents with stronger robustness guarantees.
Second, \cref{def:SRE} inherits both the fixed point argument of Nash equilibria and the worst-case flavor of security strategies. In doing so, strategically robust equilibria interpolate, in a principled way, between these two notions. As conceptually illustrated in~\cref{fig:equilibria}, when $\varepsilon=0$, the ambiguity sets collapse to their center and we therefore recover the definition of Nash equilibrium, while, when $\varepsilon\to\infty$, the ambiguity sets include all distributions so that~\cref{def:SRE} reduces to \mbox{that of security strategies.}
Finally, a special word of caution is necessary when dealing with 2-player zero-sum finite games, for which it is well-known that Nash equilibria and security strategies coincide. Limitedly to this specific setting, it is therefore immediate to conclude that the notion of strategically robust equilibrium also coincides with them, and is independent of the size of the ambiguity sets. 

\vspace*{-1mm}

\subsection{Existence of Strategically Robust Equilibria}
In this section, we show that, under minimal assumptions on the action sets, payoffs, and ambiguity sets, strategically robust equilibria exist. Naturally, in the limit cases $\varepsilon=0$ (Nash equilibrium) and $\varepsilon\rightarrow\infty$ (security strategies), this claim follows directly from classical results in game theory upon introducing the following assumption.\footnote{A weaker assumption is sufficient, i.e., that the agents' action spaces are compact Polish spaces (which we use in the proofs).}

\begin{assumption}[Action spaces and payoffs]\label{ass:action-utilities}
Either of the following holds:

\begin{enumerate}[label={\textbf{\theassumption.\Alph*:}},
  ref={assumption~\theassumption.\Alph*}]
\item \label{assump-A}%
    The action spaces $\{\actionSpacePlayerI\}_{i=1}^N$ are finite; 
\item \label{assump-B}%
    Each action space $\actionSpacePlayerI$ is a compact subset of $\;\mathbb{R}^{n}$, and each payoff function $ \utilityPlayerI$ is continuous.
\end{enumerate}
\end{assumption}

For the existence of strategically robust equilibria, we will combine the above with a second assumption, which ensures that ambiguity sets are well-behaved.

\begin{assumption}[Ambiguity set]\label{assumption:ambiguity}
    Each set-valued map $\stratPlayerIMinus\mapsto\ambiguitySet{i}{\varepsilon}{\stratPlayerIMinus}$, mapping the collection of mixed strategies $\stratPlayerIMinus$ to the ambiguity set $\ambiguitySet{i}{\varepsilon}{\stratPlayerIMinus}$, is non-empty, compact-valued, and hemicontinuous (i.e., both upper and lower hemicontinuous).\footnote{Here, hemicontinuity and compactness are intended with respect to the product of the topologies induced by narrow convergence on the space of probability measures $\probSpace{\actionSpacePlayer{j}}$ (for the domain) and the topology induced by narrow convergence on the space of probability measures $\probSpace{\actionSpacePlayerIMinus}$ (for the co-domain). Here, $\probSpace{\actionSpace{}}$ denotes the space of probability distributions over $\actionSpace{}$.}
\end{assumption}

As we will see, this assumption, which requires that ambiguity sets change continuously w.r.t. their center, holds true for different choices of ambiguity sets, including those defined through optimal transport, which we endeavor to study in this work. Interestingly, this assumption, which expands upon the classical conditions ensuring existence of Nash equilibria by merely requiring sufficient regularity of the ambiguity sets, is sufficient for existence of strategically robust equilibria.

\begin{theorem}[Existence]\label{thm:existence}
    Consider a game $\game$, a robustness level $\varepsilon\in\nonNegReals$, and suppose \crefrange{ass:action-utilities}{assumption:ambiguity} hold true. Then, a strategically robust equilibrium with robustness level $\varepsilon$ exists. 
\end{theorem}

The proof of this result and of all ensuing ones can be found in Appendix~\ref{sec:proofs}.
The argument used to show \cref{{thm:existence}} follows two main steps. We first prove continuity (w.r.t. narrow convergence of probability distributions), via Berge's Maximum Theorem together with~\cref{ass:action-utilities,assumption:ambiguity}, and concavity of the function  $(\stratPlayerI,\stratPlayerIMinus)\mapsto\min_{\stratWassersteinPlayerIMinus \in \ambiguitySet{i}{\varepsilon}{\stratPlayerIMinus}} \utilityExpMap(\stratPlayerI,\stratWassersteinPlayerIMinus)$. We can then deploy Glicksberg's fixed point theorem to prove existence of a fixed point for the best response map and so of a strategically robust equilibrium.

\subsection{Strategically Robust Equilibria via Optimal Transport and Their Existence}
In this section, we introduce ambiguity sets based on optimal transport and show that, with this choice, the existence of strategically robust equilibria is guaranteed under the \emph{very same} assumptions required to ensure the existence of Nash equilibria. Toward this goal, we begin with basic notions of optimal transport.

\paragraph{Optimal transport} Optimal transport provides us with a notion of distance between probability distributions and, thus, between mixed strategies. To ease the presentation, we first consider the case with two agents and describe the ambiguity set constructed by the first agent around $\stratPlayerB$, i.e., $\ambiguitySet{1}{\varepsilon}{\stratPlayerB}$. Toward this goal, it is instrumental to recall the optimal transport problem.

%
Let $\stratWassersteinPlayerB_1$ and $\stratWassersteinPlayerB_2$ be two mixed strategies of agent $2$, who mixes actions in $\actionSpacePlayerB$, and let $\distance^2:\actionSpacePlayerB\times\actionSpacePlayerB\to\nonNegReals$ be a distance between actions on $\actionSpacePlayerB$. Then, for $\wassnorm\geq 1$, the type-$\wassnorm$ optimal transport distance between $\stratWassersteinPlayerB_1$ and $\stratWassersteinPlayerB_2$ is defined as
\begin{equation}\label{eq:wasserstein_distance:definition}
    \wasserstein[\wassnorm](\stratWassersteinPlayerB_1,\stratWassersteinPlayerB_2)
    \coloneqq
    \left(
    \min_{\gamma\in\setPlans{\stratWassersteinPlayerB_1}{\stratWassersteinPlayerB_2}}\int_{\actionSpacePlayerB\times\actionSpacePlayerB}
    \distance^{2}(\actions_{1},\actions_{2})^\wassnorm\mathrm{d}\gamma(\actions_1,\actions_2)
    \right)^{1/\wassnorm}
\end{equation}
where $\setPlans{\stratWassersteinPlayerB_1}{\stratWassersteinPlayerB_2}$ denotes the set of probability distributions over $\actionSpacePlayer{2}\times\actionSpacePlayer{2}$ whose first marginal is $\stratWassersteinPlayerB_1$ and whose second marginal is $\stratWassersteinPlayerB_2$.
Intuitively, $\wasserstein[\wassnorm](\stratWassersteinPlayerB_1,\stratWassersteinPlayerB_2)^\wassnorm$ represents the minimum transportation cost required to move a pile of earth shaped as $\stratWassersteinPlayerB_1$ into a pile of earth shaped as $\stratWassersteinPlayerB_2$ where the cost to transport a unit amount of earth from $\actions_{1}$ to $\actions_{2}$ is given by $\distance^2(\actions_1,\actions_2)^\wassnorm$. Here, $\gamma(\actions_1,\actions_2)$ represents a ``transportation plan'', i.e., it describes the amount of earth located at $\actions_1$ that is transported to $\actions_2$.\footnote{Strictly speaking, $\gamma$ is a probability measure and the expression $\gamma(\actions_1,\actions_2)$ abuses notation. The formal statement is as follows: For sets $A,B\subset\actionSpacePlayerB$, $\gamma(A\times B)$ is the amount of earth located at $A$ that is transported to $B$. With $A=\{\actions_1\}$ and $B=\{\actions_2\}$, we obtain the mass transported from $\actions_1$ to $\actions_2$.}
In the discrete case,~\eqref{eq:wasserstein_distance:definition} reduces to a finite-dimensional linear program,
and the distance function is fully characterized by a symmetric matrix of dimension $\cardinality{\actionSpace_2}\times\cardinality{\actionSpace_2}$ with zero diagonal.
We exemplify the optimal transport problem and the choice of the distance 
in the following example.

\begin{example} First, consider $\actionSpacePlayerB=\{\actionPlayerB_1,\actionPlayerB_2 \}$. In this case, the only possible distance between actions assigns a value of $0$ to identical actions and a constant value (say $1$) to different actions. This results in an optimal transport distance coinciding with the well-known total variation distance 
~\cite{villani2009optimal}.
Second, consider $\actionSpacePlayerB=\{\actionPlayerB_1,\actionPlayerB_2,\actionPlayerB_3\}$. In this case, one can exploit the definition of distance to embed additional information. For instance, 
if we believe that the action $\actionPlayerB_1$ is closer to $\actionPlayerB_2$ than to $\actionPlayerB_3$,
we can set $\distance^{2}(\actionPlayerB_1,\actionPlayerB_2)=1$ and $\distance^{2}(\actionPlayerB_1,\actionPlayerB_3)=2$. Finally, we can set $\distance^{2}(\actionPlayerB_2, \actionPlayerB_3)=1$. It is easy to verify that the triangle inequality holds, and thus that $\distance^{2}$ is indeed a distance on $\actionSpacePlayerB$.
Third, consider a continuous action set, e.g., $\actionSpacePlayerB=\reals$. In this case, we note that any distance on $\reals$ is a valid choice. In particular, with $\distance^2=\abs{\cdot}$ we mimic the same effect discussed in the previous point by incorporating the fact that, for example, action 0 is ``more similar'' to action 1 than to action 2. 
\end{example}


\paragraph{Strategically robust game theory with optimal transport}
Armed with a notion of distance between probability distributions, we define an ambiguity set around a mixed strategy as the set containing all distributions whose distance, measured via optimal transport, is no-larger than $\varepsilon$, that is
\begin{equation*}
    \ambiguitySet{1}{\varepsilon}{\stratPlayerB}
    \coloneqq 
    \left\{
    \stratWassersteinPlayerB\in\probSpace{\actionSpacePlayerB}:
    \wasserstein[\wassnorm](\stratPlayerB,\stratWassersteinPlayerB)\leq\varepsilon
    \right\},
\end{equation*}
where $\probSpace{\actionSpacePlayerB}$ is the space of probability distributions over $\actionSpacePlayerB$.
In other words, $\ambiguitySet{1}{\varepsilon}{\stratPlayerB}$ includes all mixed strategies onto which $\stratPlayerB$ can be morphed with a transportation budget of at most $\varepsilon$.

In the general case of $N$ players, the type-$\wassnorm$ optimal transport distance between mixed strategies $\stratWassersteinPlayerIMinus_1$ and $\stratWassersteinPlayerIMinus_2$, each represented as probability distributions over $\actionSpacePlayerIMinus$, reads
\begin{equation}\label{eq:wassersteinDistance}
    \wasserstein[\wassnorm](\stratWassersteinPlayerIMinus_1,\stratWassersteinPlayerIMinus_2)
    =
    \left(
    \min_{\gamma\in\setPlans{\stratWassersteinPlayerIMinus_1}{\stratWassersteinPlayerIMinus_2}}
    \int_{\actionSpacePlayerIMinus\times\actionSpacePlayerIMinus}
    \distance^{-i}(\actionPlayerIMinus_{1},\actionPlayerIMinus_{2})^\wassnorm\mathrm{d}\gamma(\actionPlayerIMinus_{1},\actionPlayerIMinus_{2})
    \right)^{{1}/{\wassnorm}}.
\end{equation}
The ambiguity set is therefore defined by 
\begin{equation}\label{eq:ambiguity_set:definition}
\ambiguitySet{i}{\varepsilon}{\stratPlayerIMinus}
    =
    \left\{
    \stratWassersteinPlayerIMinus\in\probSpace{\actionSpacePlayerIMinus}:
    \wasserstein[\wassnorm](\stratWasserstein_{\stratPlayerIMinus},\stratWassersteinPlayerIMinus)\leq\varepsilon
    \right\},
\end{equation}
where the center $\stratWasserstein_{\stratPlayerIMinus}\coloneqq\stratPlayer{1}\times\ldots\times\stratPlayer{i-1}\times\stratPlayer{i+1}\times\ldots\times\stratPlayer{N}$ of the ambiguity set is the probability distribution over $\actionSpacePlayerIMinus$ obtained as the product of the distributions $\stratPlayerIMinus$.

\begin{remark}[Ambiguity set over the product space vs product of ambiguity sets]\label{remark:product ambiguity sets}
    The ambiguity set in~\eqref{eq:ambiguity_set:definition} is a ball in the space of probability distributions over $\actionSpacePlayerIMinus$ and \emph{not} the Cartesian product of $N-1$ balls each living in the space of probability distributions over $\actionSpacePlayer{j}$, $j\neq i$.
    Thus, strategically robust equilibria based on such ambiguity sets are robust also against \emph{correlated} deviations in the strategies of the other agents, and not only against independent deviations. While one could define ambiguity sets to protect only against uncorrelated deviations, this would give rise to further computational challenges as the minimization in \eqref{eq:StrategicallyRobustEquibrium} would be taken over a \emph{multilinear} function.
\end{remark}


\paragraph{Existence} After having introduced optimal transport-based ambiguity sets, we can present the main result of this section. Toward this goal, observe that, by~\cref{thm:existence}, existence of strategically robust equilibria requires compactness and, crucially, hemicontinuity of the ambiguity sets (cf.~\cref{assumption:ambiguity}). We now show that such hemicontinuity property does indeed hold for the optimal transport ambiguity set~\eqref{eq:ambiguity_set:definition}, a result which we believe may be of independent interest.

\begin{lemma}[Non-emptiness, continuity \& compactness]\label{lemma:continuity wasserstein ball}
The set-valued map $\stratPlayerIMinus \mapsto \ambiguitySet{i}{\varepsilon}{\stratPlayerIMinus}$, where $\ambiguitySet{i}{\varepsilon}{\stratPlayerIMinus}$ is defined in~\eqref{eq:ambiguity_set:definition}, 
is non-empty, compact-valued, and hemicontinuous, i.e., upper and lower hemicontinuous.  
\end{lemma}

At this point, existence of strategically robust equilibria follows readily from~\cref{thm:existence}.

\begin{corollary}[Existence]\label{corollary:existence:optimal_transport}
    Consider a game $\game$, a robustness level $\varepsilon\in\nonNegReals$, and the optimal transport ambiguity set~\eqref{eq:ambiguity_set:definition}.
    Suppose~\cref{ass:action-utilities} holds true. 
    Then, a strategically robust equilibrium with robustness level $\varepsilon$ exists. 
\end{corollary}

The importance of this result stems from the fact that, when ambiguity sets are defined in terms of optimal transport, existence of strategically robust equilibria is guaranteed under the \emph{very same} assumptions that ensure existence of Nash equilibria.

\section{Static $N$-Player Games with Finite Action Spaces}\label{sec:FiniteNPlayerGames}
In this section, we focus on $N$-player games with finite action spaces and study strategically robust equilibria with optimal transport ambiguity sets.
For this class of problems, we (i) study the computational complexity of strategically robust equilibria, (ii) show how existing approaches for computing Nash equilibria can be directly employed for their computation, and (iii) illustrate, through numerical simulations, the role of robustness in a variety of finite-action games, including congestion games and the pedestrian game introduced in~\cref{sec:motivating example}.
In these settings,~\cref{corollary:existence:optimal_transport} readily establishes existence of a strategically robust equilibrium for any robustness level $\varepsilon\in\nonNegReals$. Therefore, the focus of this section is largely algorithmic. Toward this goal, we begin by reformulating each agent's best response map.

\subsection{Reformulation of Strategically Robust Best Response}

We begin by considering the strategically robust best response map of agent $i$, denoted by $\stratRobustBestResponseMapPlayer{i}$ and defined as the set-valued map 
\begin{equation}\label{eq:nPlayerStratRobustBR}
\begin{aligned}
    \stratPlayerIMinus &\mapsto
    \stratRobustBestResponseMapPlayer{i}(\stratPlayerIMinus)\coloneqq 
    \argmax_{\stratPlayerI \in \strategySpacePlayerI} \min_{\stratWassersteinPlayerIMinus \in  \ambiguitySet{i}{\varepsilon}{\stratPlayerIMinus}}
    \expectedValue{(\actionPlayerI,\actionPlayerIMinus)\sim(\stratPlayerI,\stratWassersteinPlayerIMinus)}{\utilityArgsPlayerI}
\end{aligned}
\end{equation}
and immediately observe that, contrary to the case of mixed Nash equilibria where a best response can be computed by ranking the payoff accrued by each individual action (or, equivalently, by solving a linear program), solving \eqref{eq:nPlayerStratRobustBR} is significantly more challenging. This is apparent as agents now need to resolve a ``max-min'' problem. More precisely, the computation of the best response amounts to the solution of a so-called distributionally robust optimization problem, i.e., an optimization problem where decisions need to be optimal with respect to the worst-case probability distribution within a given ambiguity set. 
Interestingly, using duality tools for distributionally robust optimization, we can reformulate~\eqref{eq:nPlayerStratRobustBR} as a single-level optimization problem.

\begin{proposition}[Strategically Robust Best Response]\label{prop:finite:strategically robust best response}
The strategically robust best response map $\stratRobustBestResponseMapPlayer{i}$ can be reformulated as follows:
\begin{equation}\label{eq:finite:strategically robust best response:reformulation}
\begin{aligned}
    \stratRobustBestResponseMapPlayer{i}(\stratPlayerIMinus)
    &=
    \argmax_{\stratPlayerI \in \strategySpacePlayerI}
    \max_{\lambda^i \in \nonNegReals}
    -\lambda^i\varepsilon^\wassnorm
    +
    \expectedValue{\actionPlayerIMinus\sim\stratWassersteinPlayer{}_{\stratPlayerIMinus}}
    {\min_{\surActionPlayerIMinus\in\actionSpacePlayer{-i}}
    \left\{
    \expectedValue{\actionPlayerI\sim\stratPlayerI}{\utilityPlayer{i}(\actionPlayerI,\surActionPlayerIMinus)}
    + \lambda^i \distance^{-i}(\actionPlayerIMinus,\surActionPlayerIMinus)^\wassnorm
    \right\}},
\end{aligned}
\end{equation}
where, as above, $\stratWasserstein_{\stratPlayerIMinus}\coloneqq\stratPlayer{1}\times\ldots\times\stratPlayer{i-1}\times\stratPlayer{i+1}\times\ldots\times\stratPlayer{N}$ is the product of the distributions in $\stratPlayerIMinus$.
\end{proposition}

Notably, with~\eqref{eq:finite:strategically robust best response:reformulation}, the strategically robust best response can be computed by evaluating the expectation of a \emph{regularized} payoff w.r.t. the \emph{center} of the ambiguity set.
The expression for regularized payoff can be interpreted as resulting from a fictitious adversary who can select arbitrary actions $\surActionPlayerIMinus\in\actionSpacePlayer{-i}$ to minimize the payoff but incurs a price of $\lambda^i\distance^{-i}(\actionPlayerIMinus,\surActionPlayerIMinus)$ when deviating from the nominal action $\actionPlayerIMinus$. 
Moreover, we highlight that the optimization problem~\eqref{eq:finite:strategically robust best response:reformulation} is a finite convex program. In fact, via a standard epigraphic reformulation, it can even be reformulated as a linear program. 
We will use this fact to design computational methods for strategically robust equilibria.

\subsection{The Computational Complexity of Strategically Robust Equilibria}

We now study the computational complexity of strategically robust equilibria.
While it is well known that the computation of $\delta$-approximate Nash equilibria in $N$-player games is \ppad{}-complete~\citep{daskalakis2009complexity} (already for $N=2$), one may expect strategically robust equilibria to be more difficult to compute. Indeed, strategically robust equilibria emerge as fixed points of a best response map whose evaluation involves solving a distributionally robust optimization problem. 
On the contrary, we will show that, perhaps surprisingly, the computational complexity of strategically robust equilibria lies also in \ppad{}, and thus is no harder than that of Nash equilibria.

Formally, we define the problem of finding a strategically robust equilibrium as follows: Given an $N$-player game $\game$ with finite action spaces, a rational robustness level $\varepsilon\geq 0$, and a rational $\delta>0$, compute a $\delta$-approximate strategically robust equilibrium with robustness level $\varepsilon$, i.e., a tuple of strategies $(\optStratPlayer{1}, \dots, \optStratPlayer{N})$ so that 
    $\min_{\stratWassersteinPlayerIMinus\in\ambiguitySet{i}{\varepsilon}{\optStratPlayerIMinus}}
    \utilityExpArgsPlayer{i}{\optStratPlayerI}{\stratWassersteinPlayerIMinus}
    \geq
    \min_{\stratWassersteinPlayerIMinus\in\ambiguitySet{i}{\varepsilon}{\optStratPlayerIMinus}} 
    \utilityExpArgsPlayer{i}{\stratPlayerI}{\stratWassersteinPlayerIMinus}-\delta$
    for all $\stratPlayerI \in \strategySpacePlayerI$ and all $i\in\{1,\ldots,N\}.$
Our main result indicates that this problem is no harder than computing $\delta$-approximate Nash equilibria.

\begin{theorem}[Computational Complexity]\label{th:computationalComplexity}
    The computational complexity of strategically robust equilibria in $N$-player games lies in \ppad{}. 
\end{theorem}

The proof of this result leverages three main ingredients: (i) \cref{prop:finite:strategically robust best response} to rewrite the min-max optimization problem for the best response as a single optimization problem, using tools from distributionally robust optimization; (ii) a polynomial-time reformulation of the strategically robust game as an instance of a concave game; and (iii) a recent result on the computational complexity of concave games~\citep{papadimitriou2022computational}.

\begin{remark}
The standard proof of the computational complexity of Nash equilibria relies on a proof of existence of Nash equilibria via Brouwer fixed point theorem and on the study of the associated computational complexity~\citep{daskalakis2009complexity}. A closer inspection of this proof of existence reveals the need for the following fundamental property of Nash equilibria: A set of strategies forms a Nash equilibrium if (and only if) any deviation to a \emph{pure} strategy does not yield a higher payoff. This property fails to hold for strategically robust equilibria. Thus, we cannot resort to the classic argument via Brouwer to prove existence and, consequently, we cannot use the techniques of~\citep{daskalakis2009complexity} to establish their computational complexity. 
\end{remark}

Finally, we have already observed that, when $\varepsilon\to+\infty$, strategically robust equilibria coincide with security strategies, which can be computed in polynomial time via linear programming. We therefore conjecture that strategic robustness may even facilitate the computation of strategically robust equilibria. We leave this question to future research.

\subsection{The Computation of Strategically Robust Equilibria}\label{subsec:finite:computation of sre}

We now provide tools to compute strategically robust equilibria. Specifically, we show that, just as for (mixed) Nash equilibria, computing strategically robust equilibria in an $N$-player game amounts to solving a multilinear complementarity problem. In the 2-player case, this reduces to a \emph{linear} complementarity problem.\footnote{The statement continues to be true for so-called polymatrix games, where the influence of the selection of a strategy by one agent on the payoff of another is always the same, regardless of the strategies of the other agents~\citep{howson1972equilibria}. Accordingly, our results for 2-player games readily extend to polymatrix games.}
This way, we can readily use solvers for (multi)linear complementarity problems to compute strategically robust equilibria; e.g., see~\citep{sturmfels2002solving,van1987simplicial}.
This is not the only way to compute strategically robust equilibria. For instance, with the reformulation of the best-response in~\cref{prop:finite:strategically robust best response}, we can use homotopy-based methods~\citep{eaves1972homotopies}, also employed to compute mixed Nash equilibria, to compute strategically robust equilibria.

Our general recipe for the reformulation as a multilinear complementarity problem, inspired by its counterpart for Nash equilibria, is as follows. First, we notice that, thanks to an epigraphic reformulation, the strategically robust best response~\eqref{eq:finite:strategically robust best response:reformulation} can be cast as a linear program. Second, we replace each best response with its KKT conditions. Third, we ``stack'' the KKT conditions of each best response to obtain the multilinear complementarity problem, which simplifies to a \emph{linear} complementarity problem for the 2-player case. 

\newcommand\dualLambda{\tau^i}
\newcommand\dualProbabilityNonnegative{\omega^i(\actionPlayerI)}
\newcommand\dualProbabilityNormalization{\kappa^i}
\newcommand\dualEpigraphical{\eta^i(\actionPlayerIMinus,\surActionPlayerIMinus)}

To start, we need some additional notation. For a mixed strategy $\stratPlayerI$ for agent $i$, $\stratPlayerI(\actionPlayerI)$ is the corresponding probability of playing action $\actionPlayerI$ and, for the joint strategy $\stratWasserstein_{\stratPlayerIMinus}$ of all agents but $i$, $\stratWasserstein_{\stratPlayerIMinus}(\actionPlayerIMinus)=\prod_{j\neq i}\optStratPlayer{j}(\actionPlayer{j})$ the probability that action $\actionPlayerIMinus$ is played. 
Consider now a strategically robust equilibrium $(\optStratPlayerA,\ldots,\optStratPlayer{N})$. By definition, we have $\optStratPlayerI\in\stratRobustBestResponseMapPlayer{i}(\optStratPlayerIMinus)$.
We can now leverage~\cref{prop:finite:strategically robust best response}, together with an epigraphic reformulation, to express the best response of agent $i$ as the following linear program in the unknowns $\stratPlayerI \in \reals^{\cardinality{\actionSpacePlayerI}}$, $\lambda^i\in\reals$, $\xi^i\in\reals^{\cardinality{\actionSpacePlayerIMinus}}$:
\begin{subequations}\label{eq:finite:computation:epigraphical reformulation}
\begin{align}
\optStratPlayerI
    \in
   \stratRobustBestResponseMapPlayer{i}(\optStratPlayerIMinus)
    =
    \argmax_{\stratPlayerI \in \strategySpacePlayerI}
    \max_{\lambda^i\ge 0,~\xi^i}
    &
    - \varepsilon^\wassnorm \lambda^i + 
    \sum_{\actionPlayerIMinus\in\actionSpacePlayerIMinus}\stratWassersteinPlayer{}_{\optStratPlayerIMinus}(\actionPlayerIMinus)\xi^i(\actionPlayerIMinus)
    \label{eq:finite:computation:epigraphical reformulation:objective}
    \\*
    \text{s.t. }\:\, 
    &\xi^i(\actionPlayerIMinus)
    \leq 
    \sum_{\actionPlayerI\in\actionSpacePlayerI}\utilityPlayerI(\actionPlayerI,\surActionPlayerIMinus)\stratPlayerI(\actionPlayerI) + \lambda^i \distance^{-i}(\actionPlayerIMinus,\surActionPlayerIMinus)^\wassnorm & &\forall\surActionPlayerIMinus\in\actionSpacePlayerIMinus.
    \label{eq:finite:computation:epigraphical reformulation:epigraphic variable}
\end{align}
\end{subequations}
\noindent The
optimality conditions for the linear program~\eqref{eq:finite:computation:epigraphical reformulation} yield existence of dual multipliers $\dualLambda{}$ for constrains $\lambda^i\ge0$,  $\dualProbabilityNonnegative{}$ for $\stratPlayerI(\actionPlayerI)\geq0$, $\dualProbabilityNormalization{}$ for $\sum_{\actionPlayerI\in\actionSpacePlayerI}\stratPlayerI(\actionPlayerI)=1$, and $\dualEpigraphical$ for \eqref{eq:finite:computation:epigraphical reformulation:epigraphic variable} so that 
%
\begin{subequations}\label{eq:finite:computation:linear conditions}
\begin{align}
    \dualProbabilityNonnegative{}
    +
    \dualProbabilityNormalization{}
    +
    \sum_{\actionPlayerIMinus\in\actionSpacePlayerIMinus}\sum_{\surActionPlayerIMinus\in\actionSpacePlayerIMinus}
    \dualEpigraphical{} \utilityPlayer{i}(\actionPlayerI,\surActionPlayerIMinus)
    &=
    0
    & 
    &\forall\actionPlayerI\in\actionSpacePlayerI\label{eq:finite:computation:linear conditions:1}
    \\
    -\varepsilon^\wassnorm + \dualLambda{}
    +
    \sum_{\actionPlayerIMinus\in\actionSpacePlayerIMinus}\sum_{\surActionPlayerIMinus\in\actionSpacePlayerIMinus}\dualEpigraphical{}\distance^{-i}(\actionPlayerIMinus,\surActionPlayerIMinus)^\wassnorm
    &=
    0\label{eq:finite:computation:linear conditions:2}
    \\
    \stratWassersteinPlayer{}_{\optStratPlayerIMinus}(\actionPlayerIMinus) - \sum_{\surActionPlayerIMinus\in\actionSpacePlayerIMinus}\dualEpigraphical{}
    &=
    0
    &
    &\forall\actionPlayerIMinus\in\actionSpacePlayerIMinus\label{eq:finite:computation:linear conditions:3}
\end{align}
\end{subequations}
and the complementary conditions (here, $x\perp y\Leftrightarrow x_iy_i=0$ for all entries of $x$ and $y$)
\begin{equation}\label{eq:finite:computation:complementary conditions}
\begin{aligned}
    0\leq\dualLambda{}&\bot\lambda^i\geq 0
    \\
    0\leq \dualProbabilityNonnegative{} &\bot\optStratPlayerI(\actionPlayerI)\geq 0
    &
    &\forall\actionPlayerI\in\actionSpacePlayerI
    \\
    0\leq 
    \dualEpigraphical{}
    &\bot
    \left(-\xi^i(\actionPlayerIMinus)
    +
    \sum_{\actionPlayerI\in\actionSpacePlayerI}\utilityPlayerI(\actionPlayerI,\surActionPlayerIMinus)\optStratPlayerI(\actionPlayerI) + \lambda^i \distance^{-i}(\actionPlayerIMinus,\surActionPlayerIMinus)^\wassnorm\right)\geq 0
    & 
    &\forall\actionPlayerIMinus,\surActionPlayerIMinus\in\actionSpacePlayerIMinus
\end{aligned}
\end{equation}
are satisfied.
Conditions~\eqref{eq:finite:computation:linear conditions:2},~\eqref{eq:finite:computation:linear conditions:3}, and \eqref{eq:finite:computation:complementary conditions} are linear in variables $\optStratPlayerI, \lambda^i,\xi^i\dualLambda{},\dualProbabilityNonnegative{},\dualProbabilityNormalization{},\dualEpigraphical{}$. Condition~\eqref{eq:finite:computation:linear conditions:3}, instead, is multilinear, since the term $\stratWassersteinPlayer{}_{\optStratPlayerIMinus}(\actionPlayerIMinus)=\prod_{j\neq i}\optStratPlayer{j}(\actionPlayer{j})$ is multilinear (i.e., linear in each $\optStratPlayer{j}$).
At this point, we can therefore ``stack'' the multilinear conditions~\eqref{eq:finite:computation:linear conditions} and the linear complementary conditions~\eqref{eq:finite:computation:complementary conditions} of each agent to obtain a multilinear complementarity problem:  
\begin{proposition}[Multilinear Complementarity Problem Reformulation]\label{prop:finite:computation:N players}
Consider a game $\game$. 
The following are equivalent: 
\begin{enumerate}
    \item The profile of strategies $(\optStratPlayerA,\ldots,\optStratPlayer{N})$ is a strategically robust equilibrium of $\mathcal{G}$.
    \item There exists $\lambda^i,\xi^i(\actionPlayerIMinus),\dualLambda{},\dualProbabilityNonnegative{},\dualProbabilityNormalization{},\dualEpigraphical{}$ so that the linear conditions~\eqref{eq:finite:computation:linear conditions:1},~\eqref{eq:finite:computation:linear conditions:2}, the multilinear condition~\eqref{eq:finite:computation:linear conditions:3}, and the linear complementary conditions~\eqref{eq:finite:computation:complementary conditions} hold for all $i\in\{1,\ldots,N\}$.
\end{enumerate}
\end{proposition}

In the 2-player case, since $\stratWassersteinPlayer{}_{\optStratPlayerIMinus}(\actionPlayerIMinus)=\optStratPlayerIMinus(\actionPlayerIMinus)$, condition~\eqref{eq:finite:computation:linear conditions:3} is linear. Thus, as for Nash equilibria, strategically robust equilibria can be computed by solving a linear \mbox{complementarity problem.}%

\subsection{Example: Pedestrian Game}\label{subsec:finite:example}

We now turn our attention back to the pedestrian game of~\cref{sec:motivating example} with the game matrices in~\cref{table:finite:pedestrian}.
We compute strategically robust equilibria by solving the linear complementarity problem arising~\cref{prop:finite:computation:N players} via the PATH solver~\citep{ferris1999interfaces}\footnote{The code to generate this and all subsequent numerical examples is available at \\ \url{https://github.com/nicolaslanzetti/strategically-robust-game-theory}.}. Let agent $1$ be the autonomous vehicle with $\actionSpacePlayerA = \{\actionPlayerA_1 = \text{\aMaintain{} (M)}, \actionPlayerA_2 = \text{\aDecelerate{} (D)}, \actionPlayerA_3 = \text{\aStop{} (S)} \}$ and agent $2$ the pedestrians with $\actionSpacePlayerB = \{\actionPlayerB_1 = \text{\aWait{} (W)}, \actionPlayerB_2 = \text{\aCross{} (C)} \}$.
We define the Wasserstein ambiguity set via the total variation cost, defined by $\distanceArgsPlayerI{\actionPlayerI_k}{\actionPlayerI_l}=0$
if $l=k$ and $\distanceArgsPlayerI{\actionPlayerI_k}{\actionPlayerI_l} =1$ if $l\neq k$.
Since the distance achieves a maximum of 1, ambiguity sets are constructed with $\varepsilon\in [0,1]$. 


\pgfplotsset{
    compat=1.15,
    myaxisSimplex/.style={
        scale=1.4,
        view={135}{30},
        axis lines=center,
        axis line style={-stealth, line width=0.8pt},
        axis on top,
        xlabel={$\mathbf{D}$},
        ylabel={$\mathbf{S}$},
        zlabel={$\mathbf{M}$},
        xmax=1.3,
        ymax=1.3,
        zmax=1.3,
        ticks=none,
        x label style={font=\scriptsize, xshift=-0pt, yshift=0pt},
        y label style={font=\scriptsize, xshift=0pt,yshift=0pt},
        z label style={font=\scriptsize, xshift=0pt, yshift=5pt},
    },
    myaxisSimplex1D/.style={
        axis lines=middle,
        axis line style={-stealth, line width=0.8pt}, 
        xmin=-0.15, xmax=1.3,
        ymin=-0.15, ymax=1.3,
        ticks=none,
        xlabel={$\mathbf{W}$},
        ylabel={$\mathbf{C}$},
        x label style={at={(axis description cs:1.05,0.05)}, anchor=north, font=\scriptsize},
        y label style={at={(axis description cs:0.05,1.05)}, anchor=east, font=\scriptsize},
    }
}

\newcommand{\drawbasics}{
\addplot3[no marks] coordinates {(0,0,0) (1.2,0,0)};
\addplot3[no marks] coordinates {(0,0,0) (0,1.2,0)};
\addplot3[no marks] coordinates {(0,0,0) (0,0,1.2)};
\filldraw[ETHgray, opacity=0.2] (axis cs:0,0,0) -- (axis cs:1,0,0) -- (axis cs:0,1,0) -- cycle;
\filldraw[ETHgray, opacity=0.2] (axis cs:0,0,0) -- (axis cs:1,0,0) -- (axis cs:0,0,1) -- cycle;
\filldraw[ETHgray, opacity=0.2] (axis cs:0,0,0) -- (axis cs:0,1,0) -- (axis cs:0,0,1) -- cycle;
\filldraw[ETHgray, opacity=0.2] (axis cs:1,0,0) -- (axis cs:0,1,0) -- (axis cs:0,0,1) -- cycle;
                
\node[label={[font=\LARGE, yshift=4pt]90:{}},inner sep=.5pt,fill=black,circle] at (axis cs:1,0,0) {};
\node[label={[font=\LARGE, yshift=4pt]90:{}},inner sep=.5pt,fill=black,circle] at (axis cs:0,1,0) {};
\node[label={[font=\LARGE, xshift=4pt]0:{}},inner sep=.5pt,fill=black,circle] at (axis cs:0,0,1) {};
}

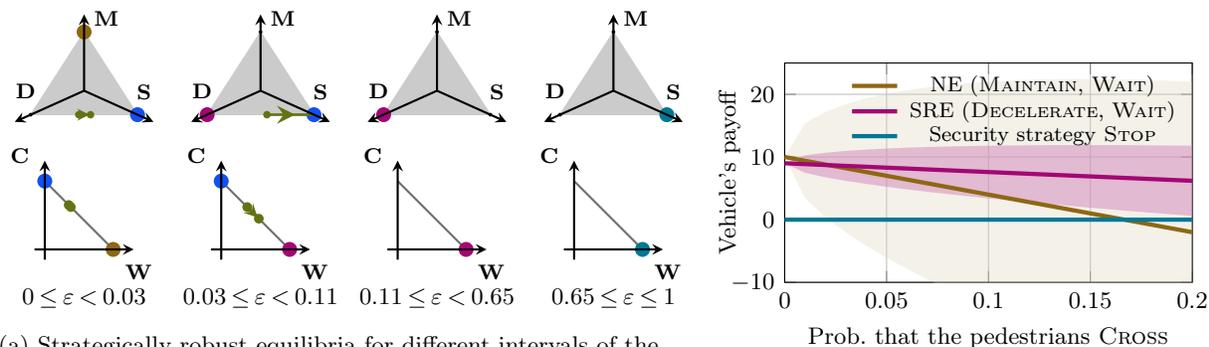
\begin{figure}[tb]
    \begin{subfigure}[b]{0.53\textwidth}
    \centering
    \begin{tikzpicture}
        \begin{groupplot}[
            group style={
                group size=4 by 2,
                horizontal sep=0.5cm,
                vertical sep=0.1cm,
            },
            width=2.9cm,
            height=2.9cm,
            axis on top,
            enlargelimits=false
        ]
        
        \nextgroupplot[myaxisSimplex]
            \drawbasics{};
            \node[inner sep=2pt,fill=ETHbronze,circle] at (axis cs:0,0,1) {}; 
            \node[inner sep=2pt,fill=ETHblue,circle] at (axis cs:0,1,0) {};
            \addplot3[no marks, ultra thick, ETHgreen, line width=1pt, mid arrow short] coordinates {(0.55, 0.45, 0) (0.44, 0.56, 0)};
            \node[inner sep=1pt,fill=ETHgreen,circle] at (axis cs:0.44,0.56,0) {};
            \node[inner sep=1pt,fill=ETHgreen,circle] at (axis cs:0.55,0.45,0) {};

        \nextgroupplot[myaxisSimplex]
            \drawbasics{};
            \node[inner sep=2pt,fill=ETHpurple,circle] at (axis cs:1,0,0) {}; 
            \node[inner sep=2pt,fill=ETHblue,circle] at (axis cs:0,1,0) {}; 
            \node[inner sep=1pt,fill=ETHgreen,circle] at (axis cs:0.44,0.56,0) {};
            \addplot3[no marks, ultra thick, ETHgreen, line width=1pt, mid arrow] coordinates {(0.44, 0.56, 0) (0, 1, 0)};    

        \nextgroupplot[myaxisSimplex]
            \drawbasics{};
            \node[inner sep=2pt,fill=ETHpurple,circle] at (axis cs:1,0,0) {}; 

        \nextgroupplot[myaxisSimplex]
            \drawbasics{};
            \node[inner sep=2pt,fill=ETHpetrol,circle] at (axis cs:0,1,0) {}; 

        \nextgroupplot[myaxisSimplex1D, title={\footnotesize $0 \leq \varepsilon < 0.03$},title style={yshift=-2.3cm}]
            \addplot[no marks, thick, ETHgray] coordinates {(0,1) (1,0)}; 
            \node[inner sep=2pt,fill=ETHbronze,circle] at (axis cs:1,0) {}; 
            \node[inner sep=2pt,fill=ETHblue,circle] at (axis cs:0,1) {}; 
            \addplot[no marks, thick, ETHgreen, line width=2pt] coordinates {(0.35,0.65) (0.38,0.62)};
            \node[inner sep=1.3pt,fill=ETHgreen,circle] at (axis cs:0.35,0.65) {};
            \node[inner sep=1.3pt,fill=ETHgreen,circle] at (axis cs:0.38,0.62) {};

        \nextgroupplot[myaxisSimplex1D, title={\footnotesize $0.03 \leq \varepsilon < 0.11$},title style={yshift=-2.3cm}]
            \addplot[no marks, thick, ETHgray] coordinates {(0,1) (1,0)}; 
            \node[inner sep=2pt,fill=ETHpurple,circle] at (axis cs:1,0) {}; 
            \node[inner sep=2pt,fill=ETHblue,circle] at (axis cs:0,1) {}; 
            \node[inner sep=1.3pt,fill=ETHgreen,circle] at (axis cs:0.38,0.62) {};
            \node[inner sep=1.3pt,fill=ETHgreen,circle] at (axis cs:0.55,0.45) {};
            \addplot[no marks, ETHgreen, line width=1.2pt, mid arrow short] coordinates {(0.38,0.62) (0.55,0.45)};

        \nextgroupplot[myaxisSimplex1D,title={\footnotesize $0.11 \leq \varepsilon < 0.65$},title style={yshift=-2.3cm}]
            \addplot[no marks, thick, ETHgray] coordinates {(0,1) (1,0)}; 
            \node[inner sep=2pt,fill=ETHpurple,circle] at (axis cs:1,0) {}; 

        \nextgroupplot[myaxisSimplex1D,title={\footnotesize $0.65 \leq \varepsilon \leq 1$},title style={yshift=-2.3cm}]
            \addplot[no marks, thick, ETHgray] coordinates {(0,1) (1,0)}; 
            \node[inner sep=2pt,fill=ETHpetrol,circle] at (axis cs:1,0) {};   
        \end{groupplot}
    \end{tikzpicture}
    \caption{Strategically robust equilibria for different intervals of the robustness parameter $\varepsilon$. The upper row shows the equilibrium strategies for the autonomous vehicle on the two-dimensional simplex; the lower row shows the equilibrium strategy for the family on the one-dimensional simplex. Strategies with the same color form an equilibrium (e.g., (S, C) and (M, W) for $\varepsilon\in[0,0.03]$). When the equilibrium changes as $\varepsilon$ is varied, we indicate this with an arrow.}
    \label{fig:pedestrian_strategies}
    \end{subfigure}
    \hfill 
    \begin{subfigure}[b]{0.45\textwidth}
    \centering
    \pgfplotstableread[col sep=comma]{results/pedestrian.csv}\datatable
    \begin{tikzpicture}
        \begin{axis}[
        width=7cm, height=4.5cm,
        xlabel={\footnotesize Prob. that the pedestrians \aCross},
        ylabel={\footnotesize Vehicle's payoff},
        ylabel near ticks,
        ylabel style={yshift=-1em},
        grid=both,
        legend style={draw=none,fill=none},
        xmin=0, ymin=-10, xmax=0.2, ymax=25,
        xtick={0, 0.05, 0.1, 0.15, 0.2}, 
        xticklabels={0, 0.05, 0.1, 0.15, 0.2}, 
        ]
        \addplot[ETHbronze,ultra thick] table[x=f, y expr=\thisrow{meanNash}] {\datatable};
        \addplot[name path=us_top,ETHbronze,thin,draw=none,forget plot] table[x=f, y expr=\thisrow{nashPlus}] {\datatable};
        \addplot[name path=us_down,ETHbronze,thin,draw=none,forget plot] table[x=f, y expr=\thisrow{nashMinus}] {\datatable};
        \addplot[fill=ETHbronze!30,fill opacity=0.3,forget plot] fill between[of=us_top and us_down];

        \addplot[ETHpurple, ultra thick] table[x=f, y expr=\thisrow{meanRobust}] {\datatable};
        \addplot[name path=us_top,ETHpurple, thin, draw=none,forget plot] table[x=f, y expr=\thisrow{robustPlus}] {\datatable};
        \addplot[name path=us_down,ETHpurple, thin,draw=none,forget plot] table[x=f, y expr=\thisrow{robustMinus}] {\datatable};
        \addplot[fill=ETHpurple!50,fill opacity=0.5,forget plot] fill between[of=us_top and us_down];

        \addplot[ETHpetrol, ultra thick] table[x=f, y expr=\thisrow{meanSecure}] {\datatable};

        \addlegendentry{NE (\textsc{Maintain}, \textsc{Wait})}
        \addlegendentry{SRE (\textsc{Decelerate}, \textsc{Wait})}
        \addlegendentry{Security strategy \textsc{Stop}}
    \end{axis}
    \end{tikzpicture}
    \caption{Vehicle's payoff attained by playing the Nash equilibrium, strategically robust equilibrium, and security strategies, as a function of the probability that the family decides to cross the road (deviating from their Nash equilibrium strategy \aWait). The shaded area represents the expected payoff plus/minus one standard deviation to illustrate the agent's risk.}
    \label{fig:pedestrianGame:performance}
    \end{subfigure}
    \caption{Strategically robust equilibrium and vehicle's payoff in the pedestrian game.}
\end{figure}

In~\cref{fig:pedestrian_strategies}, we study the evolution of strategically robust equilibria for agent $1$ (plots at the top) and agent $2$ (plots at the bottom) as we increase the robustness level from $\varepsilon = 0$ (no robustness, Nash equilibria) to $\varepsilon \in [0, 1]$ (full robustness, security strategies).
At $\varepsilon = 0$, we recover the three Nash equilibria: (M, W) in \textcolor{ETHbronze}{brown}, (S, C) in \textcolor{ETHblue}{blue}, and the mixed one in \textcolor{ETHgreen}{green}. Until $\varepsilon = 0.03$ the original Nash equilibria persist, with the mixed equilibrium moving linearly in the simplex. With $\varepsilon = 0.03$, (M, W) ceases to be a strategically robust equilibrium and it is replaced by (D, W) in \textcolor{ETHpurple}{purple}. The other equilibrium (S, C) and the mixed equilibrium persist, where the mixed equilibrium keeps moving linearly. 
At $\varepsilon = 0.11$, the mixed equilibrium and the (S, C) equilibrium also disappear. For $\varepsilon\in[0.11,0.65]$, the only strategically robust equilibrium is (D, W). Finally,  for $\varepsilon \geq 0.66$, the only strategically robust equilibrium is (S, W) in \textcolor{ETHpetrol}{petrol}, corresponding to the security strategies.
Interestingly, the fact that security strategies coincide with strategically robust equilibria already for $\varepsilon<1$ suggests that they are rational already when agents require sufficient level of robustness but do not assume others to be completely adversary.

Importantly, for a large spectrum of the robustness level ($\varepsilon\in[0.11,0.65]$), the only strategically robust equilibrium is (D, W).
In contrast, the Nash equilibria (M, W) and (S, C) have poor robustness properties and disappear already at low robustness levels. 
We argue that this is the natural solution of this decision-making problem and we suggest that this is what most human car drivers would adhere to. Indeed, \cref{fig:pedestrianGame:performance} shows the payoff of the autonomous vehicle when this strategically robust equilibrium strategy \aDecelerate{} is played (\textcolor{ETHpurple}{purple}), as a function of the percentage of pedestrians that do not stick to the equilibrium strategy (\aWait{}) and instead \aCross{}.
When equilibrium protocol is strictly followed ($\varepsilon$ small), then the strategically robust equilibrium performs almost as well as the Nash equilibrium strategy, that is (M, W), in \textcolor{ETHbronze}{brown}. However, already for 3\% of pedestrians deciding to \aCross{}, the strategically robust equilibrium results in both larger average payoff and lower variance. This behavior, which gets amplified at large deviations, is clearly preferable over the overly cautious security strategy (S, \textcolor{ETHpetrol}{petrol}), which constantly leads to zero payoff.



\subsection{Coordination via Robustification Effect in Examples}

We now study the effect of the robustification parameter $\varepsilon$ across three bi-matrix games.

\begin{figure}[!tb]
    \centering
    \begin{subfigure}[t]{0.66\textwidth}
    \centering
    \pgfplotstableread[col sep=comma]{results/inspection_game_paper.csv}\datatable
    \begin{tikzpicture}
        \begin{axis}[
        myAxisStyle,
        ylabel={\footnotesize Payoff of player 1},
        ]
        \addplot[ETHpurple,thick] table[x=epsilon, y expr=\thisrow{utility_p1}] {\datatable};
        \addplot[ETHbronze,thick] table[x=epsilon, y expr=\thisrow{utility_p1_ne}] {\datatable};
        
        \addplot[name path=us_top,ETHpurple,thin,draw=none] table[x=epsilon, y expr=\thisrow{utility_p1_up}] {\datatable};
        \addplot[name path=us_down,ETHpurple,thin,draw=none] table[x=epsilon, y expr=\thisrow{utility_p1_down}] {\datatable};
        \addplot[fill=ETHpurple!50,fill opacity=0.5] fill between[of=us_top and us_down];

        \addplot[name path=us_top,ETHbronze,thin,draw=none] table[x=epsilon, y expr=\thisrow{utility_p1_ne_up}] {\datatable};
        \addplot[name path=us_down,ETHbronze,thin,draw=none] table[x=epsilon, y expr=\thisrow{utility_p1_ne_down}] {\datatable};
        \addplot[fill=ETHbronze!30,fill opacity=0.3] fill between[of=us_top and us_down];

        \addlegendentry{\footnotesize SRE}
        \addlegendentry{\footnotesize NE}
        \end{axis}
    \end{tikzpicture}
    \begin{tikzpicture}
        \begin{axis}[
        myAxisStyle,
        ylabel={\footnotesize Payoff of player 2},
        ]
        \addplot[ETHpurple,thick] table[x=epsilon, y expr=\thisrow{utility_p2}] {\datatable};
        \addplot[ETHbronze,thick] table[x=epsilon, y expr=\thisrow{utility_p2_ne}] {\datatable};
        
        \addplot[name path=us_top,ETHpurple,thin,draw=none] table[x=epsilon, y expr=\thisrow{utility_p2_up}] {\datatable};
        \addplot[name path=us_down,ETHpurple,thin,draw=none] table[x=epsilon, y expr=\thisrow{utility_p2_down}] {\datatable};
        \addplot[fill=ETHpurple!50,fill opacity=0.5] fill between[of=us_top and us_down];

        \addplot[name path=us_top,ETHbronze,thin,draw=none] table[x=epsilon, y expr=\thisrow{utility_p2_ne_up}] {\datatable};
        \addplot[name path=us_down,ETHbronze,thin,draw=none] table[x=epsilon, y expr=\thisrow{utility_p2_ne_down}] {\datatable};
        \addplot[fill=ETHbronze!30,fill opacity=0.3] fill between[of=us_top and us_down];
        \end{axis}
    \end{tikzpicture}
    \caption{Inspection game.}\label{fig:inspectionGame}
    \end{subfigure}
    \hfill 
    \begin{subfigure}[t]{0.33\textwidth}
    \centering 
    \pgfplotstableread[col sep=comma]{results/free_rider_paper.csv}\datatable
    \begin{tikzpicture}
        \begin{axis}[
        myAxisStyle,
        ylabel={\footnotesize Payoff of player 1 and 2},
        ]
        \addplot[ETHpurple,thick] table[x=epsilon, y expr=\thisrow{utility_p1}] {\datatable};
        \addplot[ETHbronze,thick] table[x=epsilon, y expr=\thisrow{utility_p1_ne}] {\datatable};
        
        \addplot[name path=us_top,ETHpurple,thin,draw=none] table[x=epsilon, y expr=\thisrow{utility_p1_up}] {\datatable};
        \addplot[name path=us_down,ETHpurple,thin,draw=none] table[x=epsilon, y expr=\thisrow{utility_p1_down}] {\datatable};
        \addplot[fill=ETHpurple!50,fill opacity=0.5] fill between[of=us_top and us_down];

        \addplot[name path=us_top,ETHbronze,thin,draw=none] table[x=epsilon, y expr=\thisrow{utility_p1_ne_up}] {\datatable};
        \addplot[name path=us_down,ETHbronze,thin,draw=none] table[x=epsilon, y expr=\thisrow{utility_p1_ne_down}] {\datatable};
        \addplot[fill=ETHbronze!30,fill opacity=0.3] fill between[of=us_top and us_down];

        \addlegendentry{\footnotesize SRE}
        \addlegendentry{\footnotesize NE}
        \end{axis}
    \end{tikzpicture}
    \caption{Free-rider game.}
    \label{fig:freeRiderGame}
    \end{subfigure}
    \caption{Payoffs of players 1 and 2 for the inspection game and the free-rider game at the strategically robust equilibrium (as a function of $\varepsilon$) and the Nash equilibrium (independent of $\varepsilon$). As expected, strategically robust equilibria are robust against such deviations and, for $\varepsilon\in(0,0.5)$, they even yield larger payoffs for both agents. The solid lines are the expected players' payoff when both players play the equilibrium strategy; the shaded area denotes the range of expected payoff when the other player plays a perturbation of the equilibrium strategy and the first action (resp. second) is played with a $\delta$ larger probability (resp. smaller), with $\delta\in[-0.05,0.05]$.
    }
    \label{fig:firstTwoExamples}
\end{figure}
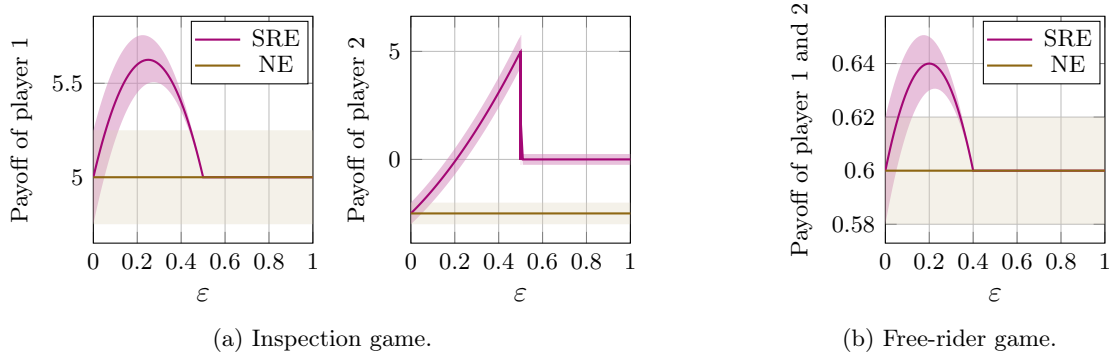

\paragraph{Inspection game}
\begin{wraptable}{r}{4.15cm}
\vspace*{-5mm}
    \centering
    {
    \setlength{\extrarowheight}{3pt}
    \begin{tabular}{C{0.8cm}|C{1.2cm}|C{1.2cm}|}
          \multicolumn{1}{c}{} & \multicolumn{1}{c}{\footnotesize I}  & \multicolumn{1}{c}{\footnotesize NI} \\[3pt]
          \cline{2-3}
          \footnotesize S & \footnotesize$(0,-5)$ & \footnotesize$(10,-10)$ \\[3pt]
          \cline{2-3}
          \footnotesize W & \footnotesize$(5,0)$ & \footnotesize$(5,5)$ \\[3pt]
          \cline{2-3}
    \end{tabular}
    }
    \vspace*{5mm}
\end{wraptable}
We consider the standard inspection game~\cite[p. 17]{fudenberg1991game}, whereby an agent (player 1) works for a principal (player 2). The agent can decide to either Shirk (S) or Work (W), whereas the principal can either Inspect (I) or not Inspect (NI), giving rise to the payoff matrix shown on the right. 
We showcase the role of $\varepsilon$ on the equilibrium payoffs in~\cref{fig:inspectionGame}. 
First, as in~\cref{subsec:finite:example}, strategically robust equilibria are more robust than Nash equilibria when agents deviate from the equilibrium play, cfr. the size of the shaded regions.
Second, we observe the anticipated \emph{coordination via robustification effect}, whereby the payoffs of \emph{both} players increase as the robustness level is increased.

\paragraph{Free-rider problem}
\begin{wraptable}{r}{4.15cm}
\vspace*{-4mm}
    \centering
    {
    \setlength{\extrarowheight}{3pt}
    \begin{tabular}{C{0.8cm}|C{1.2cm}|C{1.2cm}|}
          \multicolumn{1}{c}{} & \multicolumn{1}{c}{\footnotesize C}  & \multicolumn{1}{c}{\footnotesize NC} \\[3pt]
          \cline{2-3}
          \footnotesize C & \footnotesize$(0.6,0.6)$ & \footnotesize$(0.6,1)$ \\[3pt]
          \cline{2-3}
          \footnotesize NC & \footnotesize$(1,0.6)$ & \footnotesize$(0,0)$ \\[3pt]
          \cline{2-3}
    \end{tabular}
    }
\end{wraptable}
We now consider the standard free-rider problem, whereby each agent decides whether to contribute (C) or not (NC) to a public good. If the public good is realized, each agent receives a payoff of 1, reduced of 0.4 in the case the agent contributes. 
Also in this game, as we show in~\cref{fig:freeRiderGame}, agents benefit from adopting a robust approach: They reduce variability in their payoff while \emph{simultaneously} increasing their expected payoffs, before converging to playing the security strategy (C).

\paragraph{Taming the Braess Paradox}
\begin{wraptable}{r}{7.2cm}
\vspace*{-3mm}
    {
    \setlength{\extrarowheight}{3pt}
    \begin{tabular}{C{0.8cm}|C{1.2cm}|C{1.2cm}|C{1.2cm}|C{1.2cm}|}
          \multicolumn{1}{c}{}
          & \multicolumn{1}{c}{\footnotesize SAT}  & \multicolumn{1}{c}{\footnotesize SBT}
          & \multicolumn{1}{c}{\footnotesize SABT}  & \multicolumn{1}{c}{\footnotesize SBAT}
          \\[3pt]
          \cline{2-5}
          \footnotesize SAT
          & \cellcolor{black!20}\footnotesize$(19,19)$ & \cellcolor{black!20}\footnotesize$(14,14)$ & \footnotesize$(19,16)$ & \footnotesize$(14,19)$\\[3pt]
          \cline{2-5}
          \footnotesize SBT
          & \cellcolor{black!20}\footnotesize$(14,14)$ & \cellcolor{black!20}\footnotesize$(20,20)$ & \footnotesize$(20,17)$ & \footnotesize$(14,19)$\\[3pt]
          \cline{2-5}
          \footnotesize SABT
          & \footnotesize$(16,19)$ & \footnotesize$(17,20)$ & \footnotesize$(22,22)$ & \footnotesize$(11,19)$ \\[3pt]
          \cline{2-5}
          \footnotesize SBAT
          & \footnotesize$(19,14)$ & \footnotesize$(19,14)$ & \footnotesize$(19,11)$ & \footnotesize$(19,19)$ \\[3pt]
          \cline{2-5}
    \end{tabular}
    }
\end{wraptable}
To conclude, we consider a routing game exhibiting the Braess paradox, which we study in the two-player setting, to ease presentation.
Specifically, we consider a two-player congestion game where two agents travel from $S$ to $T$ navigating the graph shown in~\cref{fig:congestionGame:graph}.
We study two versions of the game: without bridge (feasible paths are SAT$=$S$\to$A$\to$T and SBT$=$S$\to$B$\to$T, cost highlighted in light gray) and with bridge (all four paths are feasible). 
This game yields the celebrated \emph{Breass} paradox, whereby the social cost (sum of the expected travel times) of the worst Nash equilibrium increases when the players can use the bridge, see~\cref{fig:congestionGame}. 
Consistently with the previous examples, robustness induces coordination between the agents also in this setting with both players lowering their travel time. 
Stated differently, strategically robust equilibria incur much lower price of anarchy compared to their Nash counterpart. This is an interesting aspect, which connects with much work in the algorithmic game theory literature \citep{koutsoupias1999worst}, and which we plan to explore in future work.


%

\begin{figure}[t]
    \begin{minipage}{0.44\textwidth}
    \centering
    \begin{tikzpicture}
        \node[circle,thick,draw] at (0,0) (a) {\footnotesize S}; 
        \node[circle,thick,draw] at (4,0) (b) {\footnotesize T}; 

        \node[circle,thick,draw] at (2,1.2) (up) {\footnotesize A}; 
        \node[circle,thick,draw] at (2,-1.2) (down) {\footnotesize B}; 

        \draw[-,ultra thick] (a) -- node[pos=0.7,left,yshift=0.15cm,align=center,font=\footnotesize] {\footnotesize $0.5x$} (up) -- node[pos=0.4,right,yshift=0.15cm,align=center,font=\footnotesize] {$0.9$} (b); 
        \draw[-,ultra thick] (a) -- node[pos=0.7,left,yshift=-0.15cm,align=center,font=\footnotesize] {$0.9$} (down) -- node[pos=0.4,right,yshift=-0.15cm,align=center,font=\footnotesize] {$0.5x$} (b); 

        \draw[-,ultra thick] (up) -- node[pos=.5,right,font=\footnotesize] {bridge} node[pos=.5,left,font=\footnotesize,font=\footnotesize] {0} (down); 


    \end{tikzpicture}
    \caption{Routing network with delay functions for each edge. We consider two versions of the game: without bridge, where players choose between the upper path S$\to$ A$\to$ T and the lower path S$\to$ B$\to$ T, and with bridge, where agents choose between all four paths connecting S to T.}
    \label{fig:congestionGame:graph}
    \end{minipage}
    \hfill 
    \begin{minipage}{0.55\textwidth}
    \centering
    \pgfplotstableread[col sep=comma]{results/congestion_game_paper.csv}\datatable
    \pgfplotstableread[col sep=comma]{results/congestion_game_small_paper.csv}\datatablesmall
    \begin{tikzpicture}
        \begin{axis}[
        width=8.5cm, height=4.5cm,
        xlabel=$\varepsilon$,
        ylabel={\footnotesize Cost of player 1 and 2},
        ylabel near ticks,
        xlabel near ticks,
        grid=both,
        xmin=0,xmax=1,
        legend pos=south east,
        ]
        \addplot[ETHpurple,thick] table[x=epsilon, y expr=\thisrow{utility_p1}] {\datatable};
        \addplot[ETHbronze,thick] table[x=epsilon, y expr=\thisrow{utility_p1_ne}] {\datatable};
        \addplot[black,thick,dashed] table[x=epsilon, y expr=\thisrow{utility_p1}] {\datatablesmall};
        
        \addplot[name path=us_top,ETHpurple,thin,draw=none] table[x=epsilon, y expr=\thisrow{utility_p1_up}] {\datatable};
        \addplot[name path=us_down,ETHpurple,thin,draw=none] table[x=epsilon, y expr=\thisrow{utility_p1_down}] {\datatable};
        \addplot[fill=ETHpurple!50,fill opacity=0.5] fill between[of=us_top and us_down];

        \addplot[name path=us_top,ETHbronze,thin,draw=none] table[x=epsilon, y expr=\thisrow{utility_p1_ne_up}] {\datatable};
        \addplot[name path=us_down,ETHbronze,thin,draw=none] table[x=epsilon, y expr=\thisrow{utility_p1_ne_down}] {\datatable};
        \addplot[fill=ETHbronze!30,fill opacity=0.3] fill between[of=us_top and us_down];

        \addlegendentry{\footnotesize SRE }
        \addlegendentry{\footnotesize NE }
        \end{axis}
    \end{tikzpicture}    
    \caption{
    Cost of player 1 and 2 at the strategically robust equilibrium with largest social cost in the two-player congestion games.
    In the game without bridge, the Nash equilibrium coincides with the strategically robust equilibrium, and with the security strategy (50\% on both paths, dashed line). 
    The shaded area is computed as in~\cref{fig:firstTwoExamples}.
    }
    \label{fig:congestionGame}
    \end{minipage}
\end{figure}

\section{Static $N$-Player Concave Games with Continuous Action Spaces}\label{sec:ContinuousGames}
We now focus on $N$-player game with continuous action spaces. In this setting, the space of mixed strategies $\strategySpacePlayerI$ is infinite-dimensional, which makes the computation of equilibria prohibitive. However, as this section will unveil, such challenges can be overcome by focusing on the widely-studied class of \emph{concave games} introduced by~\citet{rosen1965existence}.
We recall their definition: 

\begin{definition}[Concave Game]\label{def:convexGame}
A \emph{concave game} $\game$ is defined as an $N$-player game where
\begin{itemize}
    \item the action spaces, $\actionSpacePlayerI \subset \realVec{n}$, are compact and convex,
    \item the payoff functions, $\utilityPlayerI: \actionSpacePlayerI\times\actionSpacePlayerIMinus \goesto \reals$, are continuous,
    \item the payoff functions, $\utilityPlayerI: \actionSpacePlayerI\times\actionSpacePlayerIMinus \goesto \reals$, are concave in $\actionPlayerI$ for fixed $\actionPlayerIMinus$.
\end{itemize}
\end{definition}

Besides being theoretically attractive, concave games appear in numerous applications including economics~\citep{osborne1986price}, engineering~\citep{paccagnan2018nash}, supply chain~\citep{cachon2006game}, and transportation science~\citep{sun2007equilibrium}, to name a few. 
In what follows, we (i) show that concave games admit a pure strategically robust equilibrium, (ii) provide computational tools, which we instantiate in the case of quadratic games, and (iii) present an application to the Cournot competition.

\subsection{Existence of Pure Strategically Robust Equilibria}
While mixed equilibria in the setting of concave games are infinite-dimensional objects and are therefore difficult to describe or compute, the next result shows that \emph{pure} equilibria are guaranteed to exist in this same setting.

\begin{theorem}[Existence of Pure Strategically Robust Equilibria]\label{thm:continuous:existence pure strategically robust eq}
Consider a concave game $\game$ and a robustness level $\varepsilon\in\nonNegReals$. Then, $\mathcal{G}$ admits a pure strategically robust equilibrium with robustness level $\varepsilon$. That is, there exists $\optAction=(\optActionPlayer{1},\ldots,\optActionPlayer{N})$ so that for all $i \in \{1,\ldots,N\}$ we have
\begin{equation}\label{eq:continuous:existence pure strategically robust eq}
\begin{aligned}
    \optActionPlayer{i} &\in \argmax_{\actionPlayerI \in \actionSpacePlayerI} \min_{\stratWassersteinPlayerIMinus \in \ambiguitySet{i}{\varepsilon}{\delta_{\optActionPlayer{-i}}}} \expectedValue{\actionPlayerIMinus\sim\stratWassersteinPlayerIMinus}{\utilityArgsPlayerI},
\end{aligned}
\end{equation}
where, with slight abuse of notation, $\delta_{\optActionPlayer{-i}}$ denotes the collection of pure strategies where each agent $j\neq i$ plays the pure action $\actions_j$.
\end{theorem}

Note that, as per their definition, strategically robust equilibria (whether pure or not) remain robust against \emph{both} pure and mixed deviations from the equilibrium itself, so long as these deviations live in the Wasserstein ball $\ambiguitySet{i}{\varepsilon}{\delta_{\optActionPlayer{-i}}}$. {Note that this result is particularly appealing since pure strategies, which are convenient from an application standpoint (since no randomization is required), suffice to achieve robustness against pure and mixed play.}

\subsection{The Computation of Strategically Robust Equilibria}

Unfortunately, the existence of pure strategically robust equilibria does not suffice to yield a finite-dimensional problem. Indeed, the equilibrium condition~\eqref{eq:continuous:existence pure strategically robust eq} still involves a minimization problem over the probability space, which is infinite-dimensional.
However, we can now use the duality in the setting of distributionally robust optimization to convert~\eqref{eq:continuous:existence pure strategically robust eq} to a finite-dimensional condition.
This way, we can even reformulate a strategically robust game as a standard concave game:


\begin{proposition}[Equivalent concave game]\label{prop:equivalent game}
    Consider a concave game $\game$ and a robustness level $\varepsilon\in\posReals$. Let $\tilde{\mathcal{G}}_\varepsilon$ be the concave game defined by the actions spaces $\surActionSpacePlayer{i}\coloneqq\actionSpacePlayerI\times[0,M^i]$, with $M^i\coloneqq 2\max_{\actionPlayerI\in\actionSpacePlayerI,\actionPlayerIMinus\in\actionSpacePlayerIMinus}\abs{\utilityPlayerI(\actionPlayerI,\actionPlayerIMinus)}/\varepsilon^\wassnorm$, and the payoffs 
    \begin{equation*}
        \surUtilityPlayerI_\varepsilon((\actionPlayer{i},\lambda^i),(\actionPlayer{-i},\lambda^{-i}))
        \coloneqq 
        \min_{\surActionPlayerIMinus \in \actionSpacePlayerIMinus}
        \left\{ \utilityPlayerI(\actionPlayer{i},\surActionPlayerIMinus) + \lambda^i \distanceArgsPlayer{-i}{\actionPlayer{-i}}{\surActionPlayerIMinus}^\wassnorm\right\} - \lambda^i \varepsilon^\wassnorm.
    \end{equation*}
    Then, $\optAction=(\optActionPlayer{1},\ldots,\optActionPlayer{N})$ is a pure strategically robust equilibrium of $\mathcal{G}$ with robustness level $\varepsilon$ if and only if there exist $\bar\lambda^i\in\nonNegReals$ so that $\left((\optActionPlayer{1},\bar\lambda^1),\ldots,(\optActionPlayer{N},\bar\lambda^N)\right)$ is a Nash equilibrium of $\tilde{\mathcal{G}}_\varepsilon$.
\end{proposition}


While attractive, \cref{prop:equivalent game} requires to analyze the surrogate payoff $\surUtilityPlayerI_\varepsilon$. For general payoffs $\utilityPlayerI$, the mere evaluation of $\surUtilityPlayerI$ involves solving a general nonconvex optimization problem (the ``min'' part) over the action space $\actionSpacePlayerIMinus$, which is computationally prohibitive. 
Nonetheless, this minimization problem becomes tractable if the distance $\distance^{-i}$ is convex and the payoffs are convex in the other agents' actions. In turn, these assumptions allow us to deploy duality for convex optimization~\cite[Theorem 2]{zhen2023unified} to reformulate the minimization problem in $\surUtilityPlayerI_\varepsilon$:


\begin{corollary}[Reformulation of the surrogate payoff]\label{cor:equivalent game:convex}
Consider the concave game $\tilde{\mathcal{G}_\varepsilon}$ of~\cref{prop:equivalent game} and suppose that (i) the payoff $\utilityPlayerI$ and the distance $(\distancePlayer{-i})^\wassnorm$ are convex in $\actionPlayerIMinus$ and (ii) the action space reads $\actionSpacePlayerI = \{\actionPlayerI \in \realVec{n}: f_k^{i}(\actionPlayerIMinus) \leq 0 \text{ for all } k \in\{1,\ldots,K\}\}$ for lower semi-continuous convex $f_k^i:\reals^n\to\reals$.
The value of $\,\surUtilityPlayerI_\varepsilon((\actionPlayer{i},\lambda^i),(\actionPlayer{-i},\lambda^{-i}))$ results from the convex optimization problem
\begin{align*}
    \max_{\tau_{k}^j\in \nonNegReals, v^j, w^j, z_k^j \in \realVec{n}}
    &-\lambda \varepsilon^\wassnorm - (\utilityPlayerI)^{*}(\actionPlayerI, v) 
    - \lambda ((\distancePlayer{-i})^\wassnorm)^{*}\left(\actionPlayer{-i},\frac{w}{\lambda}\right)
    - \sum_{j=1,j\neq i}^N\sum_{k=1}^K \tau_k^j (f_k^j)^*\left(\frac{z_k^j}{\tau_k}\right)
    \\*
    \text{s.t.}\qquad 
    &v^j + w^j + \sum_{k=1}^K z_{k}^j= 0 \\
    &v=(v^1,\ldots,v^{i-1},v^{i+1},\ldots,v^N), w=(w^1,\ldots,w^{i-1},w^{i+1},\ldots,w^N)\in\realVec{(N-1)n},
\end{align*}
where $(f_k^j)^*$ is the convex conjugate of $f_k^j$, and $(\utilityPlayerI)^{*}(\actionPlayerI, v) \coloneqq\sup_{\surActionPlayerIMinus\in\reals^{N(n-1)}} v^\top\surActionPlayerIMinus-\utilityPlayerI(\actionPlayerI,\surActionPlayerIMinus)$ and $((\distancePlayer{-i})^\wassnorm)^{*}(\actionPlayerIMinus,\frac{w}{\lambda})\coloneqq \sup_{\surActionPlayerIMinus\in\reals^{N(n-1)}} \frac{w^\top}{\lambda}\surActionPlayerIMinus -\distancePlayer{-i}(\actionPlayerIMinus,\surActionPlayerIMinus)^\wassnorm$ are the convex conjugates of $\utilityPlayerI$ and $(\distancePlayer{-i})^\wassnorm$ w.r.t. their second argument. 
\end{corollary}

The reformulation of $\surUtilityPlayerI_\varepsilon$ as a maximization problem is attractive since it allows us to bypass the min-max formulation. For instance, it readily gives us access to the use of proximal responses to compute strategically robust equilibria~\cite[Section 12.6.1]{facchinei201012}, whereby $\actionPlayerI$ and $\lambda^i$ are iteratively updated to the optimal solutions of the proximal optimization problem
\begin{equation}\label{eq:continuous:proximal best response}
    \max_{\actions\in\actionSpacePlayerI,\lambda\in [0,M]}\surUtilityPlayerI_\varepsilon((\actions,\lambda),(\actionPlayer{-i},\lambda^{-i})) - \gamma \norm{\actions-\actionPlayerI}^2 - \gamma(\lambda-\lambda^i)^2,
\end{equation}
where $\gamma>0$.
Thanks to~\cref{cor:equivalent game:convex},~\eqref{eq:continuous:proximal best response} can therefore be cast as a single finite-dimensional convex problem.
We leave the study of the convergence properties of this algorithm for the general case to future research and limit ourselves to recall that fixed points of proximal operators are maximizers of $\surUtilityPlayerI_\varepsilon$ and, therefore, form strategically robust equilibria.


\subsection{Quadratic Games}
\label{subsec:quadratic-games}
We now consider the case of quadratic games, which arise in a surge of applications ranging from network science~\citep{bramoulle2014strategic} to engineering~\citep{ma2011decentralized}. Specifically, the action spaces are polytopes of the form $\actionSpacePlayerI=\{\actionPlayerI\in\reals^n:(D_k^i)^\top \actionPlayerI \leq d^i_k\text{ for all } k\in\{1,\ldots,K\}\}$ and the payoffs are the quadratic functions 
\begin{equation*}\label{eq:quadraticGame}
\begin{aligned}
    \utilityArgsPlayerI &= (\actionPlayerI)^\top Q^i \actionPlayerI + (\actionPlayerI)^\top B^i \actionPlayerIMinus + (q^i)^\top \actionPlayerI, 
\end{aligned}
\end{equation*}
where all vectors and matrices are of appropriate dimension. 
We use the type-2 Wasserstein distance with the standard Euclidean norm $\norm{\cdot}$ on $\reals^{(N-1)n}$.
In this case, the surrogate payoff results from the second-order conic program 
\begin{multline}\label{eq:quadraticGames:utility}
    \surUtilityPlayerI_\varepsilon((\actionPlayer{i},\lambda^i),(\actionPlayer{-i},\lambda^{-i}))
    =
    (\actionPlayerI)^T Q^i \actionPlayerI + (\actionPlayerI)^\top B^i\actionPlayer{-i} + (q^i)^T \actionPlayerI
    \\
    -\lambda^i \varepsilon^2 + \max_{\tau_{k}^j\in \nonNegReals} 
    \left\{
    -  \frac{1}{4\lambda^i}\norm{(B^i)^\top\actionPlayerI+\begin{bmatrix} \sum_{k=1}^K \tau^1_kD^1_k \\ \vdots \\ \sum_{k=1}^K \tau^N_kD^N_k\end{bmatrix}}^2
    +
    \sum_{k=1}^K\sum_{j=1,j\neq i}^N \tau_k^j\left((D_k^j)^\top a^j-d_k^j\right)
    \right\},
\end{multline}
where we used $f_k^j(\actionPlayer{j})=(D^j_k)^\top \actionPlayer{j}-d^j_k$ and the expressions for the convex conjugate in~\cite[Appendix 2]{kuhn2019wasserstein}. Notably,~\eqref{eq:quadraticGames:utility} coincides with the nominal payoff $\utilityArgsPlayerI$, plus an additional regularization. We further illustrate this regularization effect, which is well-known in distributionally robust optimization~\citep{gao2024wasserstein,shafieezadeh2019regularization}, in the next paragraph.

\paragraph{Regularization of games}
For simplicity, consider the unconstrained case 
in the minimization problem in $\surUtilityPlayerI_\varepsilon$ or, equivalently, the case without functions $f_k^j$ in~\cref{cor:equivalent game:convex}.
This way the maximization in~\eqref{eq:quadraticGames:utility} trivializes and one can compute the optimal $\lambda^i$ as $\lambda^{i,\star}=\norm{(B^i)^\top \actionPlayerI}/(2\varepsilon)$, yielding
\begin{equation*}
    (\actionPlayerI)^T Q^i \actionPlayerI + (\actionPlayerI)^\top B^i\actionPlayer{-i} + (q^i)^\top \actionPlayerI
    -\varepsilon\norm{(B^i)^\top\actionPlayerI},
\end{equation*}
which is the payoff $\utilityArgsPlayerI$ with the additional regularization term $\varepsilon\norm{(B^i)^\top\actionPlayerI}$. 
We can therefore interpret strategically robust equilibria as Nash equilibria of a game with \emph{regularized}~payoff.

\subsection{Coordination via Robustification in the Cournot Competition}

To conclude, we study strategic robustness in the classical Cournot competition model~\citep{cournot1838recherches} with $N$ firms competing in $T$ markets. 
We denote by $\actionPlayerI_t\geq 0$ the output quantity of firm $i$ in market $t$. The total production of each firm, $\sum_{t=1}^T\actionPlayerI_t$ cannot exceed $K^i$.
For each market $t$, consider the inverse demand curve $p_t(\sum_{i=1}^N\actionPlayer{i}_t)=\alpha_t-\beta_t\sum_{i=1}^N\actionPlayer{i}_t$ and production costs $c^i\actionPlayer{i}_t$ (with $c^i<\alpha_t$), where $\alpha_t, \beta_t$, and $c^i$ are known parameters. Accordingly, the payoff of firm $i$ reads
\begin{equation*}
    \utilityPlayer{i}(\actionPlayer{i},\actionPlayer{-i})
    =
    \sum_{t=1}^T
    (\alpha_t-c^i)\actionPlayer{i}_t
    -\beta_t(\actionPlayer{i}_t)^2
    -\beta_t\actionPlayer{i}_t\left(\sum_{j=1, j\neq i}^N\actionPlayer{j}_t\right).
\end{equation*}

\begin{figure}[tb]
    \centering
    \begin{subfigure}[b]{0.32\linewidth}
        \centering
        \footnotesize
        \renewcommand{\arraystretch}{0.85}
        \begin{tabular}{c|ccccc}
                    & $\alpha_t$ & $\beta_t$ \\ \midrule
             Market 1 & 100 & 0.8 \\
             Market 2 & 120 & 0.6 \\
             Market 3 & 110 & 0.7 \\
            \midrule
                    & $c^i$ & $K^i$ &\\ \midrule
             Firm 1 &  40 & 100\\
             Firm 2 &  45 & 120\\
             Firm 3 &  50 & 90\\
             Firm 4 &  55 & 80\\
        \end{tabular}
        \vspace*{3mm}
        \caption{Cournot game parameters. 
        }
        \label{fig:cournotGame:parameters}
    \end{subfigure}
    \hfill 
    \begin{subfigure}[b]{0.65\linewidth}
        \centering
        \pgfplotstableread[col sep=comma]{results/cournot_symmetric_paper.csv}\datatable
        \begin{tikzpicture}
        \begin{axis}[
        width=4.4cm, height=4.5cm,
        xlabel=$\varepsilon$,
        ylabel={\footnotesize Payoff of firm 1, 2, 3, 4},
        ylabel near ticks,
        xlabel near ticks,
        grid=both,
        xmin=0,ymin=0,xmax=150,
        legend pos=south west,
        ]
        \addplot[ETHpurple,thick] table[x=epsilon, y expr=\thisrow{payoff_firm_1}] {\datatable};
        \addplot[ETHbronze,thick] table[x=epsilon, y expr=\thisrow{payoff_firm_1_ne}] {\datatable};
        
        \addplot[name path=us_top,ETHpurple,thin,draw=none] table[x=epsilon, y expr=\thisrow{payoff_firm_1_std_up}] {\datatable};
        \addplot[name path=us_down,ETHpurple,thin,draw=none] table[x=epsilon, y expr=\thisrow{payoff_firm_1_std_down}] {\datatable};
        \addplot[fill=ETHpurple!50,fill opacity=0.5] fill between[of=us_top and us_down];
            
        \addplot[name path=us_top,ETHbronze,thin,draw=none] table[x=epsilon, y expr=\thisrow{payoff_firm_1_ne_std_up}] {\datatable};
        \addplot[name path=us_down,ETHbronze,thin,draw=none] table[x=epsilon, y expr=\thisrow{payoff_firm_1_ne_std_down}] {\datatable};
        \addplot[fill=ETHbronze!30,fill opacity=0.3] fill between[of=us_top and us_down];

        \addlegendentry{\footnotesize SRE}
        \addlegendentry{\footnotesize NE}
        \end{axis}
        \end{tikzpicture}
        \begin{tikzpicture}
        \begin{axis}[
        width=4.4cm, height=4.5cm,
        xlabel=$\varepsilon$,
        ylabel={\footnotesize Production of firm 1, 2, 3, 4},
        ylabel near ticks,
        xlabel near ticks,
        grid=both,
        xmin=0,ymin=0,xmax=150,
        legend pos=north east,
        ]
        \addplot[name path=zero, ETHpurple,thick,draw=none,forget plot] table[x=epsilon, y expr=0*\thisrow{out_firm_1_market_1}] {\datatable};
        \addplot[name path=aa, ETHpurple,thick,draw=none,forget plot] table[x=epsilon, y expr=\thisrow{out_firm_1_market_1}] {\datatable};
        \addplot[name path=bb, ETHpurple,thick,draw=none,forget plot] table[x=epsilon, y expr=\thisrow{out_firm_1_market_1}+\thisrow{out_firm_1_market_2}] {\datatable};
        \addplot[name path=cc, ETHpurple,thick,draw=none,forget plot] table[x=epsilon, y expr=\thisrow{out_firm_1_market_1}+\thisrow{out_firm_1_market_2}+\thisrow{out_firm_1_market_3}] {\datatable};
        
        \addplot[fill=green!50,fill opacity=0.5] fill between[of=aa and zero];
        \addplot[fill=red!50,fill opacity=0.5] fill between[of=aa and bb];
        \addplot[fill=blue!50,fill opacity=0.5] fill between[of=bb and cc];

        \addlegendentry{\footnotesize Market 1};
        \addlegendentry{\footnotesize Market 2};
        \addlegendentry{\footnotesize Market 3};
        \end{axis}
        \end{tikzpicture}
    \vspace{-.2cm}
    \caption{Identical firms (parameters of firm 1).}
    \end{subfigure}

    \vspace{3mm}
    
    \begin{subfigure}[b]{\linewidth}
    \centering
    \hspace*{5mm}
        \pgfplotstableread[col sep=comma]{results/cournot_asymmetric_paper.csv}\datatable
        \begin{tikzpicture}
        \begin{axis}[
        width=4.4cm, height=4.5cm,
        xlabel=$\varepsilon$,
        ylabel={\footnotesize Payoff of firms 1, 2},
        ylabel near ticks,
        xlabel near ticks,
        grid=both,
        xmin=0,ymin=-1,xmax=150,
        ]
        \addplot[ETHpurple,thick] table[x=epsilon, y expr=\thisrow{payoff_firm_1}] {\datatable};
        \addplot[ETHbronze,thick] table[x=epsilon, y expr=\thisrow{payoff_firm_1_ne}] {\datatable};
        \addplot[ETHpurple,thick] table[x=epsilon, y expr=\thisrow{payoff_firm_2}] {\datatable};
        \addplot[ETHbronze,thick] table[x=epsilon, y expr=\thisrow{payoff_firm_2_ne}] {\datatable};

        \addplot[name path=us_top,ETHpurple,thin,draw=none] table[x=epsilon, y expr=\thisrow{payoff_firm_1_std_up}] {\datatable};
        \addplot[name path=us_down,ETHpurple,thin,draw=none] table[x=epsilon, y expr=\thisrow{payoff_firm_1_std_down}] {\datatable};
        \addplot[fill=ETHpurple!70,fill opacity=0.5] fill between[of=us_top and us_down];

        \addplot[name path=us_top,ETHpurple,thin,draw=none] table[x=epsilon, y expr=\thisrow{payoff_firm_2_std_up}] {\datatable};
        \addplot[name path=us_down,ETHpurple,thin,draw=none] table[x=epsilon, y expr=\thisrow{payoff_firm_2_std_down}] {\datatable};
        \addplot[fill=ETHpurple!30,fill opacity=0.5] fill between[of=us_top and us_down];


            
        \addplot[name path=us_top,ETHbronze,thin,draw=none] table[x=epsilon, y expr=\thisrow{payoff_firm_1_ne_std_up}] {\datatable};
        \addplot[name path=us_down,ETHbronze,thin,draw=none] table[x=epsilon, y expr=\thisrow{payoff_firm_1_ne_std_down}] {\datatable};
        \addplot[fill=ETHbronze!40,fill opacity=0.3] fill between[of=us_top and us_down];

        \addplot[name path=us_top,ETHbronze,thin,draw=none] table[x=epsilon, y expr=\thisrow{payoff_firm_2_ne_std_up}] {\datatable};
        \addplot[name path=us_down,ETHbronze,thin,draw=none] table[x=epsilon, y expr=\thisrow{payoff_firm_2_ne_std_down}] {\datatable};
        \addplot[fill=ETHbronze!20,fill opacity=0.3] fill between[of=us_top and us_down];


        \end{axis}
        \end{tikzpicture}
        \begin{tikzpicture}
        \begin{axis}[
        width=4.4cm, height=4.5cm,
        xlabel=$\varepsilon$,
        ylabel={\footnotesize Payoff of firms 3, 4},
        ylabel near ticks,
        xlabel near ticks,
        grid=both,
        xmin=0,ymin=-1,xmax=150,
        ]
        \addplot[ETHpurple,thick] table[x=epsilon, y expr=\thisrow{payoff_firm_3}] {\datatable};
        \addplot[ETHbronze,thick] table[x=epsilon, y expr=\thisrow{payoff_firm_3_ne}] {\datatable};
        \addplot[ETHpurple,thick] table[x=epsilon, y expr=\thisrow{payoff_firm_4}] {\datatable};
        \addplot[ETHbronze,thick] table[x=epsilon, y expr=\thisrow{payoff_firm_4_ne}] {\datatable};



        \addplot[name path=us_top,ETHpurple,thin,draw=none] table[x=epsilon, y expr=\thisrow{payoff_firm_3_std_up}] {\datatable};
        \addplot[name path=us_down,ETHpurple,thin,draw=none] table[x=epsilon, y expr=\thisrow{payoff_firm_3_std_down}] {\datatable};
        \addplot[fill=ETHpurple!70,fill opacity=0.5] fill between[of=us_top and us_down];

        \addplot[name path=us_top,ETHpurple,thin,draw=none] table[x=epsilon, y expr=\thisrow{payoff_firm_4_std_up}] {\datatable};
        \addplot[name path=us_down,ETHpurple,thin,draw=none] table[x=epsilon, y expr=\thisrow{payoff_firm_4_std_down}] {\datatable};
        \addplot[fill=ETHpurple!30,fill opacity=0.5] fill between[of=us_top and us_down];
            


        \addplot[name path=us_top,ETHbronze,thin,draw=none] table[x=epsilon, y expr=\thisrow{payoff_firm_3_ne_std_up}] {\datatable};
        \addplot[name path=us_down,ETHbronze,thin,draw=none] table[x=epsilon, y expr=\thisrow{payoff_firm_3_ne_std_down}] {\datatable};
        \addplot[fill=ETHbronze!40,fill opacity=0.3] fill between[of=us_top and us_down];

        \addplot[name path=us_top,ETHbronze,thin,draw=none] table[x=epsilon, y expr=\thisrow{payoff_firm_4_ne_std_up}] {\datatable};
        \addplot[name path=us_down,ETHbronze,thin,draw=none] table[x=epsilon, y expr=\thisrow{payoff_firm_4_ne_std_down}] {\datatable};
        \addplot[fill=ETHbronze!20,fill opacity=0.3] fill between[of=us_top and us_down];
        \end{axis}
        \end{tikzpicture}
        \begin{tikzpicture}
        \begin{axis}[
        width=4.4cm, height=4.5cm,
        xlabel=$\varepsilon$,
        ylabel={\footnotesize Production of firm 1},
        ylabel near ticks,
        grid=both,
        xmin=0,ymin=-1,xmax=150,
        ]
        \addplot[name path=zero, ETHpurple,thick,draw=none,forget plot] table[x=epsilon, y expr=0*\thisrow{out_firm_1_market_1}] {\datatable};
        \addplot[name path=aa, ETHpurple,thick,draw=none,forget plot] table[x=epsilon, y expr=\thisrow{out_firm_1_market_1}] {\datatable};
        \addplot[name path=bb, ETHpurple,thick,draw=none,forget plot] table[x=epsilon, y expr=\thisrow{out_firm_1_market_1}+\thisrow{out_firm_1_market_2}] {\datatable};
        \addplot[name path=cc, ETHpurple,thick,draw=none,forget plot] table[x=epsilon, y expr=\thisrow{out_firm_1_market_1}+\thisrow{out_firm_1_market_2}+\thisrow{out_firm_1_market_3}] {\datatable};
        
        \addplot[fill=green!50,fill opacity=0.5] fill between[of=aa and zero];
        \addplot[fill=red!50,fill opacity=0.5] fill between[of=aa and bb];
        \addplot[fill=blue!50,fill opacity=0.5] fill between[of=bb and cc];
        \end{axis}
        \end{tikzpicture}
    \vspace{-.2cm}
    \caption{Non-identical firms.}
    \end{subfigure}
    \caption{Payoffs of the firms as a function of the robustness parameter $\varepsilon$. In \textcolor{ETHpurple}{purple}, we plot the payoff at the strategically robust equilibrium.  The shaded area represents the standard deviation of such a payoff when the other firms deviate from it (via an additive zero-mean Gaussian whose standard deviation is 20\% of the equilibrium quantity). For reference, we include the outcome of the Nash equilibrium (\textcolor{ETHbronze}{brown}, independent from $\varepsilon$). Strategically robust equilibria are robust against such deviations and lead to higher payoffs. 
    }
    \vspace{-.3cm}
    \label{fig:cournotGame}
\end{figure}
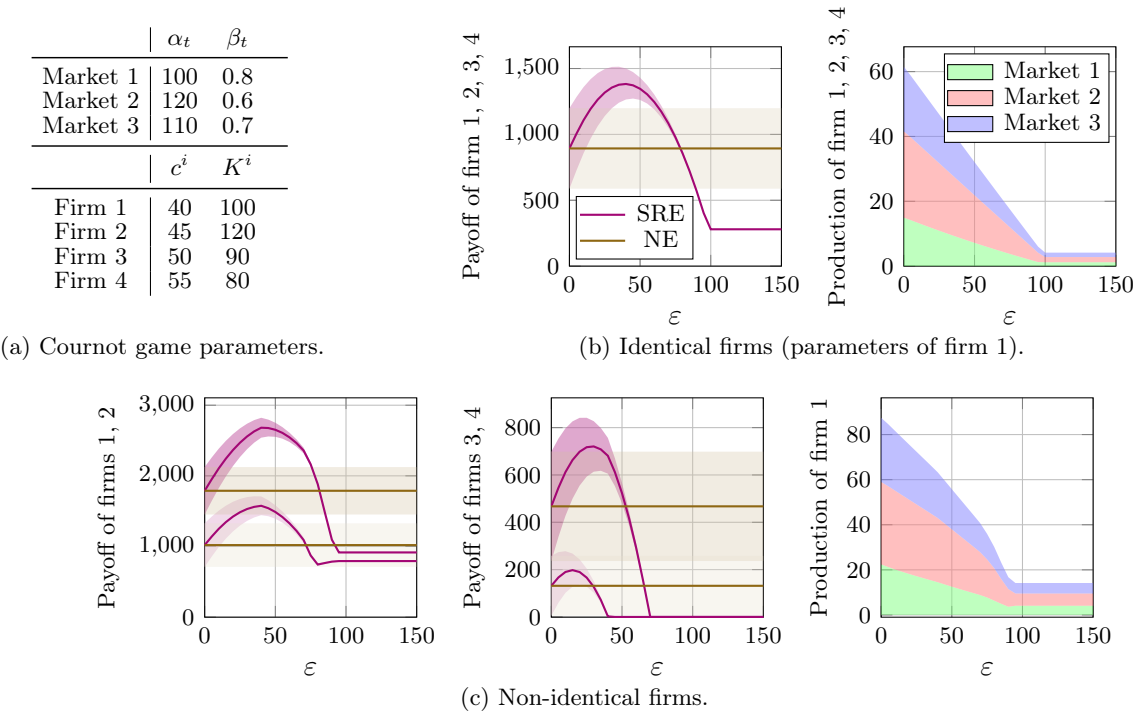

We conduct numerical experiments with $N=4$ and $T=3$ in the case of identical firms and non-identical firms, with the parameters in Table~\ref{fig:cournotGame:parameters}.
Our results, shown in~\cref{fig:cournotGame}, reveal that strategic robustness induces the firms to reduce their production in each market, to protect themselves against over-production of the other firms. 
This way, not only does each firm improve their \emph{worst-case} payoff, but, remarkably, \emph{all} firms achieve a larger \emph{nominal} payoff.
This phenomenon, which we have already observed in finite games, contrasts with standard results in robust optimization, i.e., that robustness yields lower nominal payoffs, and, once again, suggests a ``\emph{coordination via robustification}'' effect whereby strategic robustness induces more cooperative behaviors. 

For the case of two firms, one market, and no production upper bounds, we can explain these results by inspecting the regularization effect discussed in \cref{subsec:quadratic-games}. In this setting, the original game is equivalent to one where the payoff of firm $i$ is replaced by the following regularized one:
\begin{equation*}
    -\beta(\actionPlayer{i})^2 -\beta\actionPlayer{i}\actionPlayer{-i}+(\alpha-c^i)\actionPlayer{i}-\varepsilon\abs{\beta\actionPlayer{i}}
    =
    -\beta(\actionPlayer{i})^2 -\beta\actionPlayer{i}\actionPlayer{-i}+(\alpha-(c^i+\varepsilon\beta))\actionPlayer{i}.
\end{equation*}
This corresponds to the payoff of a Cournot competition with marginal cost $c^i+\varepsilon\beta$. Thus, in the Cournot competition, one can interpret strategically robust equilibria as Nash equilibria of a Cournot game with appropriately inflated production marginal costs.

\begin{APPENDICES}
\newpage

\section{Proofs}\label{sec:proofs}

\subsection{Preliminaries}

We start by recalling a notion of convergence in the space of probability measures: 

\begin{definition}[Narrow Convergence]\label{def:narrowConvergence}
Let $\spaceA$ be a Polish space. We say that $\probMeasureASeq$ \emph{converges to $\probMeasureA$ narrowly}, denoted by $\probMeasureASeqEl \narrowconvergence \probMeasureA$, if for all $\phi \in \Cb{\spaceA}$
\begin{equation*}
    \int_{\spaceA} \phi(\elA) \dprobMeasureASeqElArgs{\elA} \to \int_{\spaceA} \phi(\elA) \dprobMeasureAArgs{\elA}.
\end{equation*}
\end{definition}


We also recall from~\citet[Theorem 17.31]{Aliprantis2005InfiniteDimensionalAnalyis} and~\citet[Theorem 9.17]{Sundaram1996OptimizationTheory} Berge Maximum Theorem: 

\begin{theorem}[Berge Maximum Theorem]\label{th:maximumTheorem}
Let $\spaceA,\Theta$ be Polish spaces, $f: \cartProd{\spaceA}{\Theta} \goesto \reals$ continuous and $C: \Theta \goesto \powerSet{\spaceA}$ a compact-valued correspondence such that $C(\theta) \neq \emptyset$ for all $\theta \in \Theta$. Let the value function $f^*:\Theta \goesto \reals$ be given as
\begin{equation*}
    f^*(\theta) = \sup_{\elA \in C(\theta)} f(x, \theta)
\end{equation*}
and the set of maximizers $C^*: \Theta \goesto \spaceA$ as
\begin{equation*}
    C^*(\theta) = \argmax_{\elA \in C(\theta)} f(x, \theta).
\end{equation*}
Then, the following statements hold: 
\begin{enumerate}
    \item If $C$ is hemicontinuous (i.e. both upper and lower hemicontinuous), then $f^*$ is continuous, and $C^*$ is upper hemicontinuous, non-empty, and compact-valued. 
    \item If additionally $f$ is concave in $\elA$ for each fixed $\theta$ and $C$ is convex-valued, then $C^*$ is also a convex-valued correspondence. 
\end{enumerate}
\end{theorem}

\subsection{Auxiliary Results}

\begin{lemma}[Continuity of the product distribution]\label{lemma:continuity product distribution}
Let $\spaceA$ and $\spaceB$ be compact Polish spaces. 
The map
\begin{equation}
\begin{aligned}
\times : \; \probSpace{\spaceA}\times\probSpace{\spaceA}  &\goesto \probSpace{\spaceA\times\spaceB} \\
 (\mu,\nu) &\mapsto \mu\times\nu 
\end{aligned}
\end{equation}
is continuous w.r.t. narrow convergence. 
\end{lemma}

\proof{Proof}
Let $\probMeasureASeq$ be a sequence of probability measures in $\spaceProbMeasures{\spaceA}$ with $\probMeasureASeqEl \narrowconvergence \probMeasureA$ and $\probMeasureBSeq$ a sequence in $\spaceProbMeasures{\spaceB}$ such that $\probMeasureBSeqEl \narrowconvergence \probMeasureB$.
We need to show $\mu_n\times\nu_n\narrowconvergence\mu\times\nu$. Consider initially $\phi(x,y)=\phi_1(x)\phi_2(y)$ with $\phi_1\in\Cb{\spaceA}$ and $\phi_2\in\Cb{\spaceB}$. Then, 
\begin{align*}
    \lim_{n\to\infty}\int_{\spaceA\times\spaceB}\phi(x,y)\mathrm{d}(\mu_n\times\nu_n)(x,y)
    &=
    \lim_{n\to\infty}\int_{\spaceA}\phi_1(x)\mathrm{d}\mu_n(x)\int_{\spaceA}\phi_1(x)\mathrm{d}\mu_n(x)
    \\
    &=
    \lim_{n\to\infty}\int_{\spaceA}\phi_1(x)\mathrm{d}\mu_n(x) \lim_{n\to\infty}\int_{\spaceB}\phi_2(y)\mathrm{d}\nu_n(y)
    \\
    &=
    \int_{\spaceA}\phi_1(x)\mathrm{d}\mu(x) \int_{\spaceB}\phi_2(y)\mathrm{d}\nu(y)
    \\
    &=
    \int_{\spaceA\times\spaceB}\phi(x,y)\mathrm{d}(\mu\times\nu)(x,y).
\end{align*}
By linearity, we conclude that $\int_{\spaceA\times\spaceB}\phi(x,y)\mathrm{d}(\mu_n\times\nu_n)(x,y)\to\int_{\spaceA\times\spaceB}\phi(x,y)\mathrm{d}(\mu\times\nu)(x,y)$ for all functions $\phi$ of the form $\sum_{i=1}^N\phi_{1,i}(x)\phi_{2,i}(y)$. By Stone's Weierstrass theorem (e.g., see~\cite[Lemma 7.3]{santambrogio2015optimal} for a similar argument), such functions are dense in $\Cb{\spaceA\times\spaceB}$ when $\spaceA\times\spaceB$ are (locally) compact, and so $\int_{\spaceA\times\spaceB}\phi(x,y)\mathrm{d}(\mu_n\times\nu_n)(x,y)\to\int_{\spaceA\times\spaceB}\phi(x,y)\mathrm{d}(\mu\times\nu)(x,y)$ for all bounded continuous functions $\phi:\spaceA\times\spaceB\to\reals$. Thus, $\mu_n\times\nu_n\narrowconvergence\mu\times\nu$.
\endproof  

\begin{proposition}[Continuity of optimal transport ambiguity sets]\label{prop:continuity wasserstein ball}
    Let $X$ be a compact Polish space and $\varepsilon\in\posReals$.
    The set-valued map
    \begin{equation}\label{eq:wassersteinCorrespondence}
    \begin{aligned}
    \wassersteinCorrespondence : \; \spaceProbMeasures{\spaceA} &\goesto \powerSet{\spaceProbMeasures{\spaceA}} \\
     \probMeasureA &\mapsto \ambiguitySet{}{\varepsilon}{\probMeasureA}
     \coloneqq\{\probMeasureB\in\spaceProbMeasures{\spaceA}:\wasserstein[\wassnorm](\probMeasureA,\probMeasureB)\leq\varepsilon\}
    \end{aligned}
    \end{equation}
    is non-empty, compact-valued, and hemicontinuous (i.e., both upper and lower hemicontinuous). 
\end{proposition}

\proof{Proof}
    To start, $\probMeasureA\in\ambiguitySet{}{\varepsilon}{\probMeasureA}$ and $\wassersteinCorrespondence$ is never empty. 
    We then prove upper hemicontinuity and lower hemicontinuity separately. 

    \textbf{Upper hemicontinuity and compactness:}   
    Let $\probMeasureASeq$ be a sequence of probability measures in $\spaceProbMeasures{\spaceA}$ with $\probMeasureASeqEl \narrowconvergence \probMeasureA$ and $\probMeasureBSeq$ a sequence in $\spaceProbMeasures{\spaceA}$ such that $\probMeasureBSeqEl \in \wassersteinCorrespondence(\probMeasureASeqEl)=\ambiguitySet{}{\varepsilon}{\probMeasureASeqEl}$. We need to show that (i) $\probMeasureBSeqEl$ converges narrowly (up to subsequences) to some $\probMeasureB$ and (ii) $\probMeasureB \in \wassersteinCorrespondence(\probMeasureA) =\ambiguitySet{}{\varepsilon}{\probMeasureA}$. This way, we can deploy~\cite[Theorem 17.20]{Aliprantis2005InfiniteDimensionalAnalyis} and readily establish upper hemicontinuity and compactness. 
    For (i), narrow convergence (up to subsequences) of $\probMeasureBSeqEl$ to some  $\probMeasureB$ is ensured by compactness of $\spaceProbMeasures{\spaceA}$, which is a consequence of $\spaceA$ being compact~\cite[Proposition 7.22]{Bertsekas1996SOC}. For (ii), lower semicontinuity of the Wasserstein distance w.r.t. to narrow convergence~\cite[Proposition 7.1.3]{ambrosio2005gradient} gives 
    \begin{equation*}
        \wasserstein(\probMeasureA,\probMeasureB)
        \leq 
        \liminf_{n\to\infty}\wasserstein(\probMeasureASeqEl,\probMeasureBSeqEl)
        \leq 
        \varepsilon,
    \end{equation*}
    where we used that $\probMeasureBSeqEl \in \ambiguitySet{}{\varepsilon}{\probMeasureASeqEl}$ is equivalent to $\wasserstein(\probMeasureASeqEl,\probMeasureBSeqEl)\leq \varepsilon$.
    This directly implies $\probMeasureB \in \ambiguitySet{}{\varepsilon}{\probMeasureA}$.\footnote{The statement remains valid if the optimal transport distance is defined in terms of a general transportation cost $c:\spaceA\times\spaceA\to\reals$ as long as $c$ satisfies the assumptions of~\cite[Theorem 
    5.10]{villani2009optimal} (i.e., it is lower semicontinuous and lower bounded by an $(\mu\times\nu)$-integrable function $(x,y)\mapsto c_X(x)+c_Y(y)$).}

    \textbf{Lower hemicontinuity:}
    Rather than directly proving lower hemicontinuity of $\wassersteinCorrespondence$, we prove lower hemicontinuity of the modified map 
    \begin{align*}
        \wassersteinOpenCorrespondence : \; \spaceProbMeasures{\spaceA} &\goesto \powerSet{\spaceProbMeasures{\spaceA}} \\
        \probMeasureA &\mapsto \wassersteinBallOpen{\probMeasureA}
        \coloneqq\{\probMeasureB\in\spaceProbMeasures{\spaceA}:\wasserstein[\wassnorm](\probMeasureA,\probMeasureB)<\varepsilon\},
    \end{align*}
    which maps $\mu$ to the \emph{open} optimal transport ambiguity set around $\mu$. 
    With this, we then argue that $\wassersteinCorrespondence$ is the closure (w.r.t. narrow convergence) of $\wassersteinOpenCorrespondence$ and use~\cite[Lemma 17.22]{Aliprantis2005InfiniteDimensionalAnalyis} to conclude. 
    \begin{enumerate}
        \item Lower hemicontinuity of $\wassersteinOpenCorrespondence$:
        Let $\probMeasureASeq$ be a sequence of probability measures in $\spaceProbMeasures{\spaceA}$ with $\probMeasureASeqEl \narrowconvergence \probMeasureA$ and $\probMeasureB \in \ambiguitySetOpen{}{\varepsilon}{\probMeasureA}$. We need to show that there exists a subsequence of $\probMeasureASeq$, denoted by $\probMeasureASubSeq \in \spaceProbMeasures{\spaceA}$, and a sequence $\left(\probMeasureBSeqEl[k]\right)_{k\in\naturals}\in \spaceProbMeasures{\spaceA}$ such that $\probMeasureBSeqEl[k] \narrowconvergence \probMeasureB$ and $\probMeasureBSeqEl[k] \in \wassersteinCorrespondence(\probMeasureASubSeqEl)=\ambiguitySetOpen{}{\varepsilon}{\probMeasureASubSeqEl}$. Lower hemicontinuity follows then from~\cite[Theorem 17.21]{Aliprantis2005InfiniteDimensionalAnalyis}.
        By definition of $\ambiguitySetOpen{}{\varepsilon}{\probMeasureA}$, there is $\delta\in(0,\varepsilon)$ such that $\probMeasureB \in \ambiguitySet{}{\varepsilon - \delta}{\probMeasureA}$. Let $\probMeasureASubSeq$ be an arbitrary subsequence of $\probMeasureASubSeq$ so that $\wasserstein[\wassnorm](\probMeasureA,\probMeasureASubSeqEl) < \delta$ for all $k \in \naturals$. As $\probMeasureASubSeqEl\narrowconvergence\probMeasureA$ and $X$ is compact, $\wasserstein[\wassnorm](\probMeasureA,\probMeasureASubSeqEl)\to 0$ (indeed, the distance function is continuous and bounded on $\spaceA$~\cite[Definition 6.8 and Theorem 6.9]{villani2009optimal}) and the subsequence is well-defined. Now, let
        $
        \probMeasureBSeqEl[k]
        \coloneqq 
        \probMeasureB.
        $
        Obviously, $\probMeasureBSeqEl[k]\narrowconvergence\probMeasureB$.
        We claim that $\probMeasureBSeqEl[k] \in \ambiguitySetOpen{}{\varepsilon}{\probMeasureASubSeqEl}$. Since, by construction, $\wasserstein[\wassnorm](\probMeasureA,\probMeasureASubSeqEl) < \delta$, triangle inequality~\cite[§6]{villani2009optimal} yields
        \begin{equation*}
            \wasserstein(\probMeasureBSeqEl[k],\probMeasureASubSeqEl)
            =
            \wasserstein(\probMeasureB,\probMeasureASubSeqEl)
            \leq 
            \wasserstein(\probMeasureB,\probMeasureA) + \wasserstein(\probMeasureA,\probMeasureASubSeqEl)
            <
            (\varepsilon - \delta) + \delta = \varepsilon.
        \end{equation*}
        Thus, $\probMeasureBSeqEl[k] \in \wassersteinOpenCorrespondence(\probMeasureASubSeqEl)=\ambiguitySetOpen{}{\varepsilon}{\probMeasureASubSeqEl}$ for all $k\in\naturals$. This proves that the set-valued map $\wassersteinOpenCorrespondence$ is lower hemicontinuous.

        \item The closure of $\wassersteinOpenCorrespondence$ is $\wassersteinCorrespondence$ (i.e., the closure w.r.t. narrow convergence of  $\wassersteinOpenCorrespondence(\probMeasureA)$ is $\wassersteinCorrespondence(\probMeasureA)$ for all $\probMeasureA$): We prove the two inclusions separately. 

        ``$\subset$'': 
        Let $\probMeasureB$ be in the closure (w.r.t. narrow convergence) of $ \wassersteinOpenCorrespondence(\probMeasureA)=\ambiguitySetOpen{}{\varepsilon}{\probMeasureA}$. By definition of closure, there exists a sequence  $\probMeasureBSeq$ with $\probMeasureBSeqEl\in\ambiguitySetOpen{}{\varepsilon}{\probMeasureA}$ so that $\probMeasureBSeqEl\narrowconvergence\probMeasureB$. Then, lower semicontinuity of the Wasserstein distance w.r.t. narrow convergence~\cite[Proposition 7.1.3]{ambrosio2005gradient} gives  
        \begin{equation*}
            \wasserstein(\probMeasureA,\probMeasureB)
            \leq 
            \liminf_{n\to\infty}\wasserstein(\probMeasureA,\probMeasureBSeqEl)
            \leq
            \varepsilon,
        \end{equation*}
        where we used that $\probMeasureBSeqEl\in\ambiguitySetOpen{}{\varepsilon}{\probMeasureA}$ is equivalent to $\wasserstein(\probMeasureA,\probMeasureBSeqEl)<\varepsilon$. Therefore, $\wasserstein(\probMeasureA,\probMeasureB)\leq \varepsilon$ and $\probMeasureB\in\wassersteinCorrespondence(\probMeasureA)=\ambiguitySet{}{\varepsilon}{\probMeasureA}$, proving that the closure of $\wassersteinOpenCorrespondence(\probMeasureA)$ is contained in $\wassersteinCorrespondence(\probMeasureA)$.

        ``$\supset$'': Let $\probMeasureB\in \wassersteinCorrespondence(\probMeasureA)=\ambiguitySet{}{\varepsilon}{\probMeasureA}$. If $\wasserstein(\probMeasureA,\probMeasureB)<\varepsilon$, we directly have $\probMeasureB\in \wassersteinOpenCorrespondence(\probMeasureA)=\ambiguitySetOpen{}{\varepsilon}{\probMeasureA}$ and we are done. Thus, we can assume $\wasserstein(\probMeasureA,\probMeasureB)=\varepsilon$. Consider the sequence of probability measures $\probMeasureBSeq$ defined via the standard interpolation between $\probMeasureA$ and $\probMeasureB$:
        \begin{equation*}
            \probMeasureBSeqEl \coloneqq \left(1-\frac{1}{n}\right)\probMeasureB + \frac{1}{n}\probMeasureA.
        \end{equation*}
        First, we claim $\probMeasureBSeqEl\narrowconvergence\probMeasureB$. Indeed, for any $\phi\in\Cb{X}$, we have 
        \begin{equation*}
            \int_{X}\phi(x)\mathrm{d}\probMeasureBSeqEl(x)=\left(1-\frac{1}{n}\right)\int_{X}\phi(x)\mathrm{d}\nu(x)+\frac{1}{n}\int_{X}\phi(x)\mathrm{d}\mu(x)
            \to \int_{X}\phi(x)\mathrm{d}\probMeasureB(x),
        \end{equation*}
        since both integrals are finite. 
        Second, we claim $\probMeasureBSeqEl\in\ambiguitySetOpen{}{\varepsilon}{\probMeasureA}$.
        By Kantorovich duality~\citep[Theorem 5.10]{villani2009optimal}, the $\wassnorm$ power of the type-$\wassnorm$ Wasserstein distance is the supremum of linear functions, and it is therefore convex (see also~\citet[§7.2]{santambrogio2015optimal}). Thus, 
        \begin{equation*}
            \wasserstein(\probMeasureA,\probMeasureBSeqEl)^\wassnorm
            \leq 
            \left(1-\frac{1}{n}\right)\wasserstein(\probMeasureA,\probMeasureB)^\wassnorm+\frac{1}{n}\wasserstein(\probMeasureA,\probMeasureA)^\wassnorm
            =
            \left(1-\frac{1}{n}\right)\varepsilon^\wassnorm
        \end{equation*}
        and so for all $n\in\naturals$ we have 
        \begin{equation*}
            \wasserstein(\probMeasureA,\probMeasureBSeqEl)\leq \left(1-\frac{1}{n}\right)^\frac{1}{\wassnorm}\varepsilon<\varepsilon.
        \end{equation*}
        Thus, $\probMeasureBSeqEl\in\ambiguitySetOpen{}{\varepsilon}{\probMeasureA}$ and $\probMeasureB$ belongs to the closure (w.r.t. narrow convergence)  of $\ambiguitySetOpen{}{\varepsilon}{\probMeasureA}$, and $\wassersteinCorrespondence(\probMeasureA)$ is contained in the closure of $\wassersteinOpenCorrespondence(\probMeasureA)$.
    \end{enumerate}
\endproof

\subsection{Proofs for \cref{sec:strategically_robust_game_theory}}
\label{app:proofs:existence}

\proof{Proof of~\cref{thm:existence}}
We prove the statement in the general case of compact Polish spaces $\actionSpacePlayerI$ and continuous payoffs\footnote{In particular, this covers also the discrete case, as any function $f:S\to\reals$ from a discrete set $S$ to $\reals$ is continuous.}, which covers both cases in~\cref{ass:action-utilities}. The proof unfolds in several steps. 

\textbf{Preliminaries:}
Since $\actionSpacePlayerI$ is compact for all $i$, so $\actionSpace = \prod_i\actionSpacePlayerI$ is compact too.
We now prove that the function
\begin{equation*}
\begin{aligned}
    \utilityExpPlayerI
    : \probSpace{\actionSpacePlayerI}\times \probSpace{{\textstyle\prod_{j\neq i}\actionSpacePlayer{j}}} & \to \reals
    \\
    (\stratPlayerI,\stratWassersteinPlayerIMinus) & \mapsto 
    \utilityExpArgsPlayerIW=\expectedValue{(\actionPlayerI,\actionPlayerIMinus)\sim(\stratPlayerI,\stratWassersteinPlayerIMinus)}{\utilityArgsPlayerI},
\end{aligned}
\end{equation*}
as defined in~\cref{def:SRE}, is continuous w.r.t. narrow convergence. Indeed, $\utilityExpPlayerI$ can be seen as the composition of the map $(\stratPlayerI,\stratWassersteinPlayerIMinus) \mapsto \stratPlayerI\times\stratWassersteinPlayerIMinus$, which is continuous by~\cref{lemma:continuity product distribution}, and the map $\sigma\mapsto\expectedValue{\actions\sim\sigma}{\utilityPlayerI(\actions)}$, which is continuous since the payoff is continuous and bounded (being defined over a compact set). Since the composition of continuous functions is continuous, $\utilityExpPlayerI$ is continuous.

\textbf{Properties of the worst-case payoff:}
Consider now the function 
\begin{equation*}
\begin{aligned}
    \utilityWassersteinPlayer{i}
    : \probSpace{\actionSpacePlayerI}\times {\textstyle\prod_{j\neq i}\probSpace{\actionSpacePlayer{j}}} & \to \reals
    \\
    (\stratPlayerI,\stratPlayerIMinus) & \mapsto 
    \min_{\stratWassersteinPlayerIMinus \in \ambiguitySet{i}{\varepsilon}{\stratPlayerIMinus}} \utilityExpArgsPlayerIW.
\end{aligned}
\end{equation*}
We now use Berge's Maximum Theorem to show that $\utilityWassersteinPlayer{i}$ is well-defined and study its properties.
In terms of notation of~\cref{th:maximumTheorem}, $\Theta = \prod_{j}\probSpace{\actionSpacePlayer{j}}$, $\spaceA = \prod_{j\neq i}\probSpace{\actionSpacePlayer{j}}$, $\theta = (\stratPlayerI, \stratPlayerIMinus)$, $f(\theta, \elA) = \utilityExpArgsPlayerIW$, $f^*(\theta) = \utilityWassersteinArgsPlayerI$, and $C(\theta) = \ambiguitySetMap{i}{\varepsilon}(\stratPlayerIMinus)$.
By \cref{assumption:ambiguity}, the set-valued map $\stratPlayerIMinus\mapsto\ambiguitySetMap{i}{\varepsilon}(\stratPlayerIMinus)$ is non-empty, compact-valued, and hemicontinuous. Since, as argued above, the function $\utilityExpPlayerI$ is continuous, Berge's Maximum Theorem theorem shows that $\utilityWassersteinPlayer{i}$ is well-defined (in particular, the minimum exists and it is finite) and continuous. We now prove that $\utilityWassersteinPlayer{i}$ is concave in $\stratPlayerI$ for fixed $\stratPlayerIMinus$. Indeed, for $\lambda \in [0,1]$, we have
\begin{align*}
    \utilityWassersteinArgsPlayer{i}{\lambda &\stratPlayerI_1 + (1 - \lambda)\stratPlayerI_2}{\stratPlayerIMinus}
    \\
    &=
    \min_{\stratWassersteinPlayerIMinus \in \ambiguitySet{i}{\varepsilon}{\stratPlayerIMinus}} \expectedValue{\actionPlayerI\sim\lambda \stratPlayerI_1
    + (1 - \lambda)\stratPlayerI_2}{ \expectedValue{\actionPlayerIMinus\sim\stratWassersteinPlayerIMinus}{\utilityArgsPlayerI}}
    \\
    &=
    \min_{\stratWassersteinPlayerIMinus \in \ambiguitySet{i}{\varepsilon}{\stratPlayerIMinus}}
    \lambda \expectedValue{\actionPlayerI\sim\stratPlayerI_1}{\expectedValue{\actionPlayerIMinus\sim\stratWassersteinPlayerIMinus}{\utilityArgsPlayerI}}
    + (1 - \lambda) \expectedValue{\actionPlayerI\sim\stratPlayerI_2}{\expectedValue{\actionPlayerIMinus\sim\stratWassersteinPlayerIMinus}{\utilityArgsPlayerI}}
    \\
    &\geq
    \lambda
    \min_{\stratWassersteinPlayerIMinus \in \ambiguitySet{i}{\varepsilon}{\stratPlayerIMinus}}
    \expectedValue{(\actionPlayerI,\actionPlayerIMinus)\sim(\stratPlayerI_1, \stratWassersteinPlayerIMinus)}{\utilityArgsPlayerI}
    + (1 - \lambda) \min_{\stratWassersteinPlayerIMinus \in \ambiguitySet{i}{\varepsilon}{\stratPlayerIMinus}} \expectedValue{(\actionPlayerI,\actionPlayerIMinus)\sim(\stratPlayerI_2, \stratWassersteinPlayerIMinus)}{\utilityArgsPlayerI}
    \\
    &= \lambda \utilityWassersteinArgsPlayer{i}{\stratPlayerI_1 }{\stratWassersteinPlayerIMinus} + (1 - \lambda) \utilityWassersteinArgsPlayer{i}{\stratPlayerI_2}{\stratWassersteinPlayerIMinus}.
\end{align*}
This directly establishes concavity.

\textbf{The best response map:}
The strategically robust best response of player $i$ reads
\begin{equation*}
\begin{aligned}
    \stratRobustBestResponseMapPlayerI:
    {\textstyle\prod_{j\neq i}\probSpace{\actionSpacePlayer{j}}} &\to \powerSet{\probSpace{\actionSpacePlayerI}}\\
    \stratPlayerIMinus &\mapsto \argmax_{\stratPlayerI \in \probSpace{\actionSpacePlayerI}} \utilityWassersteinArgsPlayerI.
\end{aligned}
\end{equation*}
We can now combine all these best responses in the map 
\begin{equation*}
\begin{aligned}
    \stratRobustBestResponseMap:
    {\textstyle\prod_{j}\probSpace{\actionSpacePlayer{j}}} &\to {\textstyle\prod_{j}\powerSet{\probSpace{\actionSpacePlayer{j}}}} \\
    \stratPlayerIMinus &\mapsto (\stratRobustBestResponseMapPlayer{1}(\stratPlayer{-1}),\ldots,\stratRobustBestResponseMapPlayer{N}(\stratPlayer{-N})).
\end{aligned}
\end{equation*}

\textbf{Existence of a fixed point of the best response map:}
To complete the proof, we can now follow the lines of Nash's proof for existence of a mixed Nash equilibrium in a finite $N$-player game \cite{Nash1951EquilibriaNPlayerGames}, generalizing it to the case of compact spaces. In particular, we apply Glicksberg's fixed point theorem~\cite{Glicksberg1952GeneralizedKakutaniNE} on the strategically robust best response map $\stratRobustBestResponseMap$. We proceed in several steps:
\begin{enumerate}
    \item The domain of the best response, $\prod_{j}\probSpace{\actionSpacePlayer{j}}$, is non-empty, compact, and convex.

    \item The best response map $\stratRobustBestResponseMap$ has closed graph: We only need to focus $\stratRobustBestResponseMapPlayerI$, as the statement for $\stratRobustBestResponseMap$ follows directly. Let us first prove upper hemicontinuity and compactness, which follows from Berge's Maximum Theorem, since (i) the probability space $\probSpace{\actionPlayerI}$ is non-empty and compact~\cite[Proposition 7.22]{Bertsekas1996SOC} and (ii) the function $\utilityWassersteinPlayer{i}$ is continuous. We can then deploy the Closed Graph Theorem \cite[Theorem 17.11]{Aliprantis2005InfiniteDimensionalAnalyis} to conclude that $\stratRobustBestResponseMapPlayerI$ has a closed graph.   
    
    \item Non-emptyness and convexity of $\stratRobustBestResponseMap(\stratPlayerIMinus)$: Again, we only need to focus on $\stratRobustBestResponseMapPlayerI(\stratPlayerIMinus)$, as the statement $\stratRobustBestResponseMap(\stratPlayerIMinus)$ follows directly. 
    Since $\utilityWassersteinPlayer{i}$ is continuous and the probability space over a compact set is compact, the argmax is never empty. Convexity, instead, follows again from Berge's Maximum Theorem, since (i) the probability space $\probSpace{\actionPlayerI}$ is non-empty, convex, and compact, (ii) the function $\utilityWassersteinPlayer{i}$ is continuous and concave is $\stratPlayerI$.

    \item Existence of a fixed point: Existence of a fixed point follows now from Glicksberg's fixed point theorem applied to the best response map $\stratRobustBestResponseMap$.
\end{enumerate}
Thus, there exists a fixed point $\optStrat \in \stratRobustBestResponseArgs{\optStrat}$, which is equivalent to 
\begin{equation*}
    \optStratPlayerI \in \arg \max_{\stratPlayerI\in \strategySpacePlayerI} \min_{\stratWassersteinPlayerIMinus \in \ambiguitySet{i}{\varepsilon}{\stratPlayerIMinus}} \utilityExpArgsPlayerIW ,
    \qquad \forall \; i\in\{1,\ldots,N\}.
\end{equation*}
This point is precisely a strategically robust equilibrium. 
\endproof

\proof{Proof of~\cref{lemma:continuity wasserstein ball}}
As for~\cref{thm:existence}, we prove the statement in the more general case of compact Polish spaces $\actionSpacePlayerI$, which covers both cases in~\cref{ass:action-utilities}.
To start, the map $\stratPlayerIMinus\mapsto\ambiguitySet{i}{\varepsilon}{\stratPlayerIMinus}$ can be seen as the composition of the function $\prod_{j\neq i}\probSpace{\actionSpacePlayer{j}}\ni\stratPlayerIMinus\mapsto \stratWasserstein_{\stratPlayerIMinus}\in\probSpace{\prod\actionSpacePlayer{j}}$, taking the product of the distributions in $\stratPlayerI$, with the set-valued map $\wassersteinCorrespondence$ defined in~\cref{prop:continuity wasserstein ball}. We can therefore study these two maps and their composition. We proceed in three steps: 
\begin{enumerate}
    \item We proved in~\cref{lemma:continuity product distribution} that the map $\stratPlayerIMinus\mapsto \stratWasserstein_{\stratPlayerIMinus}$ is continuous w.r.t. narrow convergence. 

    \item We proved in \cref{prop:continuity wasserstein ball} that the set-valued map $\wassersteinCorrespondence$ is non-empty, compact-valued, and hemicontinuous w.r.t. narrow convergence. 

    \item Since $\wassersteinCorrespondence$ is non-empty and compact-valued, $\stratPlayerIMinus\mapsto\ambiguitySet{i}{\varepsilon}{\stratPlayerIMinus}$ is also non-empty and compact valued. Thus, we only need to prove hemicontinuity. Since $\stratPlayerIMinus\mapsto \stratWasserstein_{\stratPlayerIMinus}$ is continuous, the induced set-valued map $\stratPlayerIMinus\mapsto \{\stratWasserstein_{\stratPlayerIMinus}\}$ is hemicontinuous~\cite[Lemma 17.5]{Aliprantis2005InfiniteDimensionalAnalyis}. Since the composition of hemicontinuous set-valued maps is hemicontinuous~\cite[Theorem 17.23]{Aliprantis2005InfiniteDimensionalAnalyis}, we conclude that $\stratPlayerIMinus\mapsto\ambiguitySet{i}{\varepsilon}{\stratPlayerIMinus}$ is hemicontinuous too. 
\end{enumerate}
\endproof

\proof{Proof of~\cref{corollary:existence:optimal_transport}}
    Without loss of generality, we assume $\varepsilon > 0$; else, strategically robust equilibria coincide with standard Nash equilibria, which are guaranteed to exist under~\cref{ass:action-utilities}. 
    Then, the result follows directly from~\cref{thm:existence} and \cref{lemma:continuity wasserstein ball}.
\endproof

\subsection{Proofs for \cref{sec:FiniteNPlayerGames}}

We prove a more general version of~\cref{prop:finite:strategically robust best response}, where the agents' actions spaces are compact Polish spaces and the payoffs are continuous. These two assumptions are in particular satisfied when the action spaces are finite: 

\begin{proposition}[Reformulation of the strategically robust best response]\label{prop:strategically robust best response}
    Let $\actionSpacePlayerI$ be compact Polish spaces and $\utilityPlayerI$ be continuous. Then, the optimization problem for the strategically robust can be reformulated as follows: 
    \begin{equation}\label{eq:strategically robust best response:reformulation}
    \begin{aligned}
        \sup_{\stratPlayerI \in \strategySpacePlayerI} &\inf_{\stratWassersteinPlayerIMinus \in \ambiguitySet{i}{\varepsilon}{\stratPlayerIMinus}} \expectedValue{(\actionPlayerI,\actionPlayerIMinus)\sim(\stratPlayerI,\stratWassersteinPlayerIMinus)}{\utilityArgsPlayerI}
        \\
        &=
        \max_{\stratPlayerI \in \strategySpacePlayerI, \lambda^i \geq 0} \expectedValue{\actionPlayerIMinus\sim\stratWasserstein_{\stratPlayerIMinus}}{\min_{\surActionPlayerIMinus \in \actionSpacePlayerIMinus} \left\{ \expectedValue{\actionPlayerI\sim\stratPlayerI}{\utilityArgsPlayer{i}{\actionPlayerI}{\surActionPlayerIMinus}} + \lambda^i \distanceArgsPlayer{-i}{\actionPlayerIMinus}{\surActionPlayerIMinus}^\wassnorm \right\} } - \lambda^i \varepsilon^\wassnorm.
    \end{aligned}
    \end{equation}
\end{proposition}

\proof{Proof}
To ease notation, we denote the value of the strategically robust best response as 
\begin{align*}
    \utilityExpPlayer{i}_{\text{SR}}(\stratPlayerIMinus)
    &\coloneqq 
    \sup_{\stratPlayerI \in \strategySpacePlayerI} \inf_{\stratWassersteinPlayerIMinus \in \ambiguitySet{i}{\varepsilon}{\stratPlayerIMinus}} \expectedValue{(\actionPlayerI,\actionPlayerIMinus)\sim(\stratPlayerI,\stratWassersteinPlayerIMinus)}{\utilityArgsPlayerI}
    \\
    &=
    - \inf_{\stratPlayerI \in \strategySpacePlayerI} \sup_{\stratWassersteinPlayerIMinus \in \ambiguitySet{i}{\varepsilon}{\stratPlayerIMinus}} \expectedValue{(\actionPlayerI,\actionPlayerIMinus)\sim(\stratPlayerI,\stratWassersteinPlayerIMinus)}{-\utilityArgsPlayerI}.
\end{align*}
Since $\actionSpace$ is a Polish space and the transport cost, $(\distance^{-i})^\wassnorm$, used in the Wasserstein distance is lower semicontinuous (by definition since it is a proper distance function), we can use \citet[Theorem 1]{blanchet2019quantifying}, to find the dual of the inner optimization problem:
\begin{align*}
    \sup_{\stratWassersteinPlayerIMinus \in \ambiguitySet{i}{\varepsilon}{\stratPlayerIMinus}}& \expectedValue{\actionPlayerIMinus\sim\stratWasserstein_{\stratPlayerIMinus}}{\expectedValue{\actionPlayerI\sim\stratPlayerI}{-\utilityArgsPlayerI}}
    \\
    &=
    \inf_{\lambda^i \geq 0} \lambda^i \varepsilon^\wassnorm + \expectedValue{\actionPlayerIMinus\sim\stratWasserstein_{\stratPlayerIMinus}}{\sup_{\surActionPlayerIMinus \in \actionSpacePlayerIMinus} \left\{\expectedValue{\,\stratPlayerI}{-\utilityArgsPlayer{i}{\actionPlayerI}{\surActionPlayerIMinus}} - \lambda^i \distanceArgsPlayer{-i}{\actionPlayerIMinus}{\surActionPlayerIMinus}^\wassnorm \right\}}.
\end{align*}
Overall, we therefore conclude
\begin{align*}
    \utilityExpPlayer{i}_{\text{SR}}(\stratPlayerIMinus)
    &=
    \sup_{\stratPlayerI \in \strategySpacePlayerI} \inf_{\stratWassersteinPlayerIMinus \in \ambiguitySet{i}{\varepsilon}{\stratPlayerIMinus}} \expectedValue{(\actionPlayerI,\actionPlayerIMinus)\sim(\stratPlayerI,\stratWassersteinPlayerIMinus)}{\utilityArgsPlayerI}
    \\
    &=
    - \inf_{\stratPlayerI \in \strategySpacePlayerI, \lambda^i \geq 0} \lambda^i \varepsilon^\wassnorm + \expectedValue{\actionPlayerIMinus\sim\stratWasserstein_{\stratPlayerIMinus}}{\sup_{\surActionPlayerIMinus \in \actionSpacePlayerIMinus} \left\{\expectedValue{\actionPlayerI\sim\stratPlayerI}{-\utilityArgsPlayer{i}{\actionPlayerI}{\surActionPlayerIMinus}}- \lambda^i \distanceArgsPlayer{-i}{\actionPlayerIMinus}{\surActionPlayerIMinus}^\wassnorm\right\} }
    \\
    &=
    \sup_{\stratPlayerI \in \strategySpacePlayerI, \lambda^i \geq 0} \expectedValue{\actionPlayerIMinus\sim\stratWasserstein_{\stratPlayerIMinus}}{-\sup_{\surActionPlayerIMinus \in \actionSpacePlayerIMinus} \left\{ \expectedValue{\actionPlayerI\sim\stratPlayerI}{-\utilityArgsPlayer{i}{\actionPlayerI}{\surActionPlayerIMinus}} - \lambda^i \distanceArgsPlayer{-i}{\actionPlayerIMinus}{\surActionPlayerIMinus}^\wassnorm \right\} } - \lambda^i \varepsilon^\wassnorm
    \\
    &=
    \sup_{\stratPlayerI \in \strategySpacePlayerI, \lambda^i \geq 0} \expectedValue{\actionPlayerIMinus\sim\stratWasserstein_{\stratPlayerIMinus}}{\inf_{\surActionPlayerIMinus \in \actionSpacePlayerIMinus} \left\{ \expectedValue{\actionPlayerI\sim\stratPlayerI}{\utilityArgsPlayer{i}{\actionPlayerI}{\surActionPlayerIMinus}} + \lambda^i \distanceArgsPlayer{-i}{\actionPlayerIMinus}{\surActionPlayerIMinus}^\wassnorm \right\} } - \lambda^i \varepsilon^\wassnorm.
\end{align*}
Since all spaces and compact and all functions are continuous, we can replace the suprema and infima with maxima and minima. 
\endproof

\proof{Proof of~\cref{prop:finite:strategically robust best response}}
    The result follows directly from~\cref{prop:strategically robust best response}.
\endproof


We now prove~\cref{th:computationalComplexity}. We start with two preliminary results. First, we show that the search space for dual multiplier $\lambda^i$ can without loss of generality be restricted to a compact set. As above, we prove this result in the more general case of compact Polish spaces: 

\begin{lemma}[Uniform boundedness of the dual multiplier]\label{lemma:lambda_bounded}
Consider the setting of~\cref{prop:strategically robust best response} and the strategically robust best response~\eqref{eq:strategically robust best response:reformulation}. If $\varepsilon>0$, then $\lambda^i$ can without loss of generality be restricted to a compact set; i.e.,  there exists $M^i>0$ so that 
\begin{multline*}
    \max_{\stratPlayerI \in \strategySpacePlayerI, \lambda^i \geq 0} \expectedValue{\actionPlayerIMinus\sim\stratWasserstein_{\stratPlayerIMinus}}{\min_{\surActionPlayerIMinus \in \actionSpacePlayerIMinus} \left\{ \expectedValue{\actionPlayerI\sim\stratPlayerI}{\utilityArgsPlayer{i}{\actionPlayerI}{\surActionPlayerIMinus}} + \lambda^i \distanceArgsPlayer{-i}{\actionPlayerIMinus}{\surActionPlayerIMinus}^\wassnorm \right\} } - \lambda^i \varepsilon^\wassnorm
    \\
    =
    \max_{\stratPlayerI \in \strategySpacePlayerI, \lambda^i [0,M^i]} \expectedValue{\actionPlayerIMinus\sim\stratWasserstein_{\stratPlayerIMinus}}{\min_{\surActionPlayerIMinus \in \actionSpacePlayerIMinus} \left\{ \expectedValue{\actionPlayerI\sim\stratPlayerI}{\utilityArgsPlayer{i}{\actionPlayerI}{\surActionPlayerIMinus}} + \lambda^i \distanceArgsPlayer{-i}{\actionPlayerIMinus}{\surActionPlayerIMinus}^\wassnorm \right\} } - \lambda^i \varepsilon^\wassnorm
\end{multline*}
In particular, $M^i$ can be taken to be $2\max_{\actionPlayerI\in\actionSpacePlayerI,\actionPlayerIMinus\in\actionSpacePlayerIMinus}|\utilityPlayerI(\actionPlayerI,\actionPlayerIMinus)|/\varepsilon^\wassnorm<+\infty$.
\end{lemma}

\proof{Proof}
The payoff of agent $i$ is necessarily bounded by $C\coloneqq \max_{\actionPlayerI\in\actionSpacePlayerI,\actionPlayerIMinus\in\actionSpacePlayerIMinus}|\utilityPlayerI(\actionPlayerI,\actionPlayerIMinus)|$, which is finite since the payoff function is continuous and the action spaces are compact.
Then, for any $\optStratPlayerI$ and $\bar\lambda^i$ optimal in the best response optimization problem it necessarily holds:
\begin{align*}
    -C
    \overset{\heartsuit}&{\leq}
    \min_{\stratWassersteinPlayerIMinus \in \ambiguitySet{i}{\varepsilon}{\optStratPlayerIMinus}} \utilityExpArgsPlayerIW
    \\
    \overset{\diamondsuit}&{=}
    \max_{\stratPlayerI \in \strategySpacePlayerI, \lambda^i \geq 0} \expectedValue{\actionPlayerIMinus\sim\stratWasserstein_{\stratPlayerIMinus}}{\min_{\surActionPlayerIMinus \in \actionSpacePlayerIMinus} \left\{ \expectedValue{\actionPlayerI\sim\stratPlayerI}{\utilityArgsPlayer{i}{\actionPlayerI}{\surActionPlayerIMinus}} + \lambda^i \distanceArgsPlayer{-i}{\actionPlayerIMinus}{\surActionPlayerIMinus}^\wassnorm \right\} } - \lambda^i \varepsilon^\wassnorm
    \\
    &=
    \expectedValue{\actionPlayerIMinus\sim\stratWasserstein_{\stratPlayerIMinus}}{\min_{\surActionPlayerIMinus \in \actionSpacePlayerIMinus} \left\{ \expectedValue{\actionPlayerI\sim\optStratPlayerI}{\utilityArgsPlayer{i}{\actionPlayerI}{\surActionPlayerIMinus}} + \bar\lambda^i \distanceArgsPlayer{-i}{\actionPlayerIMinus}{\surActionPlayerIMinus}^\wassnorm \right\} } - \bar\lambda^i \varepsilon^\wassnorm
    \\
    \overset{\spadesuit}&{\leq}
    \expectedValue{\actionPlayerIMinus\sim\stratWasserstein_{\stratPlayerIMinus}}{\expectedValue{\actionPlayerI\sim\optStratPlayerI}{\utilityArgsPlayer{i}{\actionPlayerI}{\actionPlayerIMinus}}} - \bar\lambda^i \varepsilon^\wassnorm
    \\
    \overset{\heartsuit}&{\leq}
    -\bar\lambda^i\varepsilon^\wassnorm + C,
\end{align*}
where $\heartsuit$ follows from the definition of $C$, $\diamondsuit$ from~\cref{prop:strategically robust best response}, and $\spadesuit$ from selecting $\surActionPlayerIMinus=\actionPlayerIMinus$.
Thus, $\bar\lambda^i\leq 2C/\varepsilon$, which is precisely the desired statement. 
\endproof 

Second, we formally define $\delta$-approximate strategically robust equilibria\footnote{We use $\delta$ instead of $\varepsilon$, since $\varepsilon$ is already reserved for the radius of the ambiguity set.}:

\begin{definition}
Given a game $\game$ and $\delta\in \nonNegReals$, a tuple of strategies $(\optStratPlayer{1}, \dots, \optStratPlayer{N})$ is a \emph{$\delta$-approximate strategically robust equilibrium} with robustness level $\varepsilon\in\nonNegReals$ if we have 
\begin{equation*}
    \min_{\stratWassersteinPlayerIMinus\in\ambiguitySet{i}{\varepsilon}{\optStratPlayerIMinus}}
    \utilityExpArgsPlayer{i}{\optStratPlayerI}{\stratWassersteinPlayerIMinus}
    \geq
    \min_{\stratWassersteinPlayerIMinus\in\ambiguitySet{i}{\varepsilon}{\optStratPlayerIMinus}} 
    \utilityExpArgsPlayer{i}{\stratPlayerI}{\stratWassersteinPlayerIMinus}-\delta
    \quad\text{for all }\stratPlayerI \in \strategySpacePlayerI \text{ and all } i\in\{1,\ldots,N\}.
\end{equation*}
\end{definition}

We are now ready to prove~\cref{th:computationalComplexity}:

\proof{Proof of~\cref{th:computationalComplexity}}
Without loss of generality, we assume $\varepsilon>0$; else, strategically robust equilibria coincide with standard Nash equilibria, whose computation is well-known to be \ppad{}-complete and therefore to lie in \ppad{}.

Our proof consists of a reformulation as a standard concave game and on recent results on the complexity of concave games~\cite[Section 4]{papadimitriou2022computational}. In particular, we consider the concave game $\tilde{\mathcal{G}}$ where each agent selects $\stratPlayerI\in\strategySpacePlayerI$ (i.e., the probability over their actions) and $\lambda^i\geq 0$ (i.e., the dual multiplier) and the payoff is
\begin{equation}\label{eq:complexity proof:surrogate utility:definition}
    \tilde u^{i}((p^i,\lambda^i),(p^{-i},\lambda^{-i}))
    \coloneqq 
    - \lambda^i \varepsilon^\wassnorm  + 
    \expectedValue{\actionPlayerIMinus\sim\stratWassersteinPlayer{}_{\stratPlayerIMinus}}{
    \min_{\surActionPlayerIMinus\in\actionSpacePlayer{-i}}
    \left\{
    \expectedValue{\actionPlayerI\sim\stratPlayerI}{\utilityPlayer{i}(\actionPlayerI,\surActionPlayerIMinus)} + \lambda^i \distance^{-i}(\actionPlayerIMinus,\surActionPlayerIMinus)^\wassnorm
    \right\}}.
\end{equation}
In particular, the payoff of agent $i$ is independent of the dual multipliers $\lambda^{-i}$ of other players.
More specifically, the concave game $\tilde G$ is defined as follows: 
\begin{itemize}
    \item 
    The strategy domain of agent $i$ is $R^i\coloneqq\strategySpacePlayerI\times[0,M^i]$, where $M^i$ is the upper bound on $\lambda^i$ given by \cref{lemma:lambda_bounded}. Note also that $M^i$ is polynomially computable in the size of the game matrices, as it involves finding the maximum of $\cardinality{\actionSpacePlayerI}\cardinality{\actionSpacePlayerIMinus}$ many elements. 

    \item The payoff of each agent is $\tilde u^{i}$. Since the minimum in the definition of the payoff function is over a finite set of size $\cardinality{\actionSpacePlayer{-i}}$ and the minimum of linear functions is concave, $\tilde u^{i}$ is continuous over $R^1\times\ldots\times R^N$ and concave in $\stratPlayerI$ for fixed $\stratPlayerIMinus$.

    \item The convex constraint is simply the Cartesian product of the strategy spaces; i.e., $S=R^1\times\ldots\times R^N$.
\end{itemize}

We now specify the computational representation of the payoff functions and constraints. 

\textbf{Payoff function:} 
We represent the payoff functions via a linear arithmetic circuit, as predicated in~\cite[Definition 4.6]{papadimitriou2022computational}.
To start, notice that the evaluation of $\tilde u^i$ only involves the operations $\{+,-,\times,\max,\min\}$ (in particular, the minimum in the definition of the payoff function is over a finite set of size $\cardinality{\actionSpacePlayer{-i}}$) and that the payoff $\tilde u^i$ can be expressed as an arithmetic circuit whose size is polynomial in the size of the game matrices. In particular, the payoff $\tilde u^i$ is polynomially computable in the size of the game matrices.
We now utilize~\cite[Theorem E.2]{fearnley2022complexity} to show that we can efficiently approximate, up to arbitrary precision, $\tilde u_i$ by a linear arithmetic circuit.
Since the size of the arithmetic circuit corresponding to $\tilde u^i$ grows polynomially with the size of the game matrices, it suffices to prove that $\tilde u^i$ is Lipschitz (w.r.t. the standard type 2 Euclidean norm) over the strategy spaces $R^1\times\ldots\times R^N$ with a Lipschitz constant bounded by a quantity that grows polynomially in the size of the game matrices. 
To do so, we show  $\tilde u^i$ is Lipschitz w.r.t. to the type 1 Euclidean norm, to ease the use of triangle inequality, and then invoke equivalence of the norms to establish a bound on the Lipschitz constant (w.r.t. the standard type 2 Euclidean norm).
First, we consider the function $(\stratPlayerI,\lambda^i)\mapsto f_{1,\actionPlayerIMinus,\surActionPlayerIMinus}(\stratPlayerI,\lambda^i)\coloneqq\expectedValue{\actionPlayerI\sim\stratPlayerI}{\utilityPlayer{i}(\actionPlayerI,\surActionPlayerIMinus)} + \lambda^i \distance^{-i}(\actionPlayerIMinus,\surActionPlayerIMinus)^\wassnorm$.
Being linear, $f_{1,\actionPlayerIMinus,\surActionPlayerIMinus}$ is differentiable and each entry of its gradient is uniformly bounded by $C_0^i\coloneqq\max_{\actionPlayerI\in\actionSpacePlayerI,\actionPlayerIMinus\in\actionSpacePlayerIMinus}\max\{|\utilityPlayerI(\actionPlayerI,\actionPlayerIMinus)|,\distance^{-i}(\actionPlayerIMinus,\actionPlayerIMinus)^\wassnorm\}$, which implies that the type 1 Euclidean norm of the gradient is bounded by $C_1^i\coloneqq C_0^i(\cardinality{\actionSpacePlayerI}+1)$. 
Thus, for all $\actionPlayerIMinus,\surActionPlayerIMinus\in\actionSpacePlayerIMinus$, the function $f_{1,\actionPlayerIMinus,\surActionPlayerIMinus}$ is (i) Lipschitz (w.r.t. the type 1 Euclidean norm) with Lipschitz constant bounded by $C_1^i$ and (ii) uniformly bounded by $C_0^i(1+M^i)$, where $M^i$ is the upper bound on $\lambda^i$ given by \cref{lemma:lambda_bounded}. Note that all bounds are uniform and, in particular, independent of $\actionPlayerIMinus$ and $\surActionPlayerIMinus$.
Second, we focus on the function
$(\stratPlayerI,\lambda^i)\mapsto f_{2,\actionPlayerIMinus}(\stratPlayerI,\lambda^i)\coloneqq\min_{\surActionPlayerIMinus\in\actionSpacePlayer{-i}}f_{1,\actionPlayerIMinus,\surActionPlayerIMinus}(\stratPlayerI,\lambda^i)=\min_{\surActionPlayerIMinus\in\actionSpacePlayer{-i}}\left\{\expectedValue{\actionPlayerI\sim\stratPlayerI}{\utilityPlayer{i}(\actionPlayerI,\surActionPlayerIMinus)} + \lambda^i \distance^{-i}(\actionPlayerIMinus,\surActionPlayerIMinus)^\wassnorm\right\}$.
Being the minimum of $\cardinality{\actionSpacePlayerIMinus}$ functions each with Lipschitz constant bounded by $C_1^i$, we have that $f_{2,\actionPlayerIMinus}$ is (i) Lipschitz (w.r.t. the type 1 Euclidean norm) with a Lipschitz constant bounded by $C_2^i\coloneqq\cardinality{\actionSpacePlayerIMinus}C_1^i$ and (ii) uniformly bounded by $C_0^i(1+M^i)$.
Note that, as above, all bounds are independent of $\actionPlayerIMinus$.
Third, we consider the function
$(\stratPlayerI,\lambda^i,\stratWassersteinPlayer{}_{\stratPlayerIMinus})\mapsto
f_3(\stratPlayerI,\lambda^i,\stratWassersteinPlayer{}_{\stratPlayerIMinus})\coloneqq 
-\lambda^i\varepsilon^s+\expectedValue{\actionPlayerIMinus\sim\stratWassersteinPlayer{}_{\stratPlayerIMinus}}{f_{2,\actionPlayerIMinus}(\stratPlayerI,\lambda^i)}=
-\lambda^i\varepsilon^s+\expectedValue{\actionPlayerIMinus\sim\stratWassersteinPlayer{}_{\stratPlayerIMinus}}{\min_{\surActionPlayerIMinus\in\actionSpacePlayer{-i}}\left\{\expectedValue{\actionPlayerI\sim\stratPlayerI}{\utilityPlayer{i}(\actionPlayerI,\surActionPlayerIMinus)} + \lambda^i \distance^{-i}(\actionPlayerIMinus,\surActionPlayerIMinus)^\wassnorm\right\}}$.
Since the sum of Lipschitz functions is Lipschitz (with Lipschitz constant bounded by the sum of the Lipschitz constants, scaled by the ) and the multiplication of bounded Lipschitz functions is Lipschitz (with Lipschitz constant bounded by the sum of the Lipschitz constant scaled by the uniform bound on the functions), $f_3$ is Lipschitz (w.r.t. the type 1 Euclidean norm). 
In particular, its Lipschitz constant is bounded by
$C_3^i\coloneqq\varepsilon^s+\cardinality{\actionSpacePlayerIMinus}\max\{1, C_0^i(1+M^i)\}(C_2^i+1)$,
where $\max\{1, C_0^i(1+M^i)\}$ is a bound on $f_{2,\actionPlayerIMinus}$ and each entry of $\stratWassersteinPlayer{}_{\stratPlayerIMinus}$, and $C_2^i+1$ is the sum of (the bounds of) the Lipschitz constants of $f_{2,\actionPlayerIMinus}$ and the function extracting each entry of $\stratWassersteinPlayer{}_{\stratPlayerIMinus}$.
Fourth, consider the function $(p^i,\lambda^i,p^{-i},\lambda^{-i})\mapsto f_4(p^i,\lambda^i,p^{-i},\lambda^{-i})=(p^i,\lambda^i,\stratWassersteinPlayer{}_{\stratPlayerIMinus})$.
Since $f_4$ is continuously differentiable, each entry of its Jacobian is pointwise bounded by 1 (being the product of probabilities), and its Jacobian is of size $(1+\cardinality{\actionSpacePlayerI}+\cardinality{\actionSpacePlayerIMinus})\times(2+\sum_{i=1}^{N}\cardinality{\actionSpacePlayer{i}})$, we have the type 1 Euclidean (induced) norm of its Jacobian is bounded by $C_4^i\coloneqq1+\cardinality{\actionSpacePlayerI}+\cardinality{\actionSpacePlayerIMinus}$.
Thus, $f_4$ is Lipschitz (w.r.t. the type 1 Euclidean norm) with Lipschitz constant bounded by $C_4^i$.
Fifth, $\tilde u^i$ can be seen as the composition of $f_3$ and $f_4$. Thus, $\tilde u^i$ is Lipschitz (w.r.t. the type 1 Euclidean norm) with Lipschitz constant bounded by $C_4^iC_3^i$.
Sixth, by the equivalence between the type 1 and type 2 Euclidean norms, $\tilde u^i$ is Lipschitz (w.r.t. the standard Euclidean norm) with Lipschitz constant bounded by $C^i\coloneqq \sqrt{2+\cardinality{\actionSpacePlayerI}+\cardinality{\actionSpacePlayerIMinus}}C_4^iC_3^i$
To conclude, we observe that all constants $C^i_1, \ldots, C^i_4$ grow polynomially with the size of the game matrices and so the bound on the Lipschitz constant of $\tilde u^i$, given by $C^i$, also grows polynomially with the size of game matrices. 
By~\cite[Theorem E.2]{fearnley2022complexity}, we can therefore construct, in polynomial time in the size of the game matrices, a linear arithmetic circuit to represent $\tilde u_i$.\footnote{In what follows, we suppose that $\tilde u_i$ is exactly representable via an arithmetic circuit. Our analysis continues to hold with an additional ``precision term'' in all inequalities below.}

\textbf{Convex constraint:} The convex constraint $S$ is a compact polytope, being the Cartesian product of compact polytopes. Thus, we represent it as a strong separation oracle~\cite[Definition 3.11]{papadimitriou2022computational}. We now prove that $S$ is well-bounded; i.e., there exists a $a_0\in S$ so that a ball of non-zero radius around $a_0$ is contained in $S$. This property is in particular necessary to deploy \texttt{ConcaveGame with SO Problem} in~\cite{papadimitriou2022computational}.
In general, since $\strategySpacePlayerI$ is defined in terms of the equality constraint $\sum_{\actionPlayerI\in\actionSpacePlayerI}\stratPlayerI(\actionPlayerI)=1$, this cannot hold. Nonetheless, we can see $\strategySpacePlayerI$ as the image of $\tilde\strategySpacePlayerI\coloneqq \{(\stratPlayerI(\actionPlayerI_1),\ldots,\stratPlayerI(\actionPlayerI_{\cardinality{\actionSpacePlayerI}-1}))\in\reals^{\cardinality{\actionSpacePlayerI}-1}:\stratPlayerI(\actionPlayerI_k)\geq 0, \sum_{k=1}^{\cardinality{\actionSpacePlayerI}-1}\stratPlayerI(\actionPlayerI_k)\leq 1\}\subset\reals^{\cardinality{\actionSpacePlayerI}-1}$ through the homomorphism $\chi$ defined by $\tilde\strategySpacePlayerI\ni \stratPlayerI\mapsto \chi(\stratPlayerI)\coloneqq (\stratPlayerI(\actionPlayerI_1),\ldots,\stratPlayerI(\actionPlayerI_{\cardinality{\actionSpacePlayerI}-1}),1-\sum_{k=1}^{\cardinality{\actionSpacePlayerI}-1}\stratPlayerI(\actionPlayerI_k))\in\strategySpacePlayerI$. It is clear that $\tilde\strategySpacePlayerI$ is convex and compact. Moreover, it contains a ball around some point $a_0$ (e.g., a ball of radius $\frac{1}{\cardinality{\actionSpacePlayerI}}$ centered at $(\frac{1}{\cardinality{\actionSpacePlayerI}},\ldots,\frac{1}{\cardinality{\actionSpacePlayerI}})\in\reals^{\cardinality{\actionSpacePlayerI}-1}$).
Since $\chi$ is polynomially computable and Lipschitz with Lipschitz constant bounded by $\cardinality{\actionSpacePlayerIMinus}^2$, and the composition of Lipschitz functions is Lipschitz with Lipschitz constant bounded by the product of the two Lipschitz constants, the discussion above on the payoff functions continues to hold (or, alternatively, $\chi$ is directly representable as a linear arithmetic circuit and the ``composition'' of linear arithmetic circuits is a linear arithmetic circuit).

We can now use the \texttt{ConcaveGame with SO Problem} in~\cite{papadimitriou2022computational}, whose computational properties are in \ppad{}, to obtain a $\delta$-approximate equilibrium\footnote{Since we have access to a strong separation oracle for the set $S$, we do not need to consider ``approximations'' of set the $S$ (by shrinking and enlarging it via small balls of radius $\eta$) in our definition of approximate equilibrium, as in the more general \cite[Definition 4.4]{papadimitriou2022computational}.} of the concave game $\tilde G$. We denote this by $((\optStratPlayer{1},\bar\lambda^1),\ldots,(\optStratPlayer{N},\bar\lambda^N))$. We now prove that $(\optStratPlayer{1},\ldots,\optStratPlayer{N})$ is an $\delta$-approximate strategically robust equilibrium. By definition of $\delta$-approximate equilibrium of the concave game, for all agents $i\in\{1,\ldots,N\}$ and all $\stratPlayerI$ and $\lambda^i$, 
\begin{equation*}
    \tilde u^{i}((\optStratPlayerI,\bar\lambda^i),(\optStratPlayerIMinus,\bar\lambda^{-i}))
    \geq 
    - \lambda^i\varepsilon^\wassnorm  + 
    \expectedValue{\actionPlayerIMinus\sim\stratWassersteinPlayer{}_{\optStratPlayerIMinus}}{
    \min_{\surActionPlayerIMinus\in\actionSpacePlayer{-i}}
    \left\{
    \expectedValue{\actionPlayerI\sim\stratPlayerI}{\utilityPlayer{i}(\actionPlayerI,\surActionPlayerIMinus)} + \lambda^i \distance^{-i}(\actionPlayerIMinus,\surActionPlayerIMinus)^\wassnorm
    \right\}}
    -\delta.
\end{equation*}
In particular, 
\begin{equation}\label{eq:complexity proof:approximate equilibrium:2}
\begin{aligned}
    \tilde u^{i}((\optStratPlayerI,\bar\lambda^i),(\optStratPlayerIMinus,\bar\lambda^{-i}))
    \geq 
    &\sup_{\lambda^i\in [0,M^i]}
    - \lambda^i\varepsilon^\wassnorm 
    \\
    &+ 
    \expectedValue{\actionPlayerIMinus\sim\stratWassersteinPlayer{}_{\optStratPlayerIMinus}}{\min_{\surActionPlayerIMinus\in\actionSpacePlayer{-i}}
    \left\{\expectedValue{\actionPlayerI\sim\stratPlayerI}{\utilityPlayer{i}(\actionPlayerI,\surActionPlayerIMinus)} + \lambda^i \distance^{-i}(\actionPlayerIMinus,\surActionPlayerIMinus)^\wassnorm\right\}}
    -\delta.
\end{aligned}
\end{equation}
We are now ready to show that $(\optStratPlayer{1},\ldots,\optStratPlayer{N})$ is a $\delta$-approximate strategically robust equilibrium. Let $\stratPlayerI$ be an arbitrary strategy for agent $i$. Then, 
\begin{align*}
    \min_{\stratWassersteinPlayerIMinus\in\ambiguitySet{i}{\varepsilon}{\optStratPlayerIMinus}}
    &\utilityExpArgsPlayer{i}{\stratPlayerI}{\stratWassersteinPlayerIMinus}
    \\
    \overset{\eqref{eq:finite:strategically robust best response:reformulation}}&=
    \max_{\lambda^i\geq 0}
    - \lambda^i \varepsilon^\wassnorm  + 
    \expectedValue{\actionPlayerIMinus\sim\stratWassersteinPlayer{}_{\optStratPlayerIMinus}}{
    \min_{\surActionPlayerIMinus\in\actionSpacePlayer{-i}}
    \left\{\expectedValue{\actionPlayerI\sim\stratPlayerI}{\utilityPlayer{i}(\actionPlayerI,\surActionPlayerIMinus)} + \lambda^i \distance^{-i}(\actionPlayerIMinus,\surActionPlayerIMinus)^\wassnorm\right\}}
    \\
    \overset{\heartsuit}&{=}
    \max_{\lambda^i\in[0,M^i]}
    - \lambda^i \varepsilon^\wassnorm  + 
    \expectedValue{\actionPlayerIMinus\sim\stratWassersteinPlayer{}_{\optStratPlayerIMinus}}{
    \min_{\surActionPlayerIMinus\in\actionSpacePlayer{-i}}
    \left\{\expectedValue{\actionPlayerI\sim\stratPlayerI}{\utilityPlayer{i}(\actionPlayerI,\surActionPlayerIMinus)} + \lambda^i \distance^{-i}(\actionPlayerIMinus,\surActionPlayerIMinus)^\wassnorm\right\}}
    \\
    \overset{\eqref{eq:complexity proof:approximate equilibrium:2}}&{\leq}
    \tilde u^{i}((\optStratPlayerI,\bar\lambda^i),(\optStratPlayerIMinus,\bar\lambda^{-i})) + \delta
    \\
    &\leq 
    \max_{\lambda^i\geq 0}  \tilde u^{i}((\optStratPlayerI,\lambda^i),(\optStratPlayerIMinus,\bar\lambda^{-i})) + \delta
    \\
    \overset{\eqref{eq:complexity proof:surrogate utility:definition}}&{=}
     \max_{\lambda^i\geq 0}
    - \lambda^i \varepsilon^\wassnorm  + 
    \expectedValue{\actionPlayerIMinus\sim\stratWassersteinPlayer{}_{\optStratPlayerIMinus}}{
    \min_{\surActionPlayerIMinus\in\actionSpacePlayer{-i}}
    \left\{\expectedValue{\actionPlayerI\sim\optStratPlayerI}{\utilityPlayer{i}(\actionPlayerI,\surActionPlayerIMinus)} + \lambda^i \distance^{-i}(\actionPlayerIMinus,\surActionPlayerIMinus)^\wassnorm\right\}} + \delta
    \\
    \overset{\eqref{eq:finite:strategically robust best response:reformulation}}&=
    \min_{\stratWassersteinPlayerIMinus\in\ambiguitySet{i}{\varepsilon}{\optStratPlayerIMinus}} 
    \utilityExpArgsPlayer{i}{\optStratPlayerI}{\stratWassersteinPlayerIMinus} + \delta,
\end{align*}
where $\heartsuit$ follows from~\cref{lemma:lambda_bounded}.
Since the argument holds for all agents, $(\optStratPlayer{1},\ldots,\optStratPlayer{N})$ is an $\delta$-approximate strategically robust equilibrium.
Overall, since the computational properties of \texttt{ConcaveGame with SO Problem} are \ppad{}-complete~\cite[Theorem 4.9]{papadimitriou2022computational} and \texttt{ConcaveGame with SO Problem} can be used to find an approximate strategically robust equilibrium, the computational complexity of strategically robust equilibria lies in \ppad{}.
\endproof

\proof{Proof of~\cref{prop:finite:computation:N players}}
    The statement follows directly from the necessity and sufficiency of the KKT conditions for the linear program~\citep{boyd2004convex} and the discussion before the proposition. 
\endproof

\subsection{Proofs for \cref{sec:ContinuousGames}}

\proof{Proof of~\cref{thm:continuous:existence pure strategically robust eq}}
Our proof is based on a reformulation of the game as a standard concave game, with appropriately chosen action spaces and payoffs, for which pure equilibria are known to exist. In particular, consider a game with the action spaces $\actionPlayerI$ and the modified payoff 
\begin{equation*}
\begin{aligned}
    \utilityWassersteinPlayer{i}:
    \actionSpacePlayerI\times\actionSpacePlayerIMinus &\to \reals
    \\
    (\actionPlayerI,\actionPlayer{-i})
    &\mapsto
    \inf_{\stratWassersteinPlayerIMinus \in \ambiguitySet{i}{\varepsilon}{\delta_{\optActionPlayer{-i}}}}
    \hat{\utilityMap}^i(\actionPlayerI, \stratWassersteinPlayerIMinus),
\end{aligned}
\end{equation*}
where the function $\hat{\utilityMap}^i$ is defined by 
\begin{equation*}
\begin{aligned}
    \hat{\utilityMap}^i
    :\actionSpacePlayerI\times\probSpace{\actionSpacePlayerIMinus} &\to\reals 
    \\
    (\actionPlayerI, \stratWassersteinPlayerIMinus)
    &\mapsto 
    \expectedValue{\actionPlayerIMinus\sim\stratWassersteinPlayerIMinus}{\utilityArgsPlayerI}.
\end{aligned}
\end{equation*}
Clearly, if $(\optActionPlayer{1},\ldots,\optActionPlayer{N})$ a pure Nash Equilibrium for the game with $\utilityWassersteinPlayer{i}$ is a pure strategically equilibrium of the game $\mathcal{G}$. In what follows, we show that the game with actions spaces $\actionSpacePlayerI$ and payoffs $\utilityWassersteinPlayer{i}$ is itself a concave game and, thus, it possesses a pure Nash equilibrium. Since the action spaces are, by assumption, $\actionSpacePlayerI$ compact and convex, it suffices to prove that the modified payoff $\utilityWassersteinPlayer{i}$ is continuous and concave in $\actionPlayerI$. We do so in several steps. 

\textbf{Properties of $\hat{\utilityMap}^i$:}
We start by studying the properties of $\hat{\utilityMap}^i$, in particular (joint) continuity and concavity in $\actionPlayerI$ for fixed $\stratWassersteinPlayerIMinus$. We proceed in several steps. 
\begin{enumerate}
    \item The family of functions $\utilityPlayerI(\cdot,\actionPlayerIMinus)$ (i.e., parametrized by $\actionPlayerIMinus$) is uniformly equicontinuous: 
    Since $\utilityPlayerI$ is continuous and each $\actionSpacePlayerI$ is compact, $\utilityArgsPlayerI$ is uniformly continuous on $\cartProd{\actionSpacePlayerI}{\actionSpacePlayerIMinus}$ by the Heine-Cantor-Theorem. Thus, for all $\varepsilon>0$, there exists $\delta > 0$ such that $\abs{\utilityArgsPlayer{i}{\actionPlayerI_1}{\actionPlayerIMinus_1} - \utilityArgsPlayer{i}{\actionPlayerI_2}{\actionPlayerIMinus_2}} < \varepsilon$ for all $\actionPlayerI_1,\actionPlayerI_2\in\actionSpacePlayerI$ and all  $\actionPlayerIMinus_1,\actionPlayerIMinus_2\in \actionSpacePlayerIMinus$ that are $\delta$-close (e.g., w.r.t. to the distance on $\actionSpacePlayerI\times\actionSpacePlayerIMinus$ resulting from the sum of the distances on the individual spaces). 
    We can now take $\actionPlayerIMinus_1 = \actionPlayerIMinus_2 = \actionPlayerIMinus$. Then, for all $\varepsilon > 0$, there exists $\delta > 0$ such that $\abs{\utilityArgsPlayer{i}{\actionPlayerI_1}{\actionPlayerIMinus} - \utilityArgsPlayer{i}{\actionPlayerI_2}{\actionPlayerIMinus}} < \varepsilon$ for all $\actionPlayerIMinus \in \actionSpacePlayerIMinus$ and for all $\actionPlayerI_1, \actionPlayerI_2 \in \actionPlayerI$ that are $\delta$-close. This is precisely equicontinuity.

    \item For all $\actionPlayerI\in\actionSpacePlayerI$, the map $\stratWassersteinPlayerIMinus\mapsto\hat{\utilityMap}^i(\actionPlayerI, \stratWassersteinPlayerIMinus)$ is continuous w.r.t. narrow convergence: 
    Since $\utilityPlayerI$ is continuous and bounded (being defined on a compact set), the statement follows directly from the definition of narrow convergence. 

    \item The function $\hat{\utilityMap}^i$ is (jointly) continuous on $\cartProd{\actionSpacePlayerI}{\probSpace{\actionSpacePlayerIMinus}}$:
    Consider the sequence $(\actionPlayerI_n, \stratWassersteinPlayerIMinus_n)_{n\in\naturals} \in \cartProd{\actionSpacePlayerI}{\probSpace{\actionSpacePlayerIMinus}}$ such that $\actionPlayerI_n\to\actionPlayerI$ and $\stratWassersteinPlayerIMinus_n\narrowconvergence\stratWassersteinPlayerIMinus$ as $n\to\infty$. 
    Given $\delta > 0$, choose $N > 0$ such that $\abs{\utilityArgsPlayer{i}{\actionPlayerI_n}{\actionPlayerIMinus} - \utilityArgsPlayer{i}{\actionPlayerI}{\actionPlayerIMinus}} < \varepsilon$ for all $n > N$ and all $\actionPlayerIMinus \in \actionSpacePlayerIMinus$, which exist by equicontinuity shown above. We have
    \begin{align*}
        \limsup_{n\to\infty}\, &\expectedValue{\actionPlayerIMinus\sim\stratWassersteinPlayerIMinus_n}{\utilityArgsPlayer{i}{\actionPlayerI_n}{\actionPlayerIMinus}}
        \\
        &\leq
        \limsup_{n\to\infty} \expectedValue{\actionPlayerIMinus\sim\stratWassersteinPlayerIMinus_n}{\utilityArgsPlayer{i}{\actionPlayerI}{\actionPlayerIMinus} + \abs{\utilityArgsPlayer{i}{\actionPlayerI_n}{\actionPlayerIMinus} -\utilityArgsPlayer{i}{\actionPlayerI}{\actionPlayerIMinus}}}
        \\
        &=
        \lim_{n \goesto \infty} \expectedValue{\actionPlayerIMinus\sim\stratWassersteinPlayerIMinus_n}{\utilityArgsPlayer{i}{\actionPlayerI}{\actionPlayerIMinus}} + \limsup_{n\to\infty} \expectedValue{\actionPlayerIMinus\sim\stratWassersteinPlayerIMinus_n}{\abs{\utilityArgsPlayer{i}{\actionPlayerI}{\actionPlayerIMinus} - \utilityArgsPlayer{i}{\actionPlayerI_n}{\actionPlayerIMinus}}}
        \\
        &<
        \expectedValue{\actionPlayerIMinus\sim\stratWassersteinPlayerIMinus}{\utilityArgsPlayer{i}{\actionPlayerI}{\actionPlayerIMinus}} + \delta,
    \end{align*}
    where we also used continuity of $\hat{\utilityMap}^i$ in $\stratWassersteinPlayerIMinus$.
    Similarly,
    \begin{align*}
        \liminf_{n\to\infty} \expectedValue{\actionPlayerIMinus\sim\stratWassersteinPlayerIMinus_n}{\utilityArgsPlayer{i}{\actionPlayerI_n}{\actionPlayerIMinus}}
        & \geq
        \expectedValue{\actionPlayerIMinus\sim\stratWassersteinPlayerIMinus_n}{\utilityArgsPlayer{i}{\actionPlayerI}{\actionPlayerIMinus} - \abs{\utilityArgsPlayer{i}{\actionPlayerI}{\actionPlayerIMinus} - \utilityArgsPlayer{i}{\actionPlayerI_n}{\actionPlayerIMinus}}}
        \\
        &>
       \expectedValue{\actionPlayerIMinus\sim\stratWassersteinPlayerIMinus}{\utilityArgsPlayer{i}{\actionPlayerI}{\actionPlayerIMinus}} - \delta.
    \end{align*}
    Thus,
    \begin{align*}
        \expectedValue{\actionPlayerIMinus\sim\stratWassersteinPlayerIMinus}{\utilityArgsPlayer{i}{\actionPlayerI}{\actionPlayerIMinus}} - \varepsilon
        &<
        \liminf_{n\to\infty} \expectedValue{\actionPlayerIMinus\sim\stratWassersteinPlayerIMinus_n}{\utilityArgsPlayer{i}{\actionPlayerI_n}{\actionPlayerIMinus}}
        \\
        &\leq
        \limsup_{n\to\infty} \expectedValue{\actionPlayerIMinus\sim\stratWassersteinPlayerIMinus_n}{\utilityArgsPlayer{i}{\actionPlayerI_n}{\actionPlayerIMinus}}
        \\
        &\leq
        \expectedValue{\actionPlayerIMinus\sim\stratWassersteinPlayerIMinus}{\utilityArgsPlayer{i}{\actionPlayerI}{\actionPlayerIMinus}} + \varepsilon.
    \end{align*}
    Since this holds for all $\delta>0$, we can let $\delta\to 0$ to show that the $\liminf$ and $\limsup$ coincide. Therefore, $\lim_{n \goesto \infty} \expectedValue{\actionPlayerIMinus\sim\stratWassersteinPlayerIMinus_n}{\utilityArgsPlayer{i}{\actionPlayerI_n}{\actionPlayerIMinus}} = \expectedValue{\actionPlayerIMinus\sim\stratWassersteinPlayerIMinus}{\utilityArgsPlayer{i}{\actionPlayerI}{\actionPlayerIMinus}}$, which shows joint continuity.

    \item The function $\actionPlayerI\mapsto\hat{\utilityMap}^i(\actionPlayerI,\stratWassersteinPlayerIMinus)$ is concave for all $\stratWassersteinPlayerIMinus\in\stratWassersteinPlayerIMinus$:
    The statement follows from concavity of $\utilityPlayerI$ and linearity of the expectation. 
\end{enumerate}

\textbf{Properties of the ambiguity set $\ambiguitySet{i}{\varepsilon}{\delta_{\actionPlayer{-i}}}$:}
We now establish that the set-valued map $\actionPlayerIMinus\mapsto\ambiguitySet{i}{\varepsilon}{\delta_{\actionPlayer{-i}}}$ is non-empty, compact-valued, and hemicontinuous. In~\cref{lemma:continuity wasserstein ball}, we prove the map $\stratPlayerIMinus\mapsto\ambiguitySet{i}{\varepsilon}{\delta_{\actionPlayer{-i}}}$ is non-empty, compact-valued, and hemicontinuous (w.r.t. narrow convergence).
Thus, we only need to prove that the map $\actionPlayerIMinus\mapsto\delta_{\actionPlayer{-i}}\coloneqq(\delta_{\actionPlayer{1}},\ldots,\delta_{\actionPlayer{N}})$ is continuous. We can then leverage the fact that the induced set-valued map $\actionPlayerIMinus\mapsto\{\delta_{\actionPlayer{-i}}\}$ is hemicontinuous~\cite[Lemma 17.5]{Aliprantis2005InfiniteDimensionalAnalyis} and the composition of hemicontinuous set-valued maps is hemicontinuous~\cite[Theorem 17.23]{Aliprantis2005InfiniteDimensionalAnalyis}.
To prove continuity of~$\actionPlayerIMinus\mapsto\delta_{\actionPlayer{-i}}$, it suffices to prove that the map $\actionPlayer{j}\mapsto\delta_{\actionPlayer{j}}$ is continuous, since this readily implies continuity of $\actionPlayerIMinus\mapsto\delta_{\actionPlayer{j}}$ (for $j\neq i$) and therefore of $\actionPlayerIMinus\mapsto\delta_{\actionPlayerIMinus}$. We prove continuity via sequences. Let $(\actionPlayer{j}_n)_{n\in\naturals}\subset\actionSpacePlayer{j}$ so that $\actionPlayer{j}_n\to\actionPlayer{j}\in\actionSpacePlayer{j}$. We now prove that $\delta_{\actionPlayer{j}_n}\narrowconvergence\delta_{\actionPlayer{j}}$.
Let $\phi\in\Cb{\actionSpacePlayer{j}}$. Then, by continuity of $\phi$, we have
\begin{equation*}
    \lim_{n\to\infty}\int_{\actionSpacePlayer{j}}\phi(\actions)\mathrm{d}\delta_{\actionPlayer{j}_n}(\actions)
    =
    \lim_{n\to\infty}\phi(\actionPlayer{j}_n)
    =
    \phi(\actionPlayer{j})
    =
    \int_{\actionSpacePlayer{j}}\phi(\actions)\mathrm{d}\delta_{\actionPlayer{j}}(\actions).
\end{equation*}
Thus, $\delta_{\actionPlayer{j}_n}\narrowconvergence\delta_{\actionPlayer{j}}$, which establishes the desired continuity result.

\textbf{Properties of $\utilityWassersteinPlayer{i}$:}
We are now ready to establish continuity and concavity of $\utilityWassersteinPlayer{i}$. In particular, this entails studying its minimization problem. 
\begin{enumerate}
    \item Continuity:
    We leverage again Berge's Maximum Theorem. 
    In terms of the notation, we have $\Theta = \actionSpacePlayerI\times\actionSpacePlayerIMinus$, $\theta = (\actionPlayerI, \actionPlayerIMinus)$, $\spaceA = \probSpace{\actionSpacePlayerIMinus}$, $C = \ambiguitySet{i}{\varepsilon}{\delta_{\cdot}}$, $f^* = \utilityWassersteinPlayer{i}$ and $f = \hat{\utilityMap}^i$. Since $\hat{\utilityMap}^i$ is continuous on $\cartProd{\actionSpace}{\probSpace{\actionSpacePlayerIMinus}}$ and $\actionPlayerIMinus\mapsto\ambiguitySet{i}{\varepsilon}{\delta_{\actionPlayerIMinus}}$ is a compact-valued, non-empty, and continuous correspondence. From Berge's Maximum Theorem (\cref{th:maximumTheorem}) it then follows that $\utilityWassersteinPlayer{i}$ is continuous on $\actionSpace$ and that the infimum can be replaced by a minimum.
    
    \item Concavity of the function $\actionPlayerI\mapsto\utilityWassersteinPlayer{i}(\actionPlayerI, \actionPlayerIMinus)$ for all $\actionPlayerIMinus\in\actionSpacePlayerIMinus$ follows from concavity $\actionPlayerI\mapsto\hat{\utilityMap}^i(\actionPlayerI,\stratWassersteinPlayerIMinus)$ for all $\stratWassersteinPlayerIMinus\in\strategySpacePlayerIMinus$, shown above, and the infimum of concave functions being concave.
\end{enumerate}

\textbf{Existence of a strategically robust equilibrium:}
We therefore recover a standard (non-strategically robust) game, where $\utilityPlayerI$ is replaced with $\utilityWassersteinPlayer{i}$.
Thus, the well-known result on the existence of a pure Nash equilibrium in a concave game applies (e.g., see~\citet[Thereom 1]{rosen1965existence} and~\citet[Theorem~1.2]{Fudenberg1991GameTheory}). 
\endproof

\proof{Proof of~\cref{prop:equivalent game}}
    The proof follows directly from~\cref{thm:continuous:existence pure strategically robust eq}, where we used~\cref{prop:strategically robust best response} to reformulate the inner minimization in~\eqref{eq:continuous:existence pure strategically robust eq}.
    Thus, we only need to show that the resulting game is a concave game. We do it in three steps.
    \begin{enumerate}
        \item Compactness and convexity of $\surActionSpacePlayer{i}$: The compactness and convexity of action spaces $\surActionSpacePlayer{i}$ follows from the compactness and convexity of $\actionSpacePlayerI$ and the uniform boundedness of dual multiplier $\lambda^i$, shown in~\cref{lemma:lambda_bounded}.

        \item Continuity of $\surUtilityPlayerI$: The modified payoff function $\surUtilityPlayerI$ results from the minimum of a continuous function over a compact set. Thus, again by Berge's Maximum Theorem (\cref{th:maximumTheorem}), it is continuous. 

        \item Concavity of $\surUtilityPlayerI$: The modified payoff function $\surUtilityPlayerI$ results from the pointwise minimum of functionals that concave in $(\actionPlayerI,\lambda^i)$ (in fact, even linear in $\lambda^i$). As the minimum of concave functions is concave, we conclude. 
    \end{enumerate}
\endproof

\proof{Proof of~\cref{cor:equivalent game:convex}}
    The proof follows directly from~\citet[Theorem 2]{zhen2023unified}.
\endproof 

\section{Additional Results for the Pedestrian Game}\label{app:pedestrian game:additional results}

In this section, we include the mixed Nash equilibrium in the analysis of the pedestrian game of~\cref{sec:motivating example}. In~\cref{fig:pedestrianGame:performance:MNEgreen}, we show the performance of all equilibria, in particular also the mixed Nash equilibrium, as the probability that the pedestrians deviate towards \textsc{Cross} increases from the equilibrium value. The mixed Nash equilibrium immediately leads to a negative payoff and is therefore dominated by the security strategy \textsc{Stop}. From a practical perspective, it can therefore be discarded. 

\begin{figure}[h]
    \centering
    
    \pgfplotstableread[col sep=comma]{results/pedestrian.csv}\datatable
    \begin{tikzpicture}[baseline=(current bounding box.north)]
        \begin{axis}[
        width=10cm, height=7cm,
        xlabel={\footnotesize Prob. that the pedestrians deviate towards \aCross},
        ylabel={\footnotesize Vehicle's payoff},
        ylabel near ticks,
        grid=both,
        legend style={draw=none,fill=none},
        xmin=0, ymin=-10, xmax=0.2, ymax=25,
        xtick={0, 0.05, 0.1, 0.15, 0.2}, 
        xticklabels={0, 0.05, 0.1, 0.15, 0.2}, 
        ]
        \addplot[ETHbronze,ultra thick] table[x=f, y expr=\thisrow{meanNash}] {\datatable};
        \addplot[name path=us_top,ETHbronze,thin,draw=none,forget plot] table[x=f, y expr=\thisrow{nashPlus}] {\datatable};
        \addplot[name path=us_down,ETHbronze,thin,draw=none,forget plot] table[x=f, y expr=\thisrow{nashMinus}] {\datatable};
        \addplot[fill=ETHbronze!30,fill opacity=0.3,forget plot] fill between[of=us_top and us_down];

        \addplot[ETHpurple, ultra thick] table[x=f, y expr=\thisrow{meanRobust}] {\datatable};
        \addplot[name path=us_top,ETHpurple, thin, draw=none,forget plot] table[x=f, y expr=\thisrow{robustPlus}] {\datatable};
        \addplot[name path=us_down,ETHpurple, thin,draw=none,forget plot] table[x=f, y expr=\thisrow{robustMinus}] {\datatable};
        \addplot[fill=ETHpurple!50,fill opacity=0.5,forget plot] fill between[of=us_top and us_down];

        \addplot[ETHpetrol, ultra thick] table[x=f, y expr=\thisrow{meanSecure}] {\datatable};

        \addplot[ETHgreen, ultra thick] table[x=f, y expr=\thisrow{meanMNE}] {\datatable};
        \addplot[name path=us_top,ETHgreen, thin, draw=none,forget plot] table[x=f, y expr=\thisrow{MNEPlus}] {\datatable};
        \addplot[name path=us_down,ETHgreen, thin,draw=none,forget plot] table[x=f, y expr=\thisrow{MNEMinus}] {\datatable};
        \addplot[fill=ETHgreen!70,fill opacity=0.7,forget plot] fill between[of=us_top and us_down];

        \addlegendentry{NE (\textsc{Maintain}, \textsc{Wait})}
        \addlegendentry{SRE (\textsc{Decelerate}, \textsc{Wait})}
        \addlegendentry{Security strategy \textsc{Stop}}
        \addlegendentry{MNE} 
        
    \end{axis}
    \end{tikzpicture}
    \caption{Payoff attained by the vehicle by playing the pure Nash equilibrium, mixed Nash equilibrium, strategically robust equilibrium, and security strategies, as a function of the probability that the family decides to deviate from their equilibrium strategy in favor of crossing the road. The shaded area represents the expected payoff plus/minus the standard deviation to illustrate the risk that the player is taking.}
    \label{fig:pedestrianGame:performance:MNEgreen}
\end{figure}
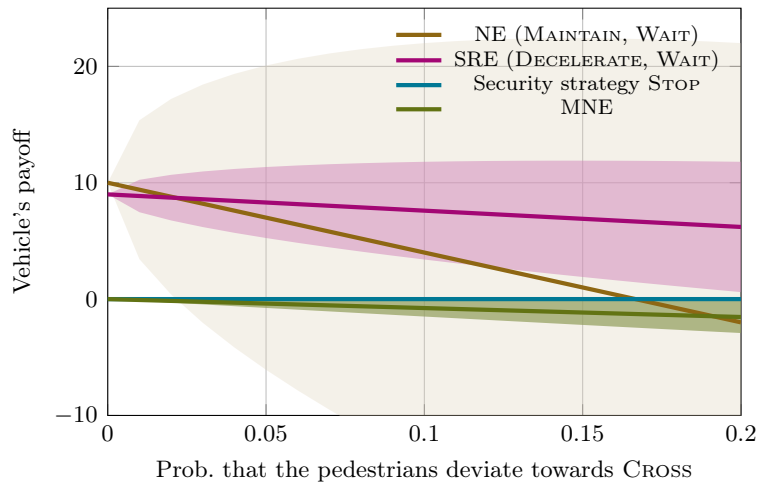

\end{APPENDICES}


\bibliographystyle{informs2014} 
\bibliography{references}

\begin{thebibliography}{80}
\providecommand{\natexlab}[1]{#1}
\providecommand{\url}[1]{\texttt{#1}}
\providecommand{\urlprefix}{URL }

\bibitem[{Aghassi \protect\BIBand{} Bertsimas(2006)}]{aghassi2006robust}
Aghassi M, Bertsimas D (2006) Robust game theory. \emph{Mathematical programming} 107(1-2):231--273.

\bibitem[{Aliprantis \protect\BIBand{} Border(2006)}]{Aliprantis2005InfiniteDimensionalAnalyis}
Aliprantis CD, Border KC (2006) \emph{Infinite Dimensional Analysis A Hitchhiker's Guide 3rd Edition} (Springer Berlin), 3 edition.

\bibitem[{Ambrosio et~al.(2005)Ambrosio, Gigli, \protect\BIBand{} Savar{\'e}}]{ambrosio2005gradient}
Ambrosio L, Gigli N, Savar{\'e} G (2005) \emph{Gradient flows: in metric spaces and in the space of probability measures} (Springer Science \& Business Media).

\bibitem[{Bagchi \protect\BIBand{} Paul(2014)}]{bagchi2014optimal}
Bagchi A, Paul JA (2014) Optimal allocation of resources in airport security: Profiling vs. screening. \emph{Operations Research} 62(2):219--233.

\bibitem[{Balseiro et~al.(2024)Balseiro, Besbes, \protect\BIBand{} Castro}]{balseiro2024mechanism}
Balseiro SR, Besbes O, Castro F (2024) Mechanism design under approximate incentive compatibility. \emph{Operations Research} 72(1):355--372.

\bibitem[{Beja(1992)}]{beja1992imperfect}
Beja A (1992) Imperfect equilibrium. \emph{Games and Economic Behavior} 4(1):18--36.

\bibitem[{Bertsekas \protect\BIBand{} Shreve(1996)}]{Bertsekas1996SOC}
Bertsekas DP, Shreve SE (1996) \emph{Stochastic Optimal Control: The Discrete-Time Case}, volume~5 (Athena Scientific).

\bibitem[{Bich(2019)}]{bich2019strategic}
Bich P (2019) Strategic uncertainty and equilibrium selection in discontinuous games. \emph{Journal of Economic Theory} 183:786--822.

\bibitem[{Blanchet \protect\BIBand{} Murthy(2019)}]{blanchet2019quantifying}
Blanchet J, Murthy K (2019) Quantifying distributional model risk via optimal transport. \emph{Mathematics of Operations Research} 44(2):565--600.

\bibitem[{Boyd \protect\BIBand{} Vandenberghe(2004)}]{boyd2004convex}
Boyd SP, Vandenberghe L (2004) \emph{Convex optimization} (Cambridge university press).

\bibitem[{Bramoull{\'e} et~al.(2014)Bramoull{\'e}, Kranton, \protect\BIBand{} D'Amours}]{bramoulle2014strategic}
Bramoull{\'e} Y, Kranton R, D'Amours M (2014) Strategic interaction and networks. \emph{American Economic Review} 104(3):898--930.

\bibitem[{Cachon \protect\BIBand{} Netessine(2006)}]{cachon2006game}
Cachon GP, Netessine S (2006) Game theory in supply chain analysis. \emph{Models, methods, and applications for innovative decision making} 200--233.

\bibitem[{Camerer(2011)}]{camerer2011behavioral}
Camerer CF (2011) \emph{Behavioral game theory: Experiments in strategic interaction} (Princeton university press).

\bibitem[{Camerer et~al.(2004)Camerer, Ho, \protect\BIBand{} Chong}]{camerer2004cognitive}
Camerer CF, Ho TH, Chong JK (2004) A cognitive hierarchy model of games. \emph{The Quarterly Journal of Economics} 119(3):861--898.

\bibitem[{Cominetti et~al.(2009)Cominetti, Correa, \protect\BIBand{} Stier-Moses}]{cominetti2009impact}
Cominetti R, Correa JR, Stier-Moses NE (2009) The impact of oligopolistic competition in networks. \emph{Operations Research} 57(6):1421--1437.

\bibitem[{Cournot(1838)}]{cournot1838recherches}
Cournot AA (1838) \emph{Recherches sur les principes math{\'e}matiques de la th{\'e}orie des richesses} (L. Hachette).

\bibitem[{Crespi et~al.(2017)Crespi, Radi, \protect\BIBand{} Rocca}]{crespi2017robust}
Crespi GP, Radi D, Rocca M (2017) Robust games: theory and application to a cournot duopoly model. \emph{Decisions in economics and finance} 40(1-2):177--198.

\bibitem[{Crespi et~al.(2025)Crespi, Radi, \protect\BIBand{} Rocca}]{crespi2025insights}
Crespi GP, Radi D, Rocca M (2025) Insights on the theory of robust games. \emph{Computational Economics} 65(2):717--761.

\bibitem[{Daskalakis et~al.(2009)Daskalakis, Goldberg, \protect\BIBand{} Papadimitriou}]{daskalakis2009complexity}
Daskalakis C, Goldberg PW, Papadimitriou CH (2009) The complexity of computing a {N}ash equilibrium. \emph{Communications of the ACM} 52(2):89--97.

\bibitem[{Dow \protect\BIBand{} Werlang(1994)}]{dow1994nash}
Dow J, Werlang SRdC (1994) Nash equilibrium under knightian uncertainty: breaking down backward induction. \emph{Journal of Economic Theory} 64(2):305--324.

\bibitem[{Drew \protect\BIBand{} Jean(1991)}]{Fudenberg1991GameTheory}
Drew F, Jean T (1991) \emph{Game Theory} (MIT Press).

\bibitem[{Eaves(1972)}]{eaves1972homotopies}
Eaves BC (1972) Homotopies for computation of fixed points. \emph{Mathematical Programming} 3(1):1--22.

\bibitem[{Esponda \protect\BIBand{} Pouzo(2016)}]{esponda2016berk}
Esponda I, Pouzo D (2016) Berk--{N}ash equilibrium: A framework for modeling agents with misspecified models. \emph{Econometrica} 84(3):1093--1130.

\bibitem[{Facchinei et~al.(1999)Facchinei, Jiang, \protect\BIBand{} Qi}]{facchinei1999smoothing}
Facchinei F, Jiang H, Qi L (1999) A smoothing method for mathematical programs with equilibrium constraints. \emph{Mathematical programming} 85(1):107.

\bibitem[{Facchinei \protect\BIBand{} Pang(2010)}]{facchinei201012}
Facchinei F, Pang JS (2010) 12 nash equilibria: the variational approach. \emph{Convex optimization in signal processing and communications} 443.

\bibitem[{Fearnley et~al.(2022)Fearnley, Goldberg, Hollender, \protect\BIBand{} Savani}]{fearnley2022complexity}
Fearnley J, Goldberg P, Hollender A, Savani R (2022) The complexity of gradient descent: {CLS} $=$ {PPAD} $\cap$ {PLS}. \emph{Journal of the ACM} 70(1):1--74.

\bibitem[{Ferris \protect\BIBand{} Munson(1999)}]{ferris1999interfaces}
Ferris MC, Munson TS (1999) Interfaces to path 3.0: Design, implementation and usage. \emph{Computational Optimization and Applications} 12:207--227.

\bibitem[{Fudenberg(1991)}]{fudenberg1991game}
Fudenberg D (1991) \emph{Game theory} (MIT press).

\bibitem[{Fudenberg \protect\BIBand{} Levine(1993)}]{fudenberg1993self}
Fudenberg D, Levine DK (1993) Self-confirming equilibrium. \emph{Econometrica: Journal of the Econometric Society} 523--545.

\bibitem[{Gairing \protect\BIBand{} Paccagnan(2023)}]{gairing2023congestion}
Gairing M, Paccagnan D (2023) In congestion games, taxes achieve optimal approximation. \emph{Operations Research} 72(3):966--982.

\bibitem[{Ganzfried(2023)}]{ganzfried2023safe}
Ganzfried S (2023) Safe equilibrium. \emph{2023 62nd IEEE Conference on Decision and Control (CDC)}, 5230--5236 (IEEE).

\bibitem[{Gao et~al.(2024)Gao, Chen, \protect\BIBand{} Kleywegt}]{gao2024wasserstein}
Gao R, Chen X, Kleywegt AJ (2024) Wasserstein distributionally robust optimization and variation regularization. \emph{Operations Research} 72(3):1177--1191.

\bibitem[{Gao \protect\BIBand{} Kleywegt(2023)}]{gao2023distributionally}
Gao R, Kleywegt A (2023) Distributionally robust stochastic optimization with wasserstein distance. \emph{Mathematics of Operations Research} 48(2):603--655.

\bibitem[{Gilboa \protect\BIBand{} Schmeidler(1989)}]{gilboa1989maxmin}
Gilboa I, Schmeidler D (1989) Maxmin expected utility with non-unique prior. \emph{Journal of mathematical economics} 18(2):141--153.

\bibitem[{Glicksberg(1952)}]{Glicksberg1952GeneralizedKakutaniNE}
Glicksberg IL (1952) A further generalization of the kakutani fixed theorem, with application to nash equilibrium points. \emph{Proceedings of the American Mathematical Society} 3(1):170--174.

\bibitem[{Ha et~al.(2011)Ha, Tong, \protect\BIBand{} Zhang}]{ha2011sharing}
Ha AY, Tong S, Zhang H (2011) Sharing demand information in competing supply chains with production diseconomies. \emph{Management science} 57(3):566--581.

\bibitem[{Han et~al.(2025)Han, Weissman, \protect\BIBand{} Zhou}]{han2025optimal}
Han Y, Weissman T, Zhou Z (2025) Optimal no-regret learning in repeated first-price auctions. \emph{Operations Research} 73(1):209--238.

\bibitem[{Harsanyi(1967)}]{Harsanyi1967BayesianGamesPart1}
Harsanyi JC (1967) Games with incomplete information played by "bayesian" players, {I-III. Part I.} the basic model. \emph{Management Science} 14:159--182.

\bibitem[{Harsanyi(1968)}]{Harsanyi1968BayesianGamesPart2}
Harsanyi JC (1968) Games with incomplete information played by "bayesian" players, i-iii. part ii. bayesian equilibrium points. \emph{Management Science} 14:320--334.

\bibitem[{Hobbs \protect\BIBand{} Pang(2007)}]{hobbs2007nash}
Hobbs BF, Pang JS (2007) Nash-cournot equilibria in electric power markets with piecewise linear demand functions and joint constraints. \emph{Operations Research} 55(1):113--127.

\bibitem[{Horvitz(1987)}]{horvitz1987computationConstraints}
Horvitz EJ (1987) Reasoning about beliefs and actions under computational resource constraints. \emph{Proceedings of the Third Conference on Uncertainty in Artificial Intelligence}, 429–447, UAI'87 (Arlington, Virginia, USA: AUAI Press).

\bibitem[{Howson~Jr(1972)}]{howson1972equilibria}
Howson~Jr JT (1972) Equilibria of polymatrix games. \emph{Management Science} 18(5-part-1):312--318.

\bibitem[{Kahneman \protect\BIBand{} Tversky(2013)}]{kahneman2013prospect}
Kahneman D, Tversky A (2013) Prospect theory: An analysis of decision under risk. \emph{Handbook of the fundamentals of financial decision making: Part I}, 99--127 (World Scientific).

\bibitem[{Kohlberg \protect\BIBand{} Mertens(1986)}]{kohlberg1986strategic}
Kohlberg E, Mertens JF (1986) On the strategic stability of equilibria. \emph{Econometrica: Journal of the Econometric Society} 1003--1037.

\bibitem[{Koutsoupias \protect\BIBand{} Papadimitriou(1999)}]{koutsoupias1999worst}
Koutsoupias E, Papadimitriou C (1999) Worst-case equilibria. \emph{Annual symposium on theoretical aspects of computer science}, 404--413 (Springer).

\bibitem[{Kuhn et~al.(2019)Kuhn, Esfahani, Nguyen, \protect\BIBand{} Shafieezadeh-Abadeh}]{kuhn2019wasserstein}
Kuhn D, Esfahani PM, Nguyen VA, Shafieezadeh-Abadeh S (2019) Wasserstein distributionally robust optimization: Theory and applications in machine learning. \emph{Operations research \& management science in the age of analytics}, 130--166 (INFORMS).

\bibitem[{Kuhn et~al.(2025)Kuhn, Shafiee, \protect\BIBand{} Wiesemann}]{kuhn2024distributionally}
Kuhn D, Shafiee S, Wiesemann W (2025) Distributionally robust optimization. \emph{Acta Numerica} 34:579--804.

\bibitem[{Liu et~al.(2018)Liu, Xu, Yang, \protect\BIBand{} Zhang}]{liu2018distributionally}
Liu Y, Xu H, Yang SJS, Zhang J (2018) Distributionally robust equilibrium for continuous games: Nash and stackelberg models. \emph{European Journal of Operational Research} 265(2):631--643.

\bibitem[{Loizou(2015)}]{loizou2015distributionally}
Loizou N (2015) Distributionally robust game theory. \emph{arXiv preprint arXiv:1512.03253} .

\bibitem[{Luo et~al.(1996)Luo, Pang, \protect\BIBand{} Ralph}]{luo1996mathematical}
Luo ZQ, Pang JS, Ralph D (1996) \emph{Mathematical programs with equilibrium constraints} (Cambridge University Press).

\bibitem[{Ma et~al.(2011)Ma, Callaway, \protect\BIBand{} Hiskens}]{ma2011decentralized}
Ma Z, Callaway DS, Hiskens IA (2011) Decentralized charging control of large populations of plug-in electric vehicles. \emph{IEEE Transactions on control systems technology} 21(1):67--78.

\bibitem[{Marinacci(2000)}]{marinacci2000ambiguous}
Marinacci M (2000) Ambiguous games. \emph{Games and Economic Behavior} 31(2):191--219.

\bibitem[{Mazumdar et~al.(2025)Mazumdar, Panaganti, \protect\BIBand{} Shi}]{mazumdar2025tractable}
Mazumdar E, Panaganti K, Shi L (2025) Tractable multi-agent reinforcement learning through behavioral economics. \emph{The Thirteenth International Conference on Learning Representations}.

\bibitem[{McKelvey \protect\BIBand{} Palfrey(1995)}]{mckelvey1995quantal}
McKelvey RD, Palfrey TR (1995) Quantal response equilibria for normal form games. \emph{Games and economic behavior} 10(1):6--38.

\bibitem[{Mohajerin~Esfahani \protect\BIBand{} Kuhn(2018)}]{mohajerin2018data}
Mohajerin~Esfahani P, Kuhn D (2018) Data-driven distributionally robust optimization using the {W}asserstein metric: performance guarantees and tractable reformulations. \emph{Mathematical Programming} 171(1-2):115--166.

\bibitem[{Myerson(1978)}]{myerson1978refinements}
Myerson RB (1978) Refinements of the nash equilibrium concept. \emph{International journal of game theory} 7:73--80.

\bibitem[{Nash(1951)}]{Nash1951EquilibriaNPlayerGames}
Nash J (1951) Non-cooperative games. \emph{Annals of Mathematics} 54:286--295.

\bibitem[{Netessine \protect\BIBand{} Shumsky(2005)}]{netessine2005revenue}
Netessine S, Shumsky RA (2005) Revenue management games: Horizontal and vertical competition. \emph{Management Science} 51(5):813--831.

\bibitem[{Osborne \protect\BIBand{} Pitchik(1986)}]{osborne1986price}
Osborne MJ, Pitchik C (1986) Price competition in a capacity-constrained duopoly. \emph{Journal of Economic Theory} 38(2):238--260.

\bibitem[{Paccagnan et~al.(2018)Paccagnan, Gentile, Parise, Kamgarpour, \protect\BIBand{} Lygeros}]{paccagnan2018nash}
Paccagnan D, Gentile B, Parise F, Kamgarpour M, Lygeros J (2018) Nash and {W}ardrop equilibria in aggregative games with coupling constraints. \emph{IEEE Transactions on Automatic Control} 64(4):1373--1388.

\bibitem[{Papadimitriou et~al.(2023)Papadimitriou, Vlatakis-Gkaragkounis, \protect\BIBand{} Zampetakis}]{papadimitriou2022computational}
Papadimitriou C, Vlatakis-Gkaragkounis EV, Zampetakis M (2023) The computational complexity of multi-player concave games and kakutani fixed points. \emph{Proceedings of the 24th ACM Conference on Economics and Computation}, 1045, EC '23 (New York, NY, USA: Association for Computing Machinery).

\bibitem[{Perchet(2020)}]{perchet2020finding}
Perchet V (2020) Finding robust nash equilibria. \emph{Algorithmic Learning Theory}, 725--751 (PMLR).

\bibitem[{Qu et~al.(2017)Qu, Meng, Zhou, \protect\BIBand{} Dai}]{qu2017distributionally}
Qu S, Meng D, Zhou Y, Dai Y (2017) Distributionally robust games with an application to supply chain. \emph{Journal of Intelligent \& Fuzzy Systems} 33(5):2749--2762.

\bibitem[{Renou \protect\BIBand{} Schlag(2010)}]{renou2010minimax}
Renou L, Schlag KH (2010) Minimax regret and strategic uncertainty. \emph{Journal of Economic Theory} 145(1):264--286.

\bibitem[{Rosen(1965)}]{rosen1965existence}
Rosen JB (1965) Existence and uniqueness of equilibrium points for concave n-person games. \emph{Econometrica: Journal of the Econometric Society} 520--534.

\bibitem[{Rosenthal(1989)}]{Rosenthal1989BoundedRationalityEquilibrium}
Rosenthal RW (1989) A bounded-rationality approach to the study of noncooperative games. \emph{International Journal of Game Theory} 18:273--292.

\bibitem[{Roughgarden(2010)}]{roughgarden2010algorithmic}
Roughgarden T (2010) Algorithmic game theory. \emph{Communications of the ACM} 53(7):78--86.

\bibitem[{Santambrogio(2015)}]{santambrogio2015optimal}
Santambrogio F (2015) Optimal transport for applied mathematicians. \emph{Birk{\"a}user, NY} 55(58-63):94.

\bibitem[{Selten(1975)}]{Selten1975PerfectionConcept}
Selten R (1975) Reexamination of the perfectness concept for equilibrium points in extensive games. \emph{Journal of Game Theory'} 4:25--55.

\bibitem[{Shafieezadeh-Abadeh et~al.(2019)Shafieezadeh-Abadeh, Kuhn, \protect\BIBand{} Esfahani}]{shafieezadeh2019regularization}
Shafieezadeh-Abadeh S, Kuhn D, Esfahani PM (2019) Regularization via mass transportation. \emph{Journal of Machine Learning Research} 20(103):1--68.

\bibitem[{Sherali et~al.(1983)Sherali, Soyster, \protect\BIBand{} Murphy}]{sherali1983stackelberg}
Sherali HD, Soyster AL, Murphy FH (1983) Stackelberg-nash-cournot equilibria: characterizations and computations. \emph{Operations Research} 31(2):253--276.

\bibitem[{Stahl \protect\BIBand{} Wilson(1995)}]{Stahl1995ModelsOtherPlayers}
Stahl DO, Wilson PW (1995) On players' models of other players: Theory and experimental evidence. \emph{Games and Economic Behavior} 10:218--254.

\bibitem[{Sturmfels(2002)}]{sturmfels2002solving}
Sturmfels B (2002) \emph{Solving systems of polynomial equations}. Number~97 (American Mathematical Soc.).

\bibitem[{Sun \protect\BIBand{} Gao(2007)}]{sun2007equilibrium}
Sun LJ, Gao ZY (2007) An equilibrium model for urban transit assignment based on game theory. \emph{European Journal of Operational Research} 181(1):305--314.

\bibitem[{Sundaram(1996)}]{Sundaram1996OptimizationTheory}
Sundaram RK (1996) \emph{A First Course in Optimization Theory} (Cambridge University Press).

\bibitem[{van~der Laan et~al.(1987)van~der Laan, Talman, \protect\BIBand{} Van~der Heyden}]{van1987simplicial}
van~der Laan G, Talman A, Van~der Heyden L (1987) Simplicial variable dimension algorithms for solving the nonlinear complementarity problem on a product of unit simplices using a general labelling. \emph{Mathematics of Operations Research} 12(3):377--397.

\bibitem[{Villani(2009)}]{villani2009optimal}
Villani C (2009) \emph{Optimal transport: old and new}, volume 338 (Springer).

\bibitem[{Von~Neumann \protect\BIBand{} Morgenstern(1947)}]{von1947theory}
Von~Neumann J, Morgenstern O (1947) Theory of games and economic behavior .

\bibitem[{Yekkehkhany et~al.(2020)Yekkehkhany, Murray, \protect\BIBand{} Nagi}]{yekkehkhany2020risk}
Yekkehkhany A, Murray T, Nagi R (2020) Risk-averse equilibrium for games. \emph{arXiv preprint arXiv:2002.08414} .

\bibitem[{Zhen et~al.(2025)Zhen, Kuhn, \protect\BIBand{} Wiesemann}]{zhen2023unified}
Zhen J, Kuhn D, Wiesemann W (2025) A unified theory of robust and distributionally robust optimization via the primal-worst-equals-dual-best principle. \emph{Operations Research} 73(2):862--878.

\end{thebibliography}

\end{document}